\documentclass[11pt]{article}
\usepackage{titlesec}
\usepackage{hyperref}

\titleclass{\subsubsubsection}{straight}[\subsection]

\usepackage[utf8]{inputenc}
\usepackage{amsmath,amsfonts,amssymb,stackengine,graphicx}
\title{appendix H2}
\author{sagnotti}
\date{June 2020}

\usepackage[utf8]{inputenc}
\newif\iffigs\figstrue

\usepackage[title]{appendix}
\usepackage{epsfig,latexsym}
\usepackage{hyperref}
\usepackage{amsmath}
\usepackage{verbatim}
\usepackage{color}
\usepackage{mathrsfs}
\usepackage{slashed}
\usepackage{amssymb}

\DeclareMathAlphabet{\mathpzc}{OT1}{pzc}{m}{it}

 \csname
@addtoreset\endcsname{equation}{section}

\def\gz0{\gamma^{0}}

\def\sign{\rm sign}

\def\beq{\begin{equation}}
\def\eeq{\end{equation}}
\def\bea{\begin{eqnarray}}
\def\eea{\end{eqnarray}}
\def\ba{\begin{array}}
\def\ea{\end{array}}
\def\bec{\begin{center}}
\def\ec{\end{center}}
\def\ba{\begin{align}}
\def\ena{\end{align}}

\def\12{\frac{1}{2}}

\def\pr{\partial}

\newcounter{subsubsubsection}[subsubsection]
\renewcommand\thesubsubsubsection{\thesubsubsection.\arabic{subsubsubsection}}

\titleformat{\subsubsubsection}
  {\normalfont\normalsize\bfseries}{\thesubsubsubsection}{1em}{}
\titlespacing*{\subsubsubsection}
{0pt}{3.25ex plus 1ex minus .2ex}{1.5ex plus .2ex}

\makeatletter
\renewcommand\paragraph{\@startsection{paragraph}{5}{\z@}%
  {3.25ex \@plus1ex \@minus.2ex}%
  {-1em}%
  {\normalfont\normalsize\bfseries}}
\renewcommand\subparagraph{\@startsection{subparagraph}{6}{\parindent}%
  {3.25ex \@plus1ex \@minus .2ex}%
  {-1em}%
  {\normalfont\normalsize\bfseries}}
\def\toclevel@subsubsubsection{4}
\def\toclevel@paragraph{5}
\def\toclevel@paragraph{6}
\def\l@subsubsubsection{\@dottedtocline{4}{7em}{4em}}
\def\l@paragraph{\@dottedtocline{5}{10em}{5em}}
\def\l@subparagraph{\@dottedtocline{6}{14em}{6em}}
\makeatother

\begin{document}

\begin{flushright}
{\today}
\end{flushright}

\vspace{10pt}

\begin{center}


{\Large\sc Brane--Like Solutions and Other Non--Supersymmetric Vacua}\\


\vspace{25pt}
{\sc J.~Mourad${}^{\; a}$, S.~Raucci${}^{\; b}$  \ and \ A.~Sagnotti${}^{\; b}$\\[15pt]

${}^a$\sl\small APC, UMR 7164-CNRS, Universit\'e  Paris Cit\'e \\
10 rue Alice Domon et L\'eonie Duquet \\75205 Paris Cedex 13 \ FRANCE
\\ e-mail: {\small \it
mourad@apc.univ-paris7.fr}\vspace{10
pt}

{${}^b$\sl\small
Scuola Normale Superiore and INFN\\
Piazza dei Cavalieri, 7\\ 56126 Pisa \ ITALY \\
e-mail: {\small \it salvatore.raucci@sns.it, sagnotti@sns.it}}\vspace{10pt}
}

\vspace{40pt} {\sc\large Abstract}\end{center}
\noindent
After recasting the standard charged and uncharged brane profiles in the harmonic gauge, we explore solutions with the same isometries where the potentials $V = T \,e^{\gamma\,\phi}$ of ten--dimensional non--supersymmetric strings are taken into account. Combining a detailed catalog of the possible asymptotics with some numerical results suggests that these spherically symmetric backgrounds terminate at singularities within finite proper distances.

\setcounter{page}{1}

\pagebreak

\newpage
\tableofcontents
\newpage
\baselineskip=20pt
\section[Introduction and Summary]{\sc  Introduction and Summary}\label{sec:intro}

The breaking of supersymmetry has dramatic effects in String Theory~\cite{stringtheory}, since it typically unfolds strong back-reactions on the vacuum and consequent deformations of Minkowski space. As of today, one can only address these phenomena within the low--energy effective theory, and thus follow the fate of vacua only within regions of weak string coupling and weak curvature. On the other hand, all available information about regions where one or both preceding conditions are not satisfied is at best of qualitative value. Nonetheless, exploring different scenarios even with our limited tools is crucial to build up some intuition on phenomena that are clearly of utmost interest. The three non--supersymmetric ten--dimensional strings of~\cite{so1616,susy95,sugimoto} are useful starting points, since their spectra lack the tachyonic modes that typically emerge when supersymmetry is broken in String Theory. In these cases the back-reaction on the vacuum is dominated, at weak coupling, by runaway potentials of the type
\beq
V \ = \  T \ e^{\gamma\,\phi} \ , \label{tadpole_pot}
\eeq
which we shall often refer to as ``tadpole potentials''. In the Einstein frame, which we shall largely use in the following, $\gamma=\frac{3}{2}$ for the two orientifolds~\cite{orientifolds} of~\cite{susy95} and~\cite{sugimoto}, and $\gamma=\frac{5}{2}$ for the SO(16) $\times$ SO(16) string of~\cite{so1616}, which is a variant of the heterotic strings of~\cite{heterotic}. The models of~\cite{so1616} and~\cite{susy95} are not supersymmetric, while supersymmetry is non--linearly realized~\cite{bsb_nonlinear} in the model of~\cite{sugimoto}, which provides the simplest instance of ``brane supersymmetry breaking''~\cite{bsb}.

The vacua of~\cite{dm_vacuum}, where tadpole potentials are taken into account, have exhibited several surprising features. The most relevant of these is the emergence of spontaneous compactifications where the internal space is an interval, whose boundaries host dynamical extended objects and whose size, differently from what happens in the ordinary Kaluza--Klein setting, is determined by the strength of the potentials. While regions of strong coupling and/or strong curvature are present near the ends, the resulting spacetimes are surprisingly stable and, in the orientifold case, make long--range gravity an inevitable feature of the effective lower--dimensional Minkowski spacetime~\cite{bms_18,ms23_1}. In contrast, $AdS\times S$ compactifications~\cite{gm_02,ms_16,raucci_22} can be supported by different types of fluxes, but are unstable in the presence of tadpole potentials~\cite{bms_18,ms23_1}.

Branes and orientifolds live naturally in proper vacua of the theory, and in the companion paper~\cite{mrs24_2} we address their emergence in the effective nine--dimensional Minkowski spaces that result from the Dudas--Mourad compactifications of~\cite{dm_vacuum}. This construction, however, involves solutions that depend on at least a pair of coordinates, so that the proper setup is provided by metrics of the axisymmetric type, as in~\cite{Weyl:1917gp} and~\cite{Charmousis:2003wm}.
Here we explore simpler backgrounds depending on a single coordinate, where the internal tori discussed in~\cite{ms21_1} and~\cite{ms21_2} are replaced with spheres. They can describe compactifications with spherically symmetric internal spaces in the presence of tadpoles, or even some aspects of the more general solutions considered in~\cite{mrs24_2}, within regions where spherical symmetry dominates. The inclusion of curvature complicates the background equations, which only afford simple exact solutions in limiting cases. One presents itself when the curvature dominates over the tadpole, and can encompass the exact description of charged and uncharged black holes and branes of supersymmetric theories, while another concerns cases where the tadpole potential dominates over the curvature, and can encompass the scenarios explored in~\cite{ms21_2}. When the effects of tadpole, curvature and fluxes are combined, the preceding cases allow one to build a detailed catalog of the possible asymptotics, and thus a semi--quantitative picture of the general case.

The plan of this paper is as follows. In Section~\ref{sec:harmonic} we recall the portions of the low--energy effective theories of interest and the resulting systems of equations that determine backgrounds with internal spheres depending on a single coordinate. In Section~\ref{sec:curvature} we move on to revisit standard results on uncharged black holes and branes in the harmonic gauge, while also recasting them in the more familiar isotropic coordinates. In Section~\ref{sec:charged_branes} we revisit electrically charged $p$-branes, provide the harmonic--coordinate description of the widely studied BPS solutions and of their non--BPS deformations, and compute the corresponding tensions and charges. We also clarify some normalization issues that arose in~\cite{ms22_1} and~\cite{ms23_3} in connection with self--dual fluxes.
In Section~\ref{sec:bulk_tadpole} we come to the central theme of this work, and address the combined effects of tadpole and curvature. While the resulting system of equations does not admit simple analytic solutions, we provide evidence that the tadpole or the curvature typically dominate over one another, so that a close scrutiny of the analytic results obtained when this occurs yields a detailed catalog of allowed asymptotics, which can be linked to one another only within a subset of the possible options. The numerical studies described in Section~\ref{sec:numerics} can be understood rather convincingly in these terms. An interesting feature of the resulting spacetimes is that they close, in cigar--like shapes, within finite radial distances from the origin, as in the original nine--dimensional solutions of~\cite{dm_vacuum}. The general emergence of this feature was nicely anticipated by Antonelli and Basile in~\cite{antonelli_basile}. Section~\ref{sec:bulk_tadpole_flux} addresses the most complete setting of this type, where the effects of curvature, tadpole and flux are combined.
Aside from a handful of special (unstable) $AdS \times S$ solutions discussed in~\cite{gm_02,ms_16}~\footnote{Or in~\cite{raucci_22}, if the forms are replaced by Abelian gauge fields.} and a special case that reduces to the results of Section~\ref{sec:bulk_tadpole}, we identify a richer catalog of asymptotics that are very useful to characterize a numerical exploration that we have performed within this more general setting. As a by-product, we identify two new exact solutions for the heterotic SO(16) $\times$ SO(16) theory, which emerge when the internal spheres are replaced by tori. Finally, Appendix~\ref{app:asymptotics} contains a summary of the different pairs of asymptotics that can combine with one another in the absence of form fluxes.

\section[Action Principle and Field Equations]{\sc  Action Principle and Field Equations}\label{sec:harmonic}

In the string frame, the relevant bosonic contributions to the low-energy effective action are encoded in
 \beq
 {\cal S}_S \ = \ \frac{1}{2\kappa_{10}^2} \ \int d^{10}x\,\sqrt{-\,G} \, \Big\{ e^{-2\phi}\Big[\, \mathcal{R}\, + \,4(\partial\phi)^2 \Big] \, - \, T \, e^{\,\gamma_S\,\phi}\, - \, \frac{e^{-2\,\beta_S\,\phi}}{2\,(p+2)!}\ {\cal H}_{p+2}^2 \Big\} \ , \label{eqs1}
 \eeq
where we have omitted non--Abelian gauge fields and localized sources, and where the relevant values of $p$, $\gamma_S$ and $\beta_S$ for the three ten-dimensional non-tachyonic models can be found in Table~\ref{table:tab_1}.
\begin{table}[!h]
 \begin{center}
\begin{tabular}{ ||c||c|c|c|c|c|| }
 \hline\hline
  Model & $p$ & $\beta_S$ & $\gamma_S$ & $\beta_p$ & $\gamma$ \\ [0.5ex]
  \hline\hline
   USp(32) & $(1,5)$ & $(0,0)$ & $-1$ & $(-\frac{1}{2},\frac{1}{2})$ & $\frac{3}{2}$ \\ [0.5ex]
  \hline
  U(32) & $(-\,1,1,3,5,7)$ & $(0,0,0,0,0)$ & $-1$ & $(-1,-\frac{1}{2},0,\frac{1}{2},1)$ & $\frac{3}{2}$\\ [0.5ex]
  \hline
 SO(16) $\times$ SO(16) & $(I,V)$ & $(1,-1)$ & 0 & $(\frac{1}{2},-\frac{1}{2})$ & $\frac{5}{2}$  \\ [0.5ex]
 \hline\hline
\end{tabular}
 \end{center}
 \vskip 12pt
 \caption{\small String--frame and Einstein-frame parameters for the tachyon--free ten--dimensional string models. Roman numerals refer to NS-NS branes, entries within parentheses refer to RR ones. The opposite signs of $\beta_S$ for the fundamental string and the NS fivebrane reflect the duality between the corresponding field strengths, which inverts the dilaton coupling.}\vskip 12pt
 \label{table:tab_1}
 \end{table}

In the action of eq.~\eqref{eqs1}
\beq \label{eq:alphaprime}
\kappa_{10}^2 \ = \ \frac{(2\pi\sqrt{\alpha'})^8}{4\pi} \ ,
\eeq
and in these units the tension of a Dp brane in the type II theories is~\cite{stringtheory}
\bea \label{eq:Dp_brane_tension}
{\cal T}_p\ = \ \frac{\sqrt{\pi}}{\kappa_{10}}(2\pi\sqrt{\alpha'})^{3-p}\ = \ \frac{2\pi}{(2\pi\sqrt{\alpha'})^{p+1}} \ .
\eea
Note that the quantization condition
\bea
2\,\kappa_{10}^2\,{\cal T}_p \,{\cal T}_{6-p}\ = \ 2\,\pi
\eea
holds.
In particular, the D9 brane tension can be cast in the form
\beq
{\cal T}_9 \ = \ \frac{1}{(2\pi)^4 \left(2 \pi \alpha'\right)^5} \ , \label{T9}
\eeq
and consequently the tadpole potential of the Sugimoto model~\cite{sugimoto}, which originates from a BPS orientifold O9$^{+}$ plane and 32 anti-D9 branes, has an overall strength
\bea
\frac{T \,e^{-\phi}}{2\,\kappa_{10}^2} \ = \  64 \ {\cal T}_9 \ e^{-\phi} \ \simeq \  \frac{0.041}{\left(2 \pi\alpha'\right)^5}\ e^{-\phi} \ ,
\eea
in our units, where half of the contribution originates from the branes and the other half from the orientifold. For the type--0'B string the orientifold is tensionless, and therefore the tadpole potential originates solely from its non--BPS D-branes. The vacuum amplitudes indicate that, in this case,
\bea
\frac{T\, e^{-\phi}}{2\,\kappa_{10}^2} \ = \  \frac{64}{\sqrt{2}} \ {\cal T}_9 \ e^{-\phi} \ ,
\eea
with ${\cal T}_9$ given again by eq.~\eqref{T9}, so that the overall tension of the D-branes is reduced by a factor $\sqrt{2}$ with respect to its value for the Sugimoto model. Finally,
for the heterotic $SO(16) \times SO(16)$ string the contribution originates from the torus amplitude, and was computed in~\cite{so1616}. Its overall strength,
\beq
 \frac{0.037}{\left(2 \pi\alpha'\right)^5} \ ,
\eeq
is similar to the preceding result, but there is no accompanying power of $e^{-\phi}$ in the string frame.

In the Einstein frame, which is reached by the Weyl transformation
\beq
G_{MN} \ = \ e^{\frac{\phi}{2}} \ g_{MN} \ ,
\eeq
the action becomes
 \bea
{\cal S}_E &=& \frac{1}{2\kappa_{10}^2}\int d^{10}x \, \sqrt{-g}\left[\mathcal{R}\ - \ \frac{1}{2}\ (\partial\phi)^2\ - \ T \, e^{\,\gamma\,\phi}\ - \ \frac{e^{-2\,\beta_p\,\phi}}{2\,(p+2)!}\, {\cal H}_{p+2}^2
\right] \ ,
\eea
where the values of $\beta_p$ and $\gamma$ are still defined in Table~\ref{table:tab_1}. For the 0'B theory one must add the self--duality condition for the five--form field strength
\beq{}
{\cal H}_5 \ = \ \star\,{\cal H}_{5} \ .
\eeq{}

In some of the ensuing analysis it will prove convenient to allow for generic values of $D$, thus working with
 \bea
{\cal S}_E &=& \frac{1}{2\kappa_{D}^2}\int d^{D}x \, \sqrt{-g}\left[\mathcal{R}\ - \ \frac{4}{D-2}\ (\partial\phi)^2\ - \ T \, e^{\,\gamma\,\phi}\ - \ \frac{e^{-2\,\beta_p\,\phi}}{2\,(p+2)!}\, {\cal H}_{p+2}^2  \
\right] \ , \label{eqs4}
\eea
although this extension does not concern the critical strings of direct interest for this paper. Actions of this type are relevant for non-critical strings (for which $\gamma_S=-2$) or for lower-dimensional non-supersymmetric models. In this more general $D$-dimensional case,
\beq
\beta_p \,=\, \beta_S \ - \ \frac{D-2(p+2)}{D-2} \ , \qquad
\gamma \,=\, \gamma_S \ + \ \frac{2 D}{D-2} \ ,
 \label{alphaE}
\eeq
and we define the length $\ell$ as
\beq \label{eq:kappa_D}
\kappa_D^2 = \frac{(2\pi \ell)^{D-2}}{4\pi} \ .
\eeq
For ten--dimensional strings $\ell=\sqrt{\alpha'}$, and the preceding relation thus links $\kappa_{10}$ to the string length.

As summarized in Table~\ref{table:tab_1}, for the D-branes of the ten--dimensional orientifolds, the allowed values of $\beta_p$ in eq.~\eqref{alphaE} become
\beq
\beta_p\ = \ \frac{p-3}{4} \ ,
\eeq
while for the heterotic model they are captured by
\beq
\beta_p\ = \ \frac{3-p}{4} \ .
\eeq

The portions of the Einstein--frame equations concerning the metric tensor, a $(p+1)$-form gauge field and the dilaton read
\bea
\mathcal{R}_{MN} \,-\, \frac{1}{2}\ g_{MN}\, \mathcal{R} &=& \!\!\frac{4}{D-2}\  \pr_M\phi\, \pr_N\phi\, + \, \frac{e^{\,-\,2\,\beta_p\,\phi} }{2(p+1)!}\,  \left({\cal H}_{p+2}^2\right)_{M N}  \nonumber\\
&-& \!\!\frac{1}{2}\,g_{MN}\Big[\frac{4\,(\pr\phi)^2}{D-2}+ \frac{e^{\,-\,2\,\beta_p\,\phi}}{2(p+2)!}\,{\cal H}_{p+2}^2\,+\, V(\phi)\Big] \ ,  \nonumber
\\
\frac{8}{D-2} \ \Box\phi &=& \,-\,\frac{\beta_p\, e^{\,-\,2\,\beta_p\,\phi}}{(p+2)!} \ {\cal H}_{p+2}^2 \, +\, V^\prime(\phi)  \ , \nonumber
\\
d\Big(e^{\,-\,2\,\beta_p\,\phi}\ \star\,{\cal H}_{p+2}\Big) &=& 0  \ , \label{eqsbeta}
\eea
where
\beq \label{eq:tadpole_potential}
V(\phi) \ = \ T \, e^{\gamma\,\phi} \ ,
\eeq
with the values of $\gamma$ that we discussed above.
The contributions of the $(p+2)$-form to the equations of motion involve $\left({\cal H}_{p+2}^2\right)_{MN}$, defined as
\beq
\left({\cal H}_{p+2}^2\right)_{MN} \ = \ {\left({\cal H}_{p+2}\right)}_{M M_2 \ldots M_{p+2}}\,{\left({\cal H}_{p+2}\right)}_{N N_2 \ldots N_{p+2}} \  g^{M_2 N_2}\ldots g^{M_{p+2} N_{p+2}}\ .
\eeq
Equivalently, one can work with
\bea
{\cal R}_{MN}  &=& \frac{4}{D-2}\  \pr_M\phi\, \pr_N\phi\ + \ \frac{1}{2(p+1)!}\ e^{\,-\,2\,\beta_p\,\phi} \ \left({\cal H}_{p+2}^2\right)_{M N}\nonumber\\
&+& g_{MN}\left[\,- \, \frac{(p+1)\, e^{\,-\,2\,\beta_p\,\phi}}{2(D-2)(p+2)!}\, {\cal H}_{p+2}^2 \  +\ \frac{V(\phi)}{D-2}\right] \ . \label{eqsnotlagbeta}
\eea

\subsection[\sc Vacuum Setup and Harmonic Coordinates]{\sc Vacuum Setup and Harmonic Coordinates}

Taking the tadpole potential~\eqref{eq:tadpole_potential} into account entails some complications, since Minkowski space is no more a vacuum solution, while supersymmetry is inevitably broken. Our ultimate goal is to clarify, insofar as possible, how the $p$-branes of different string theories are affected by the tadpole potential, and to this end we shall proceed in steps.

Our analysis will rest on two key ingredients. The first, the choice of a vacuum, is well captured, in all cases of interest, by the class of backgrounds
\bea \label{eq:ansatz}
ds^{\,2} &=& e^{2A(r)}\, \gamma_{\mu\nu}(x)\, dx^\mu\,dx^\nu \ + \ e^{2B(r)}\,dr^2\ + \ e^{2C(r)}\, \ell^2\,\gamma_{mn}(\xi)\, d \xi^m\,d \xi^n  \ , \nonumber \\
\phi &=& \phi(r) \ , \nonumber \\
{\cal H}_{p+2} &=& H_{p+2}\ e^{\,2\,\beta_p\phi + B +(p+1)A-(D-p-2)C}\ \sqrt{-\gamma(x)} \ dx^0 \wedge \ldots \wedge dx^p \wedge dr \ , \nonumber \\
{\cal H}_{p+1} &=& {\mathfrak{h}}_{p+1}\, \sqrt{-\gamma(x)} \ dx^0 \wedge \ldots \wedge dx^p \ ,\label{metric_sym}
\eea
where $\gamma_{\mu\nu}(x)$ and $\gamma_{mn}(\xi)$ are Lorentzian and Euclidean metrics of maximally symmetric spaces with curvatures $k' =(0,\pm 1)$ and $k=(0,\pm1)$, and of dimensions $(p+1)$ and $(D-p-2)$.
We have included the two electric form-field profiles, which are compatible with the isometries of the metric and solve the corresponding equations of motion with constant values of $H_{p+2}$ and $\mathfrak{h}_{p+1}$. We shall mainly concentrate on the first type of contribution, ${\cal H}_{p+2}$, because it is the proper one for charged branes. Magnetic profiles can be obtained from these by electric--magnetic duality.

This setup can encompass a number of vacuum configurations in the presence of the tadpole potential~\eqref{eq:tadpole_potential}: the original Dudas--Mourad vacuum of~\cite{dm_vacuum}, where $\gamma_{mn}$ is absent altogether and $k'=0$, and also some generalizations, including the vacua with internal tori considered in~\cite{ms21_1,ms21_2}, but also the $AdS \times S$ vacua of~\cite{gm_02,ms_16,raucci_22}, where $k=1$. The second ingredient is the inclusion of the extended object. When $k=0$ and the internal space is a sphere, the setup of eqs.~\eqref{metric_sym} suffices to encompass BPS $p$-branes and orientifolds in the Minkowski vacua of supersymmetric strings, where the tadpole potential is absent, but also some interesting deformations of them whose role emerged long ago in two--dimensional Conformal Field Theory~\cite{dms_cft}. All these solutions have $ISO(1,p) \times SO(D-p-1)$ isometry groups, and in all these cases $r$ parametrizes the distance from the brane. When this distance is large, all these solutions ought to approach the Minkowski vacuum, but as we shall see this is not generally the case.

The description of branes and orientifolds in the presence of the tadpole potential~\eqref{eq:tadpole_potential} requires a more general ansatz, which is the main subject of~\cite{mrs24_2}. Once one demands that brane solutions approach the vacuum at large distances, the proper setup is provided by metrics of the axisymmetric type, as in~\cite{Weyl:1917gp} and~\cite{Charmousis:2003wm}.
In this work we shall focus on the simpler and more symmetrical background ansatz in eqs.~\eqref{metric_sym}. With appropriate boundary conditions, this can capture branes without tadpoles, vacua with tadpoles or even some limiting behaviors of branes with tadpoles, in regions where the axisymmetry enhances to a full spherical symmetry.

The present setting entails some complications with respect to the vacuum solutions considered in~\cite{ms21_1,ms21_2}. The internal manifold is now curved, and in the following the internal metrics $\gamma_{mn}$ will correspond to spheres $S^{D-p-2}$. We shall concentrate on this option, leaving for future work the exploration of the  hyperbolic internal spaces, which would correspond to $k'=-1$, of curved spacetime sections and of corresponding cosmologies.

In the cases of interest that we have spelled out, the resulting equations are
{\small
 \bea
 A'' + A'\, F' \!\!\!&=& - \ \frac{T}{(D-2)} \ e^{2\,B\,+\,\gamma\,\phi}\ +\ \frac{k\,p}{\ell^2}\ e^{2(B-A)}  \nonumber \\
  &+& \!\!\!\frac{(D-p-3)}{2\,(D-2)} \ e^{2\,B\,+\,2\,\beta_p\,\phi\,-\,2(D-p-2)C } H_{p+2}^2 + \frac{(D-p-2)}{2\,(D-2)} \ e^{2\,B\,-\,2\,\beta_{p-1}\,\phi\,-\,2 (p+1) A} \mathfrak{h}_{p+1}^2  \nonumber \, ,  \\
C'' +  C '\, F'\!\!\!&=& - \ \frac{T}{(D-2)} \ e^{2\,B\,+\,\gamma\,\phi}\ +\ \frac{k'(D-p-3)}{\ell^2}\ e^{2(B-C)} \nonumber \\
   &-& \!\!\!\frac{(p+1)}{2\,(D-2)}\ e^{2\,B\,+\,2\,\beta_p\,\phi\,-\,2(D-p-2)C }  H_{p+2}^2 \,-\, \frac{p}{2\,(D-2)}\ e^{2\,B\,-\,2\,\beta_{p-1}\,\phi\,-\,2 (p+1) A} \mathfrak{h}_{p+1}^2 \, ,  \nonumber  \\
   \phi'' +  \phi'\, F' \!\!\!&=& \frac{T\,\gamma\,(D-2)}{8}\ e^{2\,B\,+\,\gamma\,\phi} \label{Eqs_back} \\ &+& \!\!\!\frac{\beta_p\,(D-2)}{8}\ e^{2\,B\,+\,2\,\beta_p\,\phi\,-\,2(D-p-2)C } H_{p+2}^2 \,+\, \frac{\beta_{p-1}\,(D-2)}{8}\ e^{2\,B\,-\,2\,\beta_{p-1}\,\phi\,-\,2 (p+1) A} \mathfrak{h}_{p+1}^2\, , \nonumber
  \eea
  }
  where
  \beq
F\ =\ (p+1)A\ -\ B\ +\ (D-p-2)C \ . \label{eqs6}
\eeq
``Initial'' conditions, or more properly boundary values, cannot be given independently for all fields, since the allowed choices must satisfy the ``Hamiltonian constraint''
 \begin{align}
&(p+1)A'[p\,A' \,+\, (D-p-2)C']\,+\, (D-p-2)C'[(D-p-3)C'+(p+1)A'] \nonumber \\
 &- \, \frac{4\,(\phi')^2}{D-2} \, + \, {T} \, e^{\, 2\,B\,+\,\gamma\,\phi}\, - \, \frac{k\,p(p+1)}{\ell^2}\ e^{2(B-A)}\, - \, \frac{k'(D-p-3)(D-p-2)}{\ell^2}\ e^{2(B-C)} \nonumber \\
 & + \, \frac{1}{2}\, e^{\,2\,\beta_p\,\phi\,+\,2\,B\,-\,2\,(D-p-2)\,C} \ H_{p+2}^2 \,-\, \frac{1}{2}\, e^{\,-\,2\,\beta_{p-1}\,\phi\,-\,2(p+1)A\,+\,2\,B} \ \mathfrak{h}_{p+1}^2  \,= \, 0 \ . \label{EqB_red}
 \end{align}

The ``harmonic gauge''
\beq
B \ = \ (p+1) A \ + \ (D-p-2) C  \label{harm_gauge}
\eeq
simplifies somewhat the preceding expressions, and we shall often use it in the following to build the solutions. However, it will prove helpful and instructive to compare the results with those obtained in the more familiar ``isotropic gauge'', in which
\beq
e^{2C(r)} \ = \ \frac{r^2}{\ell^2} \ e^{2B(r)} \ ,
\eeq
which is the standard choice for supersymmetric backgrounds with spherical symmetry.

Note that, in this fashion, the system possesses an interesting discrete symmetry: its equations are left invariant by the redefinitions
\bea
&& \left[A,C,\,p,\,k,k'\right] \ \longleftrightarrow \ \left[C,A,\,D-p-3,\,k',k \right] \ , \nonumber \\
&& \left[H_{p+2}^2,\beta_p;\mathfrak{h}_{p+1}^2,\beta_{p-1} \right] \ \longleftrightarrow \ \left[-\mathfrak{h}_{p+1}^2,- \beta_{p-1};- H_{p+2}^2,-\beta_p  \right] \label{sym_AC}
\ . \eea
This can be regarded as implementing a sort of ``electric-magnetic'' duality, and will be useful when exploring more general backgrounds with curvature also in the spacetime portion of the metric.

\section[Exact Solutions with Vanishing Charges]{\sc Exact Solutions with Vanishing Charges}\label{sec:curvature}

We can now examine the simplest exact solutions of the system~\eqref{Eqs_back} that are of interest for the present analysis of black holes and branes. They are uncharged and include the non--BPS branes of the type II theory.

The solutions belonging to this class obtain for $T=0$, $k=0$, $k'=1$, $h_{p+1}=0$ and $H_{p+2}=0$. The two cases $p=D-3$ and $p=D-2$ are degenerate, since the curvature does not play a role in them, and reduce to the solutions discussed in~\cite{ms21_1,ms21_2}. However, for $p<D-3$ the internal curvature does play a role, and in the harmonic gauge~\eqref{harm_gauge} the system implies that
\beq \label{eq:uncharged_most_general_case}
A \ = \  A_0 \,+\,A_1\,r \ , \qquad
\phi \ = \  \phi_0 \,+\, \phi_1\,r \ ,
\eeq
where $A_0$, $A_1$, $\phi_0$ and $\phi_1$ are constants.
Letting
\beq
X \ = \ (p+1) A \,+\, (D-p-3) C \ ,
\eeq
the remaining equation can be simply turned into
\beq
\left(X'\right)^2 \ = \  \frac{(D-p-3)^2}{\ell^2} \ e^{2X} \ + \ {\cal E}_x \ , \label{eq:uncharged_X_integrated_diffeq}
\eeq
where ${\cal E}_x$ is an integration constant. The Hamiltonian constraint~\eqref{EqB_red} reduces to
\beq
A_1^2 \ \frac{(p+1)(D-2)}{D-p-3} \ + \ \frac{4\,\phi_1^2}{D-2} \ = \ {\cal E}_x \ \frac{D-p-2}{D-p-3} \ , \label{ham_red}
\eeq
so that ${\cal E}_x \geq 0$, and one can set
\beq
{\cal E}_x \ = \ \frac{1}{\sigma^2} \ , \qquad A_1 \ = \ \sqrt{\frac{D-p-2}{(p+1)(D-2)}}\ \frac{\cos\alpha}{\sigma} \ , \qquad \phi_1 \ = \ \sqrt{\frac{(D-p-2)(D-2)}{4(D-p-3)}}\ \frac{\sin\alpha}{\sigma} \ , \label{ExA1phi1}
\eeq
with $\alpha$ an angular parameter. Eq.~\eqref{eq:uncharged_harmonic} is then solved by~\footnote{As discussed in~\cite{ms21_1}, the second--order differential equation for $X$ that leads to eq.~\eqref{eq:uncharged_X_integrated_diffeq} admits three types of solutions. One corresponds to eq.~\eqref{eq:uncharged_X_integrated_diffeq}, another rests on trigonometric functions and is obtained letting $\frac{1}{\sigma^2}\to -\frac{1}{\sigma^2}$, and finally a third one is recovered in the limit $\frac{1}{\sigma^2}\to 0$. In this case, the Hamiltonian constraint excludes the second option, while for the last one it implies the two conditions $A_1=\phi_1=0$, so that the end result is a flat spacetime. The resulting metric is actually the $r\to 0$ limit of the solutions obtained starting from eq.~\eqref{eq:uncharged_X_integrated_diffeq}, which captures the asymptotically flat space far away from the branes, as we shall see in the following.}
\beq
X \ = \ - \ \log\left[ \frac{\left(D-p-3\right)\sigma}{\ell}\ \sinh\left(\frac{r+r_0}{\sigma}\right)\right] \ ,
\eeq
where $r_0$ is an integration constant, which can be removed by shifting the $r$ variable.

The solution for $C$ can thus be cast in the form
\beq
C\ = \ - \ \frac{1}{D-p-3} \ \log\left[ \sigma \ \frac{D-p-3}{ \ell} \ \sinh\left({\frac{r}{\sigma}}\right) \ \right] - \ \frac{p+1}{D-p-3} (A_0 + A_1 r) \ , \label{C_bb}
\eeq
and in the harmonic gauge one can conclude that
\beq
B \ = \ - \ \frac{D-p-2}{D-p-3} \ \log\left[ \sigma \ \frac{D-p-3}{\ell} \ \sinh\left({\frac{r}{\sigma}}\right) \ \right] - \ \frac{p+1}{D-p-3} (A_0 + A_1 r) \ . \label{B_bb}
\eeq
This class of solutions without fluxes depends on four parameters: $\alpha$, $\sigma$, $\phi_0$ and $A_0$. The last one, however, can be simply removed rescaling the spacetime coordinates $x$, while also rescaling $r$ and $\sigma$ by the same factor. In detail, the relevant rescaling is
\beq
r \ \to \ \Lambda\,r \ , \qquad \sigma \ \to \ \Lambda \, \sigma \ ,
\eeq
so that the combinations $A_1\,r$ and $\phi_1\,r$ are unaffected, while
\beq
B \ \to \ B \ -\  \frac{\log\Lambda}{D-p-3} \ , \qquad C \ \to \ C \ - \ \frac{\log\Lambda}{D-p-3} \ .
\eeq
This transformation can remove $A_0$ from $B$ and $C$, and then rescaling the spacetime coordinates $x$ removes it completely from the metric, while still remaining in the harmonic gauge. Equivalently, one can simply perform the shifts
\beq
A(r) \to A(r) \,-\, \frac{D-p-3}{p+1}\lambda \ , \qquad B(r)\to B(r) \, + \, \lambda \ , \qquad C(r)\to C(r) \,+\, \lambda \ ,
\eeq
without rescaling $r$, $\sigma$ and the spacetime coordinates. In both cases, one is left with three independent parameters in the metric tensor, which reads
\bea
ds^2 &=&  \ e^{-\frac{2\,r}{R}}dx_{p+1}^2 \nonumber \ + \ e^{\frac{2(p+1)\,r}{\left(D-p-3\right)R}} \left[\frac{(D-p-3)\sigma}{\ell} \,\sinh{\left(\frac{r}{\sigma}\right)}\right]^{-\,\frac{2\left(D-p-2\right)}{D-p-3}} dr^2 \nonumber \\ &+& e^{\frac{2(p+1)\,r}{\left(D-p-3\right)R}}\left[ \frac{(D-p-3)\sigma}{\ell} \,\sinh{\left(\frac{r}{\sigma}\right)}  \right]^{-\,\frac{2}{D-p-3}} \ell^2\, d\Omega_{D-p-2}^2 \ , \label{eq:uncharged_harmonic}
\eea
where
\beq
R \ = \ - \ \sqrt{\frac{(p+1)(D-2)}{D-p-2}}\frac{\sigma}{\cos\alpha} \ .  \label{Rsigmaalpha}
\eeq
Near $r=0$, this reduces to the flat metric in the harmonic coordinates,
\beq
ds^2 \ \sim \  \ dx_{p+1}^2 \ + \ \left[\frac{(D-p-3)\,r}{\ell} \right]^{-\,\frac{2\left(D-p-2\right)}{D-p-3}} dr^2 + \left[ \frac{(D-p-3)\,r}{\ell}  \right]^{-\,\frac{2}{D-p-3}} \ell^2 \, d\Omega_{D-p-2}^2 \ , \label{flat_limit}
\eeq
which can be turned into the Minkowski metric
\beq
ds^2 \ \sim \  \ dx_{p+1}^2 \ + \ d\rho^2 \ + \ \rho^2\, d\Omega_{D-p-2}^2
\eeq
by defining a new polar variable $\rho$ as
\beq \label{eq:isotropic_coordinates_uncharged}
{\rho} \ = \ \ell\left[\frac{(D-p-3)}{\ell} \,r\right]^{-\,\frac{1}{D-p-3}}  \ .
\eeq

We can now take a closer look at a few interesting cases, in order to compare the results obtained in the harmonic gauge with those emerging in more familiar coordinate systems.

\subsection[\sc The Schwarzschild--Tangherlini Solutions]{\sc The Schwarzschild--Tangherlini Solutions}
For $p = 0$ and $D=4$, the preceding results ought to describe an uncharged spherical black hole if one decouples the dilaton, setting $\phi_1=0$. This choice would translate into the two options $\alpha=0,\pi$ in eq.~\eqref{ExA1phi1}. Eq.~\eqref{Rsigmaalpha} then implies that
\beq
R \ = \ - \ \epsilon \ \sigma \ ,
\eeq
where $\epsilon=\pm 1$, and in these harmonic coordinates the metric reads
\beq
ds^2 \ = \ - \ e^{-\frac{2\,r}{R}}dt^2\ + \ e^{\frac{2\,r}{R}} \left[\frac{R}{\ell} \sinh{\left(\frac{r}{R}\right)}\right]^{-4}dr^2\ + \ e^{\frac{2\,r}{R}} \left[\frac{R}{\ell} \sinh{\left(\frac{r}{R}\right)}\right]^{-2} \ell^2 \ d\Omega_2^2 \ .
\eeq
One can now work in the region $r>0$. There is a coordinate singularity as $r \to 0$ where, as we have seen, the solution approaches flat space, which lies at infinite distance from finite values of $r$, and there is another coordinate singularity as $r \to +\,\infty$. With positive values of $R$, the time-time component of the metric tensor vanishes as $r \to +\infty$, as is the case at the Schwarzschild horizon. In fact, the transition to the conventional Schwarzschild coordinates can be effected letting
\beq
e^{-\frac{2\,r}{R}} \ =\ 1\ + \ \epsilon \ \frac{\xi_0}{\xi} \ , \label{harm_vs_sch}
\eeq
where $0< \xi < +\infty$ if $\epsilon=1$, and $\xi_0< \xi < +\infty$ if $\epsilon=-1$, with
\beq
\xi_0\ = \  \frac{2\ell^2}{\sigma} \ .
\eeq
In this fashion, for $\epsilon=-1$ one recovers the familiar expression
\beq
ds^2 \ = \ - \left(1 \ - \ \frac{\xi_0}{\xi}\right) \ dt^2 \ + \ \frac{d\xi^2}{1 \ - \ \frac{\xi_0}{\xi}} \ + \ \xi^2\,d\Omega_2^{\,2} \ ,
\eeq
and the black--hole mass $M_{BH}$ is related to the integration constant $R$ according to
\beq
M_{BH} \ =  \ \frac{\ell^2}{R\,G_N} \  = \ \frac{8\pi \ell^2}{\kappa_4^2 \, R} \ ,
\eeq
since in four dimensions
\beq
\frac{1}{16\,\pi\,G_N} \ = \ \frac{1}{2\,\kappa_4{}^2} \ .
\eeq

When $R>0$, the mass $M_{BH}$ is positive and the harmonic coordinates only span the region outside the horizon, which lies at $\xi=\xi_0$.
On the other hand, negative values of $R$ would translate into negative masses, and thus into the presence of naked singularities, which are unacceptable for these black holes.

Confining the attention to positive masses, and thus to positive values of $R$, a similar procedure leads to the Schwarzschild-Tangherlini solutions,
\beq
ds^2 \ = \ - \left[1 \ - \ \left(\frac{\xi_0}{\xi}\right)^{D-3}\right] \ dt^2 \ + \ \frac{d\xi^2}{1 \ - \ \left(\frac{\xi_0}{\xi}\right)^{D-3}} \ + \ \xi^2\,d\Omega_{D-2}^{2} \ ,
\eeq
which describe spherical black holes in generic dimensions $D>3$, starting from the harmonic-gauge metrics
\bea
ds^2 &=& - \ e^{-\frac{2\,r}{R}}dt^2 \nonumber \ + \ e^{ \frac{2\,r}{\left(D-3\right)R}} \left[\frac{(D-3)R}{\ell} \sinh{\left(\frac{r}{R}\right)}\right]^{-\,\frac{2\left(D-2\right)}{D-3}} dr^2 \nonumber \\ &+& e^{ \frac{2\,r}{\left(D-3\right)R}} \left[\frac{(D-3)R}{\ell} \sinh{\left( \frac{r }{R}\right)}\right]^{-\,\frac{2}{D-3}} \ell^2 \, d\Omega_{D-2}^2 \ .
\eea
In these cases, the link between the coordinate systems is
\beq
e^{-\frac{2 r}{R}}\ =\ 1\ - \ 2\frac{\ell}{R}\left(\frac{\xi}{\ell}\right)^{-(D-3)} \ ,
\eeq
and there is again a horizon, which now lies at
\beq
\xi_0^{D-3}\ =\ \frac{2}{R}\, \ell^{D-2} \ ,
\eeq
and is approached as $r\to +\infty$ in harmonic coordinates. Also in this case, $R$ is related to the mass, but now according to
\beq
M_{BH}\ =\   \frac{(D-2) \, \Omega_{D-2} \, \ell^{D-2}}{(D-3) \, \kappa_D^2 \, R}  \ ,
\eeq
with
\beq
\Omega_n \ = \ \frac{2\,\pi^\frac{n+1}{2}}{\Gamma\left(\frac{n+1}{2}\right)} \ .
\eeq
the area of an $n$-sphere of unit radius. Using eq.~\eqref{eq:kappa_D}, we obtain
\beq
M_{BH}\ =\ \frac{2^{5-D} \, \pi^{\frac{5-D}{2}} \, (D-2)}{(D-3) \ \Gamma\left(\frac{D-1}{2}\right) \ R} \ .
\eeq
Negative values of $R$ bring along, again, naked singularities.

Isotropic coordinates are also of interest for black holes. Let us take a close look at them, since they are the standard choice for BPS branes. In this case the gauge condition is
\beq
e^C \ = \ \rho \ e^B \ ,
\eeq
where $\rho \geq 0$ is the radial variable, and the coordinate transformation linking the Schwarzschild coordinate $\xi$ to the isotropic coordinate $\rho$ is
\beq
1-\left(\frac{\xi_0}{\xi}\right)^{D-3} = \left[\frac{\left(\frac{\rho}{\rho_0}\right)^\frac{D-3}{2} \ -\ \left(\frac{\rho}{\rho_0}\right)^{\,-\frac{D-3}{2}}}{\left(\frac{\rho}{\rho_0}\right)^\frac{D-3}{2} \ + \ \left(\frac{\rho}{\rho_0}\right)^{\,-\frac{D-3}{2}}}\right]^2 \ .  \label{schw_iso}
\eeq
In isotropic coordinates, the Schwarzschild--Tangherlini metric thus takes the form
\bea
ds^2 &=& - \ \left[\frac{\left(\frac{\rho}{\rho_0}\right)^\frac{D-3}{2} \ -\ \left(\frac{\rho}{\rho_0}\right)^{\,-\frac{D-3}{2}}}{\left(\frac{\rho}{\rho_0}\right)^\frac{D-3}{2} \ + \ \left(\frac{\rho}{\rho_0}\right)^{\,-\frac{D-3}{2}}}\right]^2 dt^2 \nonumber \\&+& \left[\frac{\left(\frac{\rho}{\rho_0}\right)^\frac{D-3}{2} \ + \ \left(\frac{\rho}{\rho_0}\right)^{\,-\frac{D-3}{2}}}{2}\right]^{\frac{4}{D-3}} \left(\frac{\rho_0}{\rho}\right)^2 \left(d\rho^2 \ + \ \rho^2 d\,\Omega_{D-2}^2\right) \ .
\eea
Note that the transformation~\eqref{schw_iso} is a one-to-one map between the two intervals $\xi_0<\xi<\infty$ and $\rho_0 < \rho < \infty$.

\subsection[\sc Properties of Uncharged p-Branes]{\sc Properties of Uncharged $p$-Branes}

We can now move on to uncharged $p$-branes, which generalize the black--hole solutions just described and have a manifest $ISO(1,p)$ isometry. In the harmonic gauge, the starting point for their characterization is provided by the ansatz of eq.~\eqref{metric_sym}, which leads to eq.~\eqref{eq:uncharged_harmonic}.
The Hamiltonian constraint, which was solved in eq.~\eqref{ExA1phi1} in terms of the angle $\alpha$, leads to dilaton profiles that are generally linear in the $r$ coordinate
\beq
\phi \ = \ \phi_0  \ + \ \phi_1 r  \ .
\eeq

As $r \to 0$ the metric behaves as
\beq
ds^2 \ \sim \ \left(1 \ - \ \frac{2\,r}{R}\right) dx_{p+1}^2 \ + \ d\rho^2 \ + \ \rho^2 \, d\Omega_{D-p-2}^2 \ ,
\eeq
where $\rho$ and $r$ are related in eq.~\eqref{eq:isotropic_coordinates_uncharged}.
The correction that we retained suffices to determine the tension of the extended object. In general, if
\beq
ds^2 \ \sim \ \Big[1 \ + V\left(\rho\right)\Big] dx_{p+1}^2 \ + \ d\rho^2 \ + \rho^2 \, d\Omega_{D-p-2}^2 \ ,
\eeq
the linearized Einstein equations with a localized source would reduce to the Poisson equation for $V$,
\beq
\nabla_{\rho}^2\, V \ = \ 2\,\kappa_{D}^2\,\frac{D-p-3}{D-2} \ {\cal T}_p \, \delta(\vec{\rho}) \ ,
\eeq
where $\delta(\vec{\rho})$ is unambiguously defined via the Cartesian coordinates $y^i$ such that, asymptotically,
\beq
d\rho^2 \ + \ \rho^2 \, d\Omega_{D-p-2}^2 \ = \ \sum_{i=1}^{D-p-1} \left(dy^i\right)^2 \ . \label{deltarho}
\eeq
Information on the source is indeed available far from the singularity via Gauss's law, which translates into the condition
\beq
\Omega_{D-p-2}\, \rho^{D-p-2} \, \partial_{\rho} V \ = \ 2\,\kappa_{D}^2\,\frac{D-p-3}{D-2} \ {\cal T}_p \ . \label{V_tension}
\eeq
In the present case, the asymptotic link between $r$ and $\rho$ in eq.~\eqref{eq:isotropic_coordinates_uncharged} and the expression for the potential in terms of $r$ lead to
\beq
V \ = \ - \ \frac{2\,\ell}{R} \ \frac{1}{D-p-3}\ \left( \frac{\rho}{\ell}\right)^{-(D-p-3)} \ ,
\eeq
and thus determine the effective tension
\beq
{\cal T}_p \ = \ \frac{D-2}{D-p-3} \, \frac{   \Omega_{D-p-2} }{  \kappa_D^2 }  \frac{\ell^{D-p-2}}{R} \ .
\eeq
For non--vanishing values of $p$ and for $R>0$ ($R<0$) one is thus describing an uncharged extended object of positive (negative) tension.

On the other hand, the dominant behavior as $r\to\infty$,
\beq
A \ \sim \ - \ \frac{r}{R} \ , \qquad
C \ \sim  \ \frac{1}{D-p-3}\left[\frac{\left(p+1\right)r}{R} \ - \ \frac{r}{\sigma}\right]  \ , \
\eeq
determines the limiting form of the metric,
\beq \label{eq:uncharged_near_singularity}
ds^2 \ \sim \ e^{-\frac{2\,r}{R}} \,dx_{p+1}^2 \ + e^{\frac{2\,r}{D-p-3}\left(\frac{p+1}{R} \ - \ \frac{D-p-2}{\sigma}\right)} dr^2 \ + \ e^{\frac{2\,r}{D-p-3}\left(\frac{p+1}{R} \ - \ \frac{1}{\sigma}\right)} \, \ell^2 \, d\Omega_{D-p-2}^2 \ ,
\eeq
while the dilaton tends to $+\infty$ if $\phi_1>0$ and to $-\,\infty$ if $\phi_1<0$.
Letting
\beq
\Gamma=\frac{1}{D-p-3}\left(- \, \frac{p+1}{R}+\frac{D-p-2}{\sigma}\right) \ ,
\eeq
one can define the proper distance from the singularity as
\beq
\zeta=\frac{1}{\Gamma}\ e^{-\Gamma r} \ ,
\eeq
and taking eq.~\eqref{Rsigmaalpha} into account, one can also conclude that
\beq
\Gamma>0 \ ,
\eeq
irrespective of the sign of $R$. In all cases, $r\to\infty$ thus lies at a finite distance from finite non--vanishing values of $r$.
In terms of the proper distance $\zeta$ the limiting behavior of the metric is captured by
\beq
ds^2 \ \sim \ \left(\Gamma\zeta\right)^{\frac{2}{R \Gamma}} \,dx_{p+1}^2 \ + d\zeta^2 \ + \ \left(\Gamma\zeta\right)^{\frac{2}{(D-p-3)\Gamma} \left(-\,\frac{p+1}{R} \ + \ \frac{1}{\sigma}\right)} \, \ell^2 \, d\Omega_{D-p-2}^2 \ ,
\eeq
or alternatively by
\beq
ds^2 \ \sim \ \left(\Gamma\zeta\right)^{\frac{2}{R \Gamma}} \,dx_{p+1}^2 \ + d\zeta^2 \ + \ \left(\Gamma\zeta\right)^{2\,-\,\frac{2}{\Gamma\sigma}} \,\ell^2 \, d\Omega_{D-p-2}^2 \ . \label{dsalpha}
\eeq
Note that, as $\zeta \to 0$, the spheres in the directions orthogonal to the branes become infinitely large compared to their flat--space counterparts. For this reason, close to the branes the internal curvature is subleading, and these solutions approach those discussed in~\cite{ms21_1} for the case of an internal torus.
A curvature singularity is present as $r\,\to \,+\,\infty$, or $\zeta\to 0$, unless $p=0$ and $\alpha=\pi$. This is due to the fact that $\Gamma$ is always positive, since the equations of motion imply that
\beq
{\cal R}\propto (\partial \phi)^2 \propto \left(\sin\alpha\right)^2 e^{2\Gamma r} \ ,
\eeq
which diverges as $r\to\infty$, unless $\alpha=0$ or $\pi$. However, a direct computation shows that if $\alpha=0$ the scalar combination $R^{MNPQ}\,R_{MNPQ}$ diverges for any value of $p$, so that the only singularity--free case corresponds to $p=0$ and $\alpha=\pi$, and thus to the black holes with positive mass described in the previous section. In this case the limit $r\to\infty$ identifies the horizon, which shields the true singularity.
However, the presence of naked singularities is a generic feature for branes when only the two--derivative Einstein action is taken into account, which signals the need for higher--derivative corrections in order to provide a satisfactory description of these extended objects.

There are special uncharged solutions with $\phi_1=0$, and thus with a constant dilaton, which possess the attractive feature of reducing to flat spacetimes when the tension ${\cal T}_p$ vanishes. This option can be neatly characterized referring to the Noether current for the shift symmetry $\phi \to \phi \ + \ c$, where $c$ is a constant, which has only a radial component in the background,
\beq
j_r \ = \ \frac{8}{D-2} \  \partial_r \phi \ = \ \frac{8}{D-2}\  \phi_1 \ .
\eeq
A non-vanishing component of this current signals a charge flow between $r\to\infty$, the singularity, and $r=0$, where the background approaches flat space far away from the extended object, which should not occur for branes embedded in supersymmetric spacetimes. For the two classes of solutions with $\phi_1\neq 0$, the scalar profile is always singular as $r \to \infty$.

Before concluding this analysis, it is instructive to recast the metric in isotropic coordinates in the whole spacetime, performing the change of variables
\beq
\frac{2\left(D-p-3\right)\sigma}{\ell} \,\tanh\left(\frac{r}{2\,\sigma}\right) \ = \ \left(\frac{\ell}{\rho}\right)^{D-p-3} \ .
\eeq
The flat limiting behavior at large distances from the extended object is now recovered as $\rho \to \infty$, where the preceding equation reduces to eq.~\eqref{eq:isotropic_coordinates_uncharged}.
Letting now
\beq \label{eq:vrho}
v(\rho)\ = \ \tanh\left(\frac{r}{2 \, \sigma }\right)\ = \ \frac{\ell}{2(D-p-3) \sigma} \left(\frac{\ell}{\rho}\right)^{D-p-3} \  ,
\eeq
the metric takes the form
\bea \label{eq:uncharged_isotropic}
ds^2 &=& \left[\frac{1+v(\rho)}{1-v(\rho)}\right]^{-\frac{2\, \sigma}{R}} dx_{p+1}^2 \nonumber \\
&+& \left[\frac{1+v(\rho)}{1-v(\rho)}\right]^{\frac{2\,\sigma \left(p+1\right)}{R\left(D-p-3\right)}} \left[1\,-\,v^2(\rho)\right]^{\frac{2}{D-p-3}} \left(d\rho^2+ \rho^2  d\Omega_{D-p-2}^2\right) ,
\eea
while the dilaton profile reads
\beq
\phi\ =  \ \phi_0 \ + \ \phi_1 \,\sigma \log\left[\frac{1\,+\,v(\rho)}{1\,-\,v(\rho)}\right] \ .
\eeq

This type of solution was already discussed in the literature in~\cite{zhou}, and was reconsidered, in~\cite{brax}, under the spell of Sen's construction of non--BPS branes~\cite{sen}. However, the emphasis placed on this correspondence led the authors to focus on combinations of branes and anti--branes in equal numbers, to which they associated uncharged branes, without considering other options.

All the preceding results refer to the range $p < D-3$, where the internal space can be curved. In the complementary range, the internal space is flat, and the solutions reduce everywhere to those considered in~\cite{ms21_1}.

\section[Electrically Charged p-Branes]{\sc Electrically Charged $p$-Branes}  \label{sec:charged_branes}

We can now see how, in the absence of a tadpole potential, the harmonic gauge adapts itself to the description of charged branes.

\subsection[\sc Charged p-Branes]{\sc Charged $p$-Branes}

In the absence of form profiles the dilaton can be decoupled, and the metric equations in the Einstein frame reduce to those of General Relativity, as was the case in the previous section. However, the dilaton enters the kinetic terms of form fields, and therefore it plays a non--trivial role for charged branes, to which we now turn. Consequently, the corresponding solutions are not simple generalizations of the Reissner-Nordstrom black hole.

We shall continue to focus on metrics with $ISO(1,p)$ isometries, along the lines of the CFT analysis in~\cite{dms_cft}. As in the preceding sections, harmonic coordinates allow for a unified treatment of different systems where $H_{p+2} \neq 0$ but $\mathfrak{h}_{p-1}=0$. The equations with only $\mathfrak{h}_{p-1}\neq0$ but $H_{p+2}=0$ are more complicated and less relevant to the current analysis of branes, so that we shall confine our attention to the first option.

In special cases, the solutions belonging to this class preserve a number of supersymmetries. An extensive review of supersymmetric brane solutions, with details on their emergence in the literature, can be found in~\cite{duff_review}. BPS branes can be obtained more simply by solving first--order equations~\cite{Dabholkar:1990yf}, but working in the harmonic gauge we shall be able to recover these supersymmetric solutions, while also capturing some non--supersymmetric deformations.

In the conventions of eqs.~\eqref{Eqs_back}, it is now convenient to introduce the three combinations
\bea
X &=& (p+1)A\ +\ (D-p-3)C \ , \nonumber \\
Y &=& (p+1)A\ +\ \beta_p \phi \ ,  \nonumber \\
Z &=& \frac{D-2}{4}\beta_p A \ - \ \frac{D-p-3}{D-2}\phi \ , \label{eqsXYZ}
\eea
which correspond to the three exponents present in the original system~\eqref{Eqs_back}.
The metric functions $A$, $B$ and $C$ and the dilaton $\phi$ are then determined by $X$, $Y$ and $Z$ according to
\bea
A \ &=& \ \frac{4  \left(D-p-3\right)}{\Delta} \ Y +\ \frac{4  \left(D-2\right)\beta_p}{\Delta} \ Z \ , \\
B \ &=& \ \frac{D-p-2}{D-p-3}\ X  \ - \  \frac{4\left(p+1\right) }{\Delta} \ Y - \frac{4(D-2)(p+1)\beta_p}{(D-p-3)\Delta} \ Z \ , \nonumber \\
C  \ &=& \   \frac{1}{D-p-3}\  X \ - \ \frac{4\left(p+1\right) }{\Delta}\  Y\ -\frac{4(D-2)(p+1)\beta_p}{(D-p-3)\Delta} \ Z , \nonumber \\
\phi \ &=& \  \frac{{(D-2)^2 \beta_p} }{\Delta} \  Y  - \frac{4(D-2)(p+1)}{\Delta} \ Z \ ,\nonumber
\eea
where
\beq
\Delta \ = \ 4(D-p-3)(p\,+\,1) \ +\  (D-2)^2 \beta_p{}^2  \ .
\eeq
The equations for $X$ and $Y$ decouple and become
\beq
X''\ = \  \frac{(D-p-3)^2}{\ell^2} \ e^{2X} \ , \qquad
Y''\ = \  \frac{\Delta}{8(D-2)}\ H_{p+2}^2 \ e^{2Y} \ ,  \label{eqs_XYZ}
\eeq
while the equation for $Z$ is simply
\beq
Z'' \ = \ 0 \ .
\eeq
It is solved by
\beq
Z=z_0+z_1 r \ ,
\eeq
where $z_0$ and $z_1$ are integration constants, and consequently the Hamiltonian constraint becomes
\bea
0&=& \frac{2(D-p-3)^2(D-p-2)}{\ell^2}\ e^{2X}\ + \ \frac{8(D-2)(D-p-3)}{\Delta}(Y')^2 \\
&-& 2(D-p-2)(X')^2\ +\ \frac{32(D-2)(p+1)}{\Delta}\ z_1{}^2  \ - \ (D-p-3)H_{p+2}^2 \ e^{2Y} \ , \nonumber
\eea
while the second--order equations for $X$ and $Y$ can be turned into
\bea
(X')^2&=&\frac{(D-p-3)^2}{\ell^2}  \ e^{2X} + {\cal E}_x \ ,  \nonumber \\
(Y')^2&=&\frac{\Delta}{8(D-2)} H_{p+2}^2  \ e^{2Y} + {\cal E}_y \ , \label{eq:XandY}
\eea
where the two ``energies'' ${\cal E}_x$ and ${\cal E}_y$ are integration constants. The three constants $z_1$, ${\cal E}_x$ and ${\cal E}_y$ are not independent, however, since the Hamiltonian constraint links them according to
\beq
0 \ = \ \frac{8(D-2)(D-p-3)}{\Delta} \ {\cal E}_y - 2(D-p-2) \ {\cal E}_x  \ + \ \frac{32(D-2)(p+1)}{\Delta} \ z_1^2 \ . \label{ham_z2ExEy}
\eeq

The special solutions with ${\cal E}_x=0$ and ${\cal E}_y=0$ have thus $z_1=0$. They stand out for their simplicity and their physical significance, and can be cast in the form
\bea
X&=&-\log\left[\frac{D-p-3}{\ell} \ r \right] \ ,  \nonumber \\
Y&=& - \log \left|\widetilde{H}{}_{p+2}\left(r \, + \,r_1\right)\right| \ ,  \nonumber \\
Z&=&z_0 \ .
\eea
Here $r>0$ and $r_1$ is an integration constant, and
\beq \label{eq:Htilde}
\widetilde{H}{}_{p+2} \ = \ \sqrt{\frac{\Delta}{8(D-2)}} \ H_{p+2} \ ,
\eeq
where the relative factor equals one in ten dimensions. The metric becomes
\bea
ds^2 &=& \frac{dx_{p+1}^2}{\left|\widetilde{H}{}_{p+2}\left(r+r_1\right)\right|^\frac{8(D-p-3)}{\Delta}}\ e^{\frac{8(D-2)\beta_p}{\Delta} z_0} \label{metric_Q}\\  &+& \left|\widetilde{H}{}_{p+2}\left(r+r_1\right)\right|^\frac{8(p+1)}{\Delta} e^{-\frac{8(D-2)(p+1)\beta_p}{(D-p-3)\Delta}  z_0} \left(\frac{dr^2}{\left[\frac{(D-p-3)}{\ell } \ r \right]^\frac{2(D-p-2)}{D-p-3}}  \ + \ \frac{\ell^2\ d\Omega_{D-p-2}^2}{\left[\frac{(D-p-3)}{\ell } \ r \right]^\frac{2}{D-p-3}} \right) \ , \nonumber
\eea
while the dilaton and form profiles are given by
\bea
e^\phi &=&  \frac{e^{-\frac{4(D-2)(p+1)}{\Delta} z_0}}{\left|\widetilde{H}{}_{p+2}\left(r\,+\,r_1\right)\right|^\frac{\left(D-2\right)^2\beta_p}{\Delta}} \ , \nonumber \\
{\cal H}_{p+2} &=& - \  \frac{H_{p+2}}{\left[\widetilde{H}_{p+2} \left(r \, + \,r_1 \right)\right]^2} \ dx^0 \wedge \ldots \wedge dx^p \wedge dr \ . \label{phiH_Q}
\eea
We have inserted a minus sign in the definition of $H_{p+2}$ with respect to eqs.~\eqref{metric_sym}, so that positive values of $H_{p+2}$ correspond to branes with identical signs for tension and charge.

For $\beta_p\neq0$ and $r_1 \neq 0$, the dependence on $z_0$ can be completely eliminated, while also casting the metric in a form that approaches the standard Minkowski in eq.~\eqref{flat_limit} as $r \to 0$, by performing the redefinitions
\bea
r &\to& e^{\,-\,\frac{4(p+1)}{(D-2)\beta_p}\left(z_0\,-\,\frac{D-p-3}{(D-2)\beta_p} \ \log\left|\widetilde{H}_{p+2}\,r_1 \right|\right)}\, r \ , \nonumber \\
r_1 &\to& e^{\,-\,\frac{4(p+1)}{(D-2)\beta_p}\left(z_0\,-\,\frac{D-p-3}{(D-2)\beta_p} \ \log\left|\widetilde{H}_{p+2}\,r_1 \right|\right)}\, r_1 \ , \nonumber \\
x^\mu &\to& e^{\,-\,\frac{4}{(D-2)\beta_p}\left(z_0\,-\,\frac{D-p-3}{(D-2)\beta_p} \ \log\left|\widetilde{H}_{p+2}\,r_1 \right|\right)}\, x^\mu \ . \label{rescalings_1}
\eea
These results can be equivalently obtained performing the shifts
\begin{alignat}{3}
A(r) &\to A(r) \,-\, \frac{D-p-3}{p+1}\,\lambda \ , \qquad && B(r) &&\to B(r) \,+\,\lambda \ , \nonumber \\ C (r) &\to C(r) \,+\,\lambda \ , \qquad &&\phi(r) &&\to \phi(r) \,+\,\frac{D-p-3}{\beta_p}\,\lambda \ , \label{redundancy1}
\end{alignat}
with a constant
\beq
\lambda = \frac{4(D-2)(p+1)\beta_p}{(D-p-3)\Delta}z_0 \ - \ \frac{4(p+1)}{\Delta}\log\left|\widetilde{H}_{p+2} \ r_1\right| \ ,
\eeq
which also leave the system in a harmonic gauge. These steps effectively lead to the identification
\beq
z_0 = \frac{D-p-3}{(D-2)\beta_p} \log\left|\widetilde{H}_{p+2} \ r_1\right| \ .
\eeq

Letting then
\beq
e^{\phi_0} = \left|\widetilde{H}_{p+2} \ r_1\right|^{-\frac{1}{\beta_p}} \ ,
\eeq
the metric can be finally cast in the form
\bea
ds^2 &=& \left|1+\frac{r}{r_1}\right|^{-\frac{8(D-p-3)}{\Delta}} dx_{p+1}^2 \label{metric_reduced_betanot0}\\  &+& \left|1+\frac{r}{r_1}\right|^\frac{8(p+1)}{\Delta} \left(\frac{dr^2}{\left[\frac{(D-p-3)}{\ell} \ r \right]^\frac{2(D-p-2)}{D-p-3}}  \ + \ \frac{\ell^2\ d\Omega_{D-p-2}^2}{\left[\frac{(D-p-3)}{\ell} \ r \right]^\frac{2}{D-p-3}} \right) \ , \nonumber
\eea
while the dilaton and the form--field strength become
\bea
e^\phi &=& e^{\phi_0} \left|1+\frac{r}{r_1}\right|^{-\frac{(D-2)^2\beta_p}{\Delta}}  \ , \nonumber \\
{\cal H}_{p+2} &=& - \ \epsilon \ \sqrt{\frac{8(D-2)}{\Delta}} \ \frac{1}{\left|r_1\right|} \  e^{\beta_p \phi_0} \left|1+\frac{r}{r_1}\right|^{-2} \ dx^0 \wedge \ldots \wedge dx^p \wedge dr \ ,  \label{backgrounds_reduced_betanot0}
\eea
where $\epsilon$ denotes the sign of $H_{p+2}$.
As we have stressed, we are confining our attention to positive nonzero values of $r$, while allowing both positive and negative values for $r_1$. All in all, the solutions finally contain two free parameters, $\phi_0$ and $r_1$. We have left out the cases with $\beta_p=0$ and/or $r_1=0$, which require different rescalings. However, if $r_1\neq 0$ and $\beta_p=0$ the end result, as we shall see, can be simply obtained from eqs.~\eqref{metric_reduced_betanot0} and \eqref{backgrounds_reduced_betanot0},
setting $\beta_p=0$ in them.

If $r_1>0$, the allowed range for $r$ is $0<r<\infty$, while if $r_1<0$ the range is limited to the interval $0<r<|r_1|$, since we focus on solutions that approach flat space far away from the singularity, as ought to be the case for branes in a Minkowski vacuum. Otherwise, the range $|r_1|<r<\infty$ would be another viable option.

One can now redefine the independent variable according to
\beq \label{eq:D_branes_isotropic_coordinates}
\left(\frac{\ell}{\rho}\right)^{D-p-3} \ = \ \frac{(D-p-3)}{\ell} \ r\ ,
\eeq
in order to recast the metric in the isotropic gauge, and it is also convenient to let
\beq
h_p \ = \ \frac{\ell^{D-p-2}}{(D-p-3) \ r_1}\ . \label{hpr1}
\eeq
The metric, dilaton, and form--field strength then become~\footnote{No absolute values are needed in the following expressions, with our choices for the range of $r$.}
\bea \label{eq:BPS_branes_isotropic}
ds^2 &=& \left(1+\frac{h_p}{\rho^{D-p-3}}\right)^{-\frac{8(D-p-3)}{\Delta}} dx_{p+1}^2 + \left(1+\frac{h_p}{\rho^{D-p-3}}\right)^{\frac{8(p+1)}{\Delta}} \left(d\rho^2 + \rho^2\ d\Omega_{D-p-2}^2\right) \ , \nonumber \\
e^\phi &=& e^{\phi_0} \left(1+\frac{h_p}{\rho^{D-p-3}}\right)^{-\frac{(D-2)^2\beta_p}{\Delta}}  \ ,  \\
{\cal H}_{p+2} &=&  \epsilon \, \sqrt{\frac{8(D-2)}{\Delta}} \ e^{\beta_p \phi_0} \left(1+\frac{h_p}{\rho^{D-p-3}}\right)^{-2}\!\! |h_p| \, (D-p-3) \, \rho^{-(D-p-2)} \, dx^0 \wedge .. \wedge d\rho \ . \nonumber
\eea

For large values of $\rho$, the spacetime part of the metric approaches Minkowski space, up to the leading correction factor
\beq
\left(1+\frac{h_p}{\rho^{D-p-3}}\right)^{-\frac{8(D-p-3)}{\Delta}} \ \sim \  1\ - \ \frac{8(D-p-3)h_p}{\Delta} \ \rho^{-(D-p-3)} \ .
\eeq
For $h_p \neq 0$ this signals, via Gauss's theorem, the presence of a source with non--vanishing tension at $\rho=0$. One can extract the value of the tension proceeding as in eq.~\eqref{V_tension}, while also taking into account the dilaton dressing $e^{\beta_p\phi} \, {\cal T}_p$\footnote{In the string frame the D-brane tension would enter the DBI action in the combination ${\cal T} e^{-\,\phi}$, which becomes ${\cal T} e^{\beta_p\,\phi}$ in the Einstein frame. This is also the proper Einstein--frame dressing for NS5 branes, since $p-3=7-p$ in that case.}, or alternatively one can deduce it from the $\delta(\vec{\rho})$ contributions in the equations of motion. The result is
\beq \label{eq:BPS_tension}
{\cal T}_p \ = \ \tilde{\epsilon}\ e^{-\beta_p\phi_0} \ \frac{8 (D-2)(D-p-3)\left|h_p\right|}{2\kappa_D^2 \ \Delta}\ \Omega_{D-p-2}  \ ,
\eeq
where $\tilde{\epsilon}$ is the sign of $h_p$, or equivalently, in view of eq.~\eqref{hpr1}, the sign of $r_1$. Positive values of $r_1$ lead to a positive tension, while negative values lead to a negative tension. Note that the factor $e^{-\beta_p\phi_0}$ is present, since we are working in the Einstein frame.

The charge can be computed from the equation for the form field with a localized source
\beq
\frac{1}{2\kappa_D^2} \ d \ (e^{-2\beta_p\phi} \ \star {\cal H}_{p+2}) = Q_p \ \delta(\vec{\rho}) \ ,
\eeq
or from Gauss's theorem, integrating over the interior of a sphere of large radius, where
\beq
\delta\left(\vec{\rho}\right) \ = \ \prod_{a=1}^{D-p-1} \delta\left(y^a\right) d\,y^a \ , \label{deltarho1}
\eeq
as we already stressed in eq.~\eqref{deltarho}, involves all Cartesian coordinates $y^a$ that are transverse to the brane.
This leads to
\beq \label{eq:BPS_charge}
Q_p = \epsilon \ e^{-\beta_p\phi_0} \ \sqrt{\frac{8(D-2)}{\Delta}} \ \frac{(D-p-3) \ |h_p| \ \Omega_{D-p-2}}{2\kappa_D^2} \ ,
\eeq
where $\epsilon$ is the sign of $H_{p+2}$. Positive values of $\epsilon$ lead to a positive charge, while negative values leads to a negative charge. All in all
\beq
\frac{{\cal T}_p}{Q_p} \ = \ \epsilon\,\tilde{\epsilon}\, \sqrt{\frac{8(D-2)}{\Delta}} \ ,
\eeq
so that charge and tension are proportional for generic values of $D$.

In the ten--dimensional case, which is relevant for String Theory, the results for the isotropic gauge become
\bea \label{eq:ST_BPS_branes}
ds^2 &=& \left(1+\frac{h_p}{\rho^{7-p}}\right)^{-\frac{(7-p)}{8}} dx_{p+1}^2 + \left(1+\frac{h_p}{\rho^{7-p}}\right)^{\frac{(p+1)}{8}} \left(d\rho^2 + \rho^2\ d\Omega_{8-p}^2\right) \ , \nonumber \\
e^\phi &=& e^{\phi_0} \left(1+\frac{h_p}{\rho^{7-p}}\right)^{-\beta_p}  \ , \nonumber \\
{\cal H}_{p+2} &=&  \epsilon \ e^{\beta_p \phi_0} \left(1+\frac{h_p}{\rho^{7-p}}\right)^{-2} |h_p| \ (7-p) \ \rho^{-(8-p)} \ dx^0 \wedge \ldots \wedge dx^p \wedge d\rho \ ,
\eea
where
\beq
h_p \ = \ \tilde{\epsilon}\, \frac{\ell^{\,8-p}}{(7-p) \ |r_1|}\ .
\eeq
The absolute values of tension and charge of the source now coincide, while the relative sign is determined by the product $\epsilon \,\tilde{\epsilon}$, where $\epsilon$ and $\tilde{\epsilon}$ are the signs of $H_{p+2}$ and of $h_p$, or equivalently of $r_1$. In particular, for $N$ D-branes, whose tension and charge are both positive and given by eq.~\eqref{eq:Dp_brane_tension}, one can link $h_p$ to the string scale $\alpha'$ according to
\beq
h_p \ =  \ \frac{(2\pi\sqrt{\alpha'})^{7-p}}{(7-p)\Omega_{8-p}}\, N \,e^{\beta_p \phi_0} \ .
\eeq

Note that this result apparently differs from the standard one, $h_p\propto N e^{\phi_0}$. The discrepancy arises since we are using a different Einstein-frame convention, which will prove more natural when discussing branes in the presence of a tadpole potential. For supersymmetric branes, the dilaton zero--mode is usually absorbed in the Planck mass, and as a result it explicitly contributes to the form-field kinetic term. This is a natural choice for solutions that are asymptotic to the Minkowski vacuum with a constant dilaton. In this paper, however, we shall explore cases where a constant contribution to the dilaton, $\phi_0$, is not related to the asymptotic behavior, so that it is more natural not to remove it from the rest of $\phi$. The relation between the different conventions is as follows:
\beq
{\cal T}_{\mathrm{standard}} = {\cal T}_p \ e^{\frac{p-7}{4}\,\phi_0}  \ , \qquad
{\cal H}_{p+2, \mathrm{standard}} = {\cal H}_{p+2} \ e^{-\,\frac{p+1}{4}\,\phi_0} \ .
\eeq

\subsection[\sc The case b=0]{\sc The case $\beta_p=0$}
If $\beta_p=0$, the dilaton decouples, and the metric and $(p+2)$-form field strength reduce to
\bea
ds^2 &=& \frac{dx_{p+1}^2}{\left|\widetilde{H}{}_{p+2}\left(r\,+\,r_1\right)\right|^\frac{8(D-p-3)}{\Delta}} \label{metric_beta0} \nonumber\\  &+& \left|\widetilde{H}{}_{p+2}\left(r\,+\,r_1\right)\right|^\frac{8(p+1)}{\Delta}  \left(\frac{dr^2}{\left[\frac{(D-p-3)}{\ell } \ r \right]^\frac{2(D-p-2)}{D-p-3}}  \ + \ \frac{\ell^2\ d\Omega_{D-p-2}^2}{\left[\frac{(D-p-3)}{\ell } \ r \right]^\frac{2}{D-p-3}} \right) \ , \nonumber \\
{\cal H}_{p+2} &=& - \ \frac{H_{p+2}}{\left[\widetilde{H}_{p+2} \left(r \,+\, r_1 \right)\right]^2} \ dx^0 \wedge \ldots \wedge dx^p \wedge dr \ .
\eea
There is still some redundancy, however, which can be fixed rescaling $\widetilde{H}_{p+2}$ to $\frac{\widetilde{H}_{p+2}}{\left|\widetilde{H}_{p+2}\,r_1\right|}$. Alternatively, one can rescale $r$ and $x$ in such a way that
\begin{alignat}{3}
A(r) &\to A(\lambda\,r) \,+\, \frac{1}{p+1}\log\lambda \ , \qquad && B(r) &&\to B(\lambda \, r) \,+\,\log\lambda \ , \nonumber \\ C (r) &\to C(\lambda \, r)  \ , \qquad &&\phi(r) &&\to \phi(\lambda \, r) \ , \label{redundancy2}
\end{alignat}
with
\beq
\lambda=\left|\widetilde{H}_{p+2} \ r_1 \right|  \ .
\eeq
Letting now
\beq
\widetilde{r}_1 \ = \ \frac{\tilde{\epsilon}}{\left|\widetilde{H}_{p+2}\right|} \ ,
\eeq
where $\tilde{\epsilon}$ is again the sign of $r_1$, the background takes the form
\bea
ds^2 &=&\left|1\ + \ \frac{r}{\widetilde{r}_1} \right|^{-\frac{2}{p+1}} dx_{p+1}^2 \nonumber \\ &+&\left|1\ + \ \frac{r}{\widetilde{r}_1} \right|^{\frac{2}{D-p-3}} \left(\frac{dr^2}{\left[\frac{(D-p-3)}{\ell } \ r \right]^\frac{2(D-p-2)}{D-p-3}}  \ + \ \frac{\ell^2\ d\Omega_{D-p-2}^2}{\left[\frac{(D-p-3)}{\ell } \ r \right]^\frac{2}{D-p-3}} \right) \ , \nonumber \\
{\cal H}_{p+2} &=& - \ \epsilon \ \sqrt{\frac{2(D-2)}{(p+1)(D-p-3)}} \ \frac{1}{\left|\widetilde{r}_1\right|} \left(1\ + \ \frac{r}{\widetilde{r}_1} \right)^{-\,2} \ dx^0 \wedge \ldots \wedge dx^p \wedge dr \ , \label{eq:beta=0_form}\nonumber \\
e^\phi &=& e^{\phi_0} \ .
\eea
In this fashion, one ends up with expressions that are identical to those in eqs.~\eqref{metric_reduced_betanot0} and \eqref{backgrounds_reduced_betanot0} even for $\beta_p=0$, after a relabeling of the residual integration constants.
Consequently, the isotropic--gauge backgrounds in eqs.~\eqref{eq:BPS_branes_isotropic} with $\beta_p=0$ (also in $\Delta$) account for this special case, if now
\beq
h_p \ = \ \frac{\ell^{D-p-2}}{(D-p-3) \ \widetilde{r}_1}\ .
\eeq

The tension and the charge of the extended object that sources this solution are still given by eqs.~\eqref{eq:BPS_tension} and~\eqref{eq:BPS_charge}, which in terms of $\widetilde{r}_1$ read
\bea \label{eq:BPS_tension2}
{\cal T}_p &=& \tilde{\epsilon}\  \ \frac{8 (D-2)}{\Delta }\ \frac{\Omega_{D-p-2}}{2\kappa_D^2} \ \frac{\ell^{D-p-2}}{\left|\widetilde{r}_1\right|} \ , \nonumber \\
Q_p &=& \epsilon \ \sqrt{\frac{8(D-2)}{\Delta}}\ \frac{\Omega_{D-p-2}}{2\kappa_D^2}   \ \frac{\ell^{D-p-2}}{ \left|\widetilde{r}_1\right|}  \ .
\eea

For positive values of $h_p$ these solutions interpolate between flat space as $\rho\to\infty$ and smooth $AdS_{p+2}\times S^{D-p-2}$ spaces, with
\beq \label{eq:AdStimesS_radii}
R_{AdS_{p+2}} \ = \  \frac{p+1}{D-p-3}\left|h_p\right|^{\frac{1}{D-p-3}} \ , \qquad
R_{S^{D-p-2}} \ = \  \left|h_p\right|^{\frac{1}{D-p-3}}
\eeq
as $\rho \to 0$.

\subsection[\sc The D3 brane and the self--dual five-form]{\sc The D3 brane and the self--dual five-form}

In ten dimensions, the value $\beta_p=0$ corresponds to the D3 brane. In this case, however, the five--form field strength is self--dual, and thus additional care is required. In fact, adding to the form field strength in eqs.~\eqref{eq:beta=0_form} for $p=3$ and $D=10$ its Hodge dual does build a self--dual five--form, but the stress tensor is then doubled. In order to retain the original form of the metric, one can simply divide by $\sqrt{2}$ the field strength ${\cal H}_5$, so that the self--dual background of interest reads
\bea
ds^2 &=&\left|1\ + \ \frac{r}{\widetilde{r}_1} \right|^{-\frac{1}{2}} dx_4^2 \ + \ \left|1\ + \ \frac{r}{\widetilde{r}_1} \right|^{\frac{1}{2}} \left(\frac{dr^2}{\left(\frac{4\,r}{\ell } \right)^\frac{5}{2}}  \ + \ \frac{\ell^2 \ d\Omega_{5}^2}{\left(\frac{4\,r}{\ell }  \right)^\frac{1}{2}} \right) \ , \nonumber \\
{\cal H}_{5} &=& - \ \frac{\epsilon}{\sqrt{2}\left|\widetilde{r}_1\right|} \left[ \left(1\ + \ \frac{r}{\widetilde{r}_1} \right)^{-\,2}  dx^0 \wedge \ldots \wedge dx^3 \wedge dr \ + \ \text{vol}_{S^5} \right]\ , \nonumber \\
e^\phi &=& e^{\phi_0} \ ,
\eea
where $\text{vol}_{S^5}$ denotes the volume form on the unit sphere. In this case the tension is still given by eq.~\eqref{eq:BPS_tension2}, which becomes
\beq
{\cal T}_3 \ =\  \tilde{\epsilon}\ \frac{\Omega_{5}}{2\kappa_{10}^2} \frac{\ell^5}{\left|\widetilde{r}_1\right|}
\eeq
in ten dimensions, while the charge is divided by $\sqrt{2}$, and reads
\beq
Q_3 \ = \ \epsilon \  \frac{\Omega_{5}}{2\kappa_{10}^2}  \frac{\ell^5 }{\sqrt{2}\,   \left|\widetilde{r}_1\right|} \ ,
\eeq
so that the BPS condition is now
\beq
\left|{\cal T}_3 \right| \ = \ \sqrt{2} \ \left| Q_3 \right| \ .
\eeq

This relation depends on the conventions for the kinetic term of the form field, and there are different choices in the literature.
In general, if one starts from the kinetic term
\beq
S\ = \ \frac{1}{2\kappa_{10}^2}\int \frac{\chi}{2} \, {\cal H}_5 \,\wedge \,\star\, {\cal H}_5  \ ,
\eeq
where self--duality is to be imposed at the end and $\chi$ is a real parameter that reflects the choice of normalization, the form field strength of our self--dual solution becomes
\beq
{\cal H}_{5} = - \ \frac{\epsilon}{\sqrt{2 \,\chi}\,\left|\widetilde{r}_1\right|} \left[ \left(1\ + \ \frac{r}{\widetilde{r}_1} \right)^{-\,2}  dx^0 \wedge \ldots \wedge dx^3 \wedge dr \ + \ \text{vol}_{S^5} \right]\ .
\eeq
In the isotropic gauge the background then takes the form
\bea \label{eq:metric_D3}
ds^2 &=& \left(1\,+\,\frac{h_3}{\rho^4}\right)^{-\frac{1}{2}} dx_{4}^2 \ + \ \left(1\,+\,\frac{h_3}{\rho^4}\right)^{\frac{1}{2}}  \left(d\rho^2 \ +\  \rho^2 d\Omega_5^2\right) \ , \nonumber \\
{\cal H}_5 &=& \, \frac{4 \, \epsilon}{\sqrt{2 \, \chi}} \left|h_3\right| \left[\left(1\,+\,\frac{h_3}{\rho^4}\right)^{-2}\, \frac{1}{\rho^5}  \, dx^0 \wedge \ldots \wedge d\rho  \ - \ \text{vol}_{S^5} \right]\ ,
\eea
where the relative minus sign between the two terms is induced by the transformation in eq.~\eqref{eq:D_branes_isotropic_coordinates} linking the harmonic and isotropic coordinates, so that also here ${\cal H}_5=\star {\cal H}_5$.
One can define the charge according to
\beq
d \star {\cal H}_5 \ = \ d \ {\cal H}_5 \ = \  2\kappa_{10}^2  \ Q_3 \  \delta(\vec{\rho}) \ ,
\eeq
or better via Gauss's theorem, as
\beq
Q_3 \ = \ \frac{1}{2\,\kappa_{10}^2} \ \int_{S^5} \star\,{\cal H}_5 \ = \ \frac{1}{2\,\kappa_{10}^2} \ \int_{S^5} {\cal H}_5 \ .
\eeq
This would correspond to a probe D3 brane coupling
\beq
\delta S\ = \ \chi \,Q_3 \int B_4 \ . \label{d3coupling}
\eeq
The choice $\chi=2$ corresponds to the convention of~\cite{ms21_1}~\footnote{Note, however, that the probe brane coupling used in~\cite{ms22_1} differs from eq.~\eqref{d3coupling} by a factor two, so that tension and charge should rather appear in the combination ${\cal T}_3 - Q_3$. There is actually an additional factor of two, since in this selfdual case an electric coupling brings along an identical magnetic one. As a result the complete interaction potential is ${\cal T}_3\,{\cal T}_3'\,-\,2\,\chi\,Q_3\,Q_3'$,  so that the no--force condition is
precisely as demanded by eq.~\eqref{BPSchi}. These difficulties have been known for a while~\cite{GKP}, and this brief discussion should clarify the issue, correcting some factors in the example discussed in~\cite{ms23_3}, which is also a neat illustration of the idea of dynamical cobordism~\cite{dynamicalcobordism}.}, while $\chi=1$ is the standard convention for the non--self--dual case that we are using in this paper, and finally $\chi=\frac{1}{2}$ corresponds to the convention of~\cite{bkrvp}.
Note that the BPS condition is also convention--dependent, and becomes in general
\beq
\left|{\cal T}_3 \right| \ = \ \sqrt{2 \, \chi} \ \left| Q_3 \right| \ . \label{BPSchi}
\eeq

\subsection[\sc The limiting value r1=0]{\sc The limiting value $r_1=0$}

Returning to generic values of $p$ and $D$, the limiting case $r_1=0$ was left out in the preceding discussion, and we now want to address its meaning, starting from eqs.~\eqref{metric_Q} and \eqref{phiH_Q}. The rescalings
\bea
r &\to& e^{\,-\,\frac{4(p+1)}{(D-2)\beta_p}\left(z_0\,-\,\frac{D-p-3}{(D-2)\beta_p} \ \log\left|\widetilde{H}_{p+2}\,\ell \right|\right)}\, r \ , \nonumber \\
x^\mu &\to& e^{\,-\,\frac{4}{(D-2)\beta_p}\left(z_0\,-\,\frac{D-p-3}{(D-2)\beta_p} \ \log\left|\widetilde{H}_{p+2}\,\ell \right|\right)}\, x^\mu \ ,
\eea
eliminate $z_0$ and $\left|H_{p+2}\right|$, turning the background into
\bea \label{eq:BPS_near_horizons}
ds^2 &=& \left(\frac{r}{\ell}\right)^{-\frac{8(D-p-3)}{\Delta}} dx_{p+1}^2 \\  &+& \left(\frac{r}{\ell}\right)^\frac{8(p+1)}{\Delta} \left(\frac{dr^2}{\left[\frac{(D-p-3)}{\ell } \ r \right]^\frac{2(D-p-2)}{D-p-3}}  \ + \ \frac{\ell^2\ d\Omega_{D-p-2}^2}{\left[\frac{(D-p-3)}{\ell } \ r \right]^\frac{2}{D-p-3}} \right) \ , \nonumber\\
e^\phi &=&  e^{\phi_0} \left(\frac{r}{\ell}\right)^{-\frac{\left(D-2\right)^2\beta_p}{\Delta}} \ , \nonumber \\
{\cal H}_{p+2} &=& - \  \epsilon \ \sqrt{\frac{8(D-2)}{\Delta}} \ e^{\beta_p \phi_0} \ \frac{\ell}{r^2}\ dx^0 \wedge \ldots \wedge dx^p \wedge dr \ , \nonumber
\eea
where $0<r<\infty$,
\beq
e^{\phi_0} = \left|\widetilde{H}_{p+2} \, \ell\right|^{-\frac{1}{\beta_p}} \ ,
\eeq
and $\epsilon$ is the sign of $H_{p+2}$, as before.

These backgrounds are not asymptotically flat, and consequently if they capture the whole spacetime they cannot be used to describe branes. However, they describe regions close to their horizons. Formally, one could still refer to eqs.~\eqref{metric_Q} and \eqref{phiH_Q}, using a non--vanishing value of $r_1$ as a regulator, to then take the $r_1 \to 0$ limit, or equivalently the $h_p \to \infty$ limit, of eqs.~\eqref{eq:BPS_tension} and \eqref{eq:BPS_charge}. In this fashion, for generic values of $\beta_p$ tension and charge would tend to infinity in the limit, while still being proportional.  Moreover, for $\beta_p \neq 0$ there are singularities at $r=0, \infty$: the former lies at an infinite proper distance, while the latter lies at a finite proper distance. On the other hand, for $\beta_p=0$ there are no singularities, and one recovers $AdS \times S$ backgrounds, as in eq.~\eqref{eq:AdStimesS_radii}.

Let us now show in detail that these solutions capture the large--$r$ region of the original backgrounds of eqs.~\eqref{metric_Q} and \eqref{phiH_Q}. For convenience, we choose isotropic coordinates with a rescaled $\rho$ compared to eq.~\eqref{eq:D_branes_isotropic_coordinates}, and then, letting
\beq
\left(\frac{\ell}{\rho}\right)^{D-p-3} \ = \ \frac{r}{\ell} \ ,
\eeq
leads to
\bea
ds^2 &=& \left(\frac{\ell}{\rho}\right)^{-\frac{8(D-p-3)^2}{\Delta}} dx_{p+1}^2 \nonumber \\
&+& \left(\frac{\ell}{\rho}\right)^{\frac{8(D-p-3)(p+1)}{\Delta}}(D-p-3)^{-\frac{2}{D-p-3}} \left(d\rho^2 + \rho^2 d\Omega_{D-p-2}^2\right) \ ,
\eea
and
\bea
e^\phi &=&  e^{\phi_0} \left(\frac{\ell}{\rho}\right)^{-\frac{\left(D-p-3\right)\left(D-2\right)^2\beta_p}{\Delta}} \ , \nonumber \\
{\cal H}_{p+2} &=& \epsilon \ \sqrt{\frac{8(D-2)}{\Delta}} \ e^{\beta_p \phi_0} \ \frac{D-p-3}{\ell} \left(\frac{\rho}{\ell}\right)^{D-p-4}\ dx^0 \wedge \ldots \wedge dx^p \wedge d\rho\ .
\eea
These expressions capture indeed the ``near--horizon'' region of charged branes, and in fact they reduce to the near--horizon limit of eqs.~\eqref{eq:BPS_branes_isotropic} after rescaling $\rho$ and $x$ according to
\beq
\rho \ \to \   \left(\frac{\left|h_p\right|}{\ell^{D-p-3}}\right)^{\frac{4(p+1)}{(D-2)^2\beta_p^2}} \rho \ , \qquad
x^\mu \ \to \ \left(\frac{\left|h_p\right|}{\ell^{D-p-3}}\right)^{-\frac{4(D-p-3)}{(D-2)^2 \beta_p^2}} x^\mu \ ,
\eeq
and after redefining $\phi_0$ according to
\beq
\phi_0\ \to \ \phi_0 \ - \ \frac{1}{\beta_p}\log{\left|h_p\right|} \ .
\eeq

In conclusion, these solutions are effectively zooming in near the core of the branes.

\subsection[\sc BPS Branes in String Theory]{\sc BPS Branes in String Theory}
\label{sec:string_theory_BPS}

Setting $D=10$ and selecting the appropriate values for $\beta_p$ from Table~\ref{table:tab_1}, one can recover from the preceding setup the BPS branes of String Theory~\cite{susy_solitons} from eqs.~\eqref{eq:ST_BPS_branes}, and from~\eqref{eq:metric_D3} for the self--dual D3 case.
$\tilde{\epsilon}$, the sign of $r_1$, or equivalently of $h_p$, determines whether the solution has positive or negative tension. On the other hand, $\epsilon$ controls the sign of the charge.
Therefore, branes and O$^+$ planes correspond to $\tilde{\epsilon}=1$ and $\epsilon=1$, while orientifold O$^-$ planes correspond to $\tilde{\epsilon}=-1$ and $\epsilon=-1$. Anti--branes and anti--orientifolds are simply obtained from these by reverting $\epsilon$.

When sources with negative tension are present, which is the case when $\tilde{\epsilon}=-1$, the solution describes in principle two regions: the region for $0<r<|r_1|$ is the proper one to describe extended objects, since it includes an asymptotically flat limit as $r\to 0$. The tension and the charge of the extended object can be deduced, as before, from the limiting behavior of the background in the asymptotically flat region. In seeking a connection with String Theory, one should stop before the singularity is reached, in order to leave out contributions that are not under control in perturbation theory.
On the other hand, retaining  the region $|r_1| < r< \infty$ yields a solution that is not asymptotically flat. This different solution includes a pair of singularities at the ends of the relevant interval, which are separated by a finite proper distance. Orthogonal spheres shrink there to points, while the tensor lines of force go from one end to the other, in a way that is somewhat reminiscent of what happens inside a plane capacitor in Electromagnetism.

\subsection[\sc Non--BPS Branes]{\sc Non--BPS Branes}

The BPS branes that we described in the previous sections obtain if one sets to zero the two energy--like quantities ${\cal E}_x$ and ${\cal E}_y$ in eqs.~\eqref{eq:XandY}. In this section we shall address the general case, and in particular we shall derive tensions and charges of these deformed branes. For brevity, we shall refer to the ${\cal E}_x={\cal E}_y=0$ solutions as BPS for any dimension $D$, although they are actually BPS solutions only in ten--dimensional String Theory. Turning on these parameters yields solutions that do not preserve any supersymmetry in ten dimensions.

The starting point is again provided by eqs.~\eqref{eq:XandY} and by the Hamiltonian constraint
\beq \label{eq:deformed_charged_z1}
z_1^2 \ =  \ \frac{1}{4(p+1)}\left[\frac{(D-p-2)}{4(D-2)}\,\Delta \, {\cal E}_x \ - \ (D-p-3) \, {\cal E}_y\right] \ ,
\eeq
which restricts the ranges of ${\cal E}_x$ and ${\cal E}_y$ according to
\beq \label{eq:deformed_charged_ExEy}
{\cal E}_x \ \geq \ \frac{4(D-2)(D-p-3)}{\Delta (D-p-2)} \ {\cal E}_y \ .
\eeq
In order to characterize the independent solutions, it is now convenient to define a unit vector $\mathbf{n}$, while also grouping ${\cal E}_x$ and ${\cal E}_y$ into a two component vector $\boldsymbol{\mathcal{E}}$, according to
\beq
\mathbf{n} \ \equiv \ \left(n_x,n_y\right) \ = \ \left(\frac{(D-p-2) \Delta}{16(D-2)(p+1) \Sigma}  \ , \ - \ \frac{(D-p-3)}{4(p+1)\Sigma}\right) \ , \qquad \boldsymbol{\mathcal{E}} \ = \ \left({\cal E}_x,{\cal E}_y\right),
\eeq
where
\beq
\Sigma \ = \ \sqrt{\left(\frac{(D-p-2)\Delta}{16(D-2)(p+1)}\right)^2 + \left(\frac{(D-p-3)}{4(p+1)}\right)^2} \ ,
\eeq
in terms of which the Hamiltonian constraint reduces to
\beq \label{eq:Edotn}
\boldsymbol{\mathcal{E}}\cdot \mathbf{n}  \ = \ \frac{z_1^2}{\Sigma} \ .
\eeq
Consequently, one can parameterize the vector $\boldsymbol{\mathcal{E}}$ in terms of two independent constants, $z_1$ and $w$, as
\beq \label{eq:paramEnergies}
\boldsymbol{\mathcal{E}} \ = \ \frac{1}{\Sigma} \left(z_1^2 \ \mathbf{n} \ + \ w \  \mathbf{n}_{\perp} \right) \ ,
\eeq
where
\beq
\mathbf{n}_{\perp} \ = \ \left(-\ n_y \ , \ n_x \right)
\eeq
is a unit vector perpendicular to $\mathbf{n}$. Explicitly, the two components of $\boldsymbol{\mathcal{E}}$ read
\beq
{\cal E}_x \ = \ \frac{z_1^2}{\Sigma} \ n_x \ - \ \frac{w}{\Sigma}\ n_y \ , \qquad {\cal E}_y \ = \ \frac{z_1^2}{\Sigma} \ n_y \ + \  \frac{w}{\Sigma}\ n_x \ .
\eeq

The solutions for $X$ and $Y$ depend on the signs of ${\cal E}_x$ and ${\cal E}_y$ and, up to an overall translation of $r$, one can conveniently cast them in the form
\beq
X \ = \ - \log\left|\frac{D-p-3}{\ell} \ {\cal F}\left({\cal E}_x, r \right)\right| \ ,  \qquad Y \ = \ - \log\left|\widetilde{H}_{p+2} \ {\cal F}\left({\cal E}_y, r+r_1 \right)\right| \ ,
\eeq
where
\beq
{\cal F}\left(\mathcal{E}, r \right)\ = \ \begin{cases}
    \frac{1}{\sqrt{\mathcal{E}}}\sinh{\left(\sqrt{\mathcal{E}} \, r\right)} \qquad & \text{if   } \mathcal{E}>0 \ , \\
    r & \text{if   } \mathcal{E}=0 \ , \label{cases_F} \\
    \frac{1}{\sqrt{|\mathcal{E}|}}\sin{\left(\sqrt{|\mathcal{E}|} \, r\right)} & \text{if   } \mathcal{E}<0 \ .
\end{cases}
\eeq
In terms of $\cal{F}$, the general solution for the metric reads
\bea
ds^2 &=& \frac{dx_{p+1}^2}{\left|\widetilde{H}_{p+2} \ {\cal F}\left({\cal E}_y, r+r_1 \right)\right|^\frac{8(D-p-3)}{\Delta}} e^{\frac{8(D-2)\beta_p}{\Delta} (z_0+ z_1 r )} \nonumber \\  &+& \left|\widetilde{H}_{p+2} \ {\cal F}\left({\cal E}_y, r+r_1 \right)\right|^\frac{8(p+1)}{\Delta} e^{-\frac{8(D-2)(p+1)\beta_p}{(D-p-3)\Delta}  (z_0 + z_1 r )} \left(\frac{dr^2}{\left|\frac{D-p-3}{\ell} \ {\cal F}\left({\cal E}_x, r \right)\right|^\frac{2(D-p-2)}{D-p-3}}  \right.\nonumber \\
&+& \left. \ \frac{\ell^2 \ d\Omega_{D-p-2}^2}{\left|\frac{D-p-3}{\ell} \ {\cal F}\left({\cal E}_x, r \right)\right|^\frac{2}{D-p-3}} \right) \ ,
\eea
while the dilaton and the form fields are
\bea
e^\phi &=&  \frac{e^{-\frac{4(D-2)(p+1)}{\Delta} (z_0 + z_1 r )}}{\left|\widetilde{H}_{p+2} \ {\cal F}\left({\cal E}_y, r+r_1 \right)\right|^\frac{\left(D-2\right)^2\beta_p}{\Delta}} \ , \nonumber \\
{\cal H}_{p+2} &=& - \ \frac{H_{p+2}}{\left|\widetilde{H}_{p+2} \ {\cal F}\left({\cal E}_y, r+r_1 \right)\right|^2} \ dx^0 \wedge \ldots \wedge dx^p \wedge dr \ ,
\eea
where
\beq \label{eq:Htilde2}
\widetilde{H}{}_{p+2} \ = \ \sqrt{\frac{\Delta}{8(D-2)}} \ H_{p+2} \ ,
\eeq
as in eq.~\eqref{eq:Htilde}. One can eliminate $z_0$, as in previous sections, by performing the rescalings
\bea \label{eq:rescaling2}
\left(r,r_1,\left|{\cal E}_{x,y}\right|^{-\frac{1}{2}}, z_1^{-1}\right) &\to& e^{\,-\,\frac{4(p+1)}{(D-2)\beta_p}\left(z_0\,-\,\frac{D-p-3}{(D-2)\beta_p} \ \log\left|\widetilde{H}_{p+2}\,{\cal F}\left({\cal E}_y, r_1 \right) \right|\right)}\, \left(r,r_1,\left|{\cal E}_{x,y}\right|^{-\frac{1}{2}}, z_1^{-1}\right) \ , \nonumber \\
x^\mu &\to& e^{\,-\,\frac{4}{(D-2)\beta_p}\left(z_0\,-\,\frac{D-p-3}{(D-2)\beta_p} \ \log\left|\widetilde{H}_{p+2}\,{\cal F}\left({\cal E}_y, r_1 \right) \right|\right)}\, x^\mu \ ,
\eea
which generalize those in eq.~\eqref{rescalings_1}, and the background can be finally cast in the form
\bea \label{eq:deformation_backgroun_general}
ds^2 &=& \left|\frac{{\cal F}\left({\cal E}_y, r+r_1 \right)}{{\cal F}\left({\cal E}_y, r_1 \right)}\right|^{-\frac{8(D-p-3)}{\Delta}} e^{\frac{8(D-2)\beta_p}{\Delta} z_1 r }  dx_{p+1}^2 \nonumber \\  &+& \left|\frac{{\cal F}\left({\cal E}_y, r+r_1 \right)}{{\cal F}\left({\cal E}_y, r_1 \right)}\right|^{\frac{8(p+1)}{\Delta}} e^{-\frac{8(D-2)(p+1)\beta_p}{(D-p-3)\Delta} z_1 r } \left(\frac{dr^2}{\left|\frac{(D-p-3)}{\ell } \, {\cal F}\left({\cal E}_x, r \right) \right|^\frac{2(D-p-2)}{D-p-3}} \right.  \nonumber \\
&+& \left. \ \frac{\ell^2\ d\Omega_{D-p-2}^2}{\left|\frac{(D-p-3)}{\ell } \, {\cal F}\left({\cal E}_x, r \right) \right|^\frac{2}{D-p-3}} \right) \ , \nonumber \\
e^\phi &=&  e^{\phi_0}\ e^{-\frac{4(D-2)(p+1)}{\Delta} z_1 r}  \left|\frac{{\cal F}\left({\cal E}_y, r+r_1 \right)}{{\cal F}\left({\cal E}_y,r_1 \right)}\right|^{-\frac{\left(D-2\right)^2\beta_p}{\Delta}} \ ,  \\
{\cal H}_{p+2} &=& - \ \epsilon \  \sqrt{\frac{8(D-2)}{\Delta}} \ e^{\beta_p \phi_0} \ \left|{\cal F}\left({\cal E}_y, r_1 \right)\right|^{-1}   \left|\frac{{\cal F}\left({\cal E}_y, r+r_1 \right)}{{\cal F}\left({\cal E}_y, r_1 \right)}\right|^{-2} \ dx^0 \wedge \ldots \wedge dx^p \wedge dr \ , \nonumber
\eea
where
\beq
e^{\phi_0} \ = \ \left|\widetilde{H}_{p+2} \,{\cal F}\left({\cal E}_y, r_1 \right)\right|^{-\frac{1}{\beta_p}} \ .
\eeq
These solutions thus depend on the four independent real parameters $r_1$, $\phi_0$, $z_1$ and $w$, whose physical significance will be addressed shortly.

\subsubsection[\sc Different classes of solutions]{\sc Different classes of solutions}

There are actually, altogether, \emph{six} different classes of solutions, depending on the signs of ${\cal E}_x$ and ${\cal E}_y$. Indeed, out of the nine a priori possible options, the three cases $({\cal E}_x<0 \, , \, {\cal E}_y\geq0)$ and $({\cal E}_x=0 \, , \, {\cal E}_y>0)$ are excluded by the positivity of $z_1{}^2$ in eq.~\eqref{eq:Edotn}, and thus by the parametrization in eq.~\eqref{eq:paramEnergies}. Moreover, in the six remaining cases this parametrization implies the inequality
\beq
{\cal E}_x \ \geq \ \left|\frac{n_y}{n_x}\right| \  {\cal E}_y \ ,
\eeq
which restricts the possible ranges of ${\cal E}_x$ and ${\cal E}_y$. One can show that, within the relevant ranges of $D$ and $p$, $\left|n_y\right|<\left|n_x\right|$, so that when ${\cal E}_x<0$ and ${\cal E}_y<0$ the inequality
\beq
\frac{1}{\sqrt{\left|{\cal E}_y\right|}} \ < \  \frac{1}{\sqrt{\left|{\cal E}_x\right|}}
\eeq
holds,
and consequently the period of ${\cal F}\left({\cal E}_y, r+r_1 \right)$ is smaller than that of ${\cal F}\left({\cal E}_x, r \right)$.
As a result, the solutions within this class cannot interpolate between two consecutive zeros of ${\cal F}\left({\cal E}_x, r \right)$ without encountering, between them, a zero of ${\cal F}\left({\cal E}_y, r+r_1 \right)$.

Interestingly, when ${\cal E}_x$ and ${\cal E}_y$ are positive, in the region ${\cal E}_x \, r\gg 1$ the solutions approach the vacuum solutions with form fluxes of~\cite{ms21_1}. The integration constants of that paper can be mapped to the present ones by the redefinitions
\beq
\phi_1 \ =\  -\ \frac{4(D-2)(p+1)}{\Delta} \, z_1 \ , \qquad C_1 \ = \ -\ \frac{\sqrt{{\cal E}_x}}{2(D-p-3)} \ -\  \frac{4(D-2)(p+1)\beta_p}{(D-p-3)\Delta}\, z_1 \ ,
\eeq
and the function $f(r)$ of~\cite{ms21_1} corresponds to $\left|\widetilde{H}_{p+2} \ {\cal F}\left({\cal E}_y, r+r_1 \right)\right|$.
This class of solutions interpolates between flat space as $r\to 0$ and the flux vacua of~\cite{ms21_1} when ${\cal E}_x \, r\gg 1$.

All six classes of solutions allow for asymptotically flat spacetimes. This is the case when $r=0$ lies in the allowed region, so that the corresponding metrics approach eq.~\eqref{flat_limit}, as should be the case when describing extended objects embedded in a Minkowski vacuum. The range of $r$ then extends up to the first zero of ${\cal F}\left({\cal E}_y, r+r_1 \right)$, or up to $r=\infty$, if no such zero is present. When ${\cal E}_x<0$, asymptotically flat spacetimes also emerge when the range of $r$ extends from a zero of ${\cal F}\left({\cal E}_x, r \right)$ to a zero of ${\cal F}\left({\cal E}_y, r+r_1 \right)$.
In fact, with an appropriate rescaling one can turn the metric into eq.~\eqref{eq:deformation_backgroun_general}, while also shifting $r_1$ by a suitable number of periods of $\left|{\cal F}\left({\cal E}_x, r \right)\right|$. Consequently, one can always work with $r=0$ at one end of the allowed range.

There are also solutions that do not include an asymptotically flat region. These vacua, which are at times reminiscent of ``dipole--like'' setups, do not afford an interpretation related to extended objects embedded in a vacuum, and in these cases one cannot associate to them tension and charge. This is the case, for instance, for positive values of ${\cal E}_x$ and ${\cal E}_y$, if $r_1<0$ and one is working in the region $\left|r_1\right|<r<+\infty$, or for positive ${\cal E}_x$ and negative ${\cal E}_y$ when the range of $r$ is between two consecutive zeros of ${\cal F}\left({\cal E}_y, r+r_1\right)$.

\subsubsection[\sc Tensions and charges of asymptotically flat solutions]{\sc Tensions and charges of asymptotically flat solutions}

For the asymptotically flat solutions one can compute tension and charge, as in previous cases, starting from the first order expansion of the background around $r=0$. One thus finds~\footnote{We are considering the same dilaton dressing as for BPS branes, namely $e^{\beta_p \phi} {\cal T}_p$, which is instrumental to define the tension compatibly with the Einstein--frame actions for the branes of ten--dimensional string theory.}
\bea\label{eq:tension_and_charge_nonBPS}
{\cal T}_p &=& e^{-\beta_p \phi_0} \frac{8(D-2)}{\Delta}\frac{\Omega_{D-p-2}}{2\kappa_D^2} \,\ell^{\, D-p-2}\left[\frac{{\cal F}'\left({\cal E}_y, r_1\right)}{{\cal F}\left({\cal E}_y, r_1\right)} \ - \ \frac{D-2}{D-p-3}\ \beta_p z_1\right] \ , \nonumber \\
Q_p &=& \epsilon \  e^{-\beta_p \phi_0} \sqrt{\frac{8(D-2)}{\Delta}}\frac{\Omega_{D-p-2}}{2\kappa_D^2}\,\ell^{\, D-p-2}\left|{\cal F}\left({\cal E}_y, r_1\right)\right|^{-1} \ .
\eea
In the three possible cases, eq.~\eqref{cases_F} leads to
\beq
\frac{{\cal F}'\left({\cal E}_y, r_1\right)}{{\cal F}\left({\cal E}_y, r_1\right)} \ = \ \begin{cases}
    \sqrt{{\cal E}_y} \coth\left(\sqrt{{\cal E}_y} \, r_1\right) \qquad & \text{if   } \mathcal{E}_y>0 \ , \\
   \frac{1}{r_1} & \text{if   } \mathcal{E}_y=0 \ , \\
    \sqrt{\left|{\cal E}_y\right|} \cot\left(\sqrt{\left|{\cal E}_y\right|} \, r_1\right) & \text{if   } \mathcal{E}_y<0 \ .
\end{cases}
\eeq
In general, the ratio $\frac{{\cal T}_p}{Q_p}$ can take any value by tuning $z_1$ and ${\cal E}_y$, which can be chosen independently, compatibly with the Hamiltonian constraint~\eqref{eq:deformed_charged_ExEy}.
Allowing a non--zero value for $z_1$ has an interesting effect, along the lines of what we saw for uncharged branes. The original Lagrangian is invariant under a generalized shift symmetry, which affects the dilaton and the form field according to
\beq
\delta\phi \ =\  c \ , \qquad \delta {\cal B}_{p+1} \ =\  \beta_p \, c \, {\cal B}_{p+1} \ .
\eeq
The corresponding Noether current has again a non--vanishing $r$ component, which is proportional to $z_1$. In detail
\beq
j_r \ = \ - \ \frac{8}{D-2}\,\partial_r \phi \ - \ \beta \frac{e^{-2\beta_p\phi}}{(p+1)!} \ {\cal H}_{r \mu_0 \dots \mu_{p}} {\cal B}^{\mu_0 \dots \mu_p} \ = \ \frac{32 (p+1)}{\Delta} z_1 \ ,
\eeq
and the presence of this current signals again a charge flow between the asymptotically flat region and the extended object.
For the special solutions with $z_1=0$, one can make definite statements about the ratio between tension and charge. In detail
\beq
\left|\frac{{\cal T}_p}{Q_p}\right| \ = \ \sqrt{\frac{8(D-2)}{\Delta}} \left|{\cal F}'\left({\cal E}_y, r_1\right)\right| \ ,
\eeq
so that, taking eqs.~\eqref{cases_F} into account, in ten dimensions,
\beq
\left|\frac{{\cal T}_p}{Q_p}\right| \ = \ \begin{cases}
    \cosh\left(\sqrt{{\cal E}_y } \, r_1\right) \qquad & \text{if   } \mathcal{E}_y>0 \ , \\
    1 & \text{if   } \mathcal{E}_y=0 \ , \\
    \left|\cos\left(\sqrt{\left|{\cal E}_y\right|} \, r_1\right) \right| & \text{if   } \mathcal{E}_y<0 \ .
\end{cases}
\eeq
Solutions with positive ${\cal E}_y$ have $\left|{\cal T}_p\right|>\left|Q_p\right|$, those with ${\cal E}_y=0$ are the BPS ones with  $\left|{\cal T}_p\right|=\left|Q_p\right|$, and finally solutions with ${\cal E}_y<0$ have $\left|{\cal T}_p\right|<\left|Q_p\right|$.
Note also that the backgrounds with $z_1=0$ involve three independent parameters, which are in one-to-one correspondence with the tension, the charge, and the asymptotic value of the dilaton.

\subsubsection[\sc Singularities]{\sc Singularities}

In ten dimensions, in all instances with $p<7$ related to String Theory, which concern D-branes or NS5 branes, to which the formalism applies verbatim,
\beq
e^B \ \sim \ \left|{\cal F}\left({\cal E}_y, r+r_1 \right)\right|^{\frac{(p+1)}{16}} e^{-\frac{(p+1)\beta_p}{2(7-p)} z_1 r }  \left|{\cal F}\left({\cal E}_x, r \right) \right|^{-\frac{8-p}{7-p}}  \ .
\eeq
When ${\cal F}\left({\cal E}_y, r+r_1 \right)$ has a zero at some finite value of $r$, say $r^\star$, the range of $r$ terminates there. This point always lies at a finite proper distance from a generic point $r>0$, because $e^{B(r^\star)}=0$. This can happen either when $r_1<0$, in which case $r^\star=|r_1|$, or when ${\cal E}_y<0$, so that the corresponding ${\cal F}\left({\cal E}_y, r+r_1 \right)$ is a trigonometric function.
Otherwise, when the range of $r$ extends to $\infty$, the proper distance between a finite value of $r$ and $\infty$ remains finite, with the only exception of the BPS D3 brane. In fact, if ${\cal E}_x >0$ and ${\cal E}_y \geq 0$, $e^B$ tends to zero exponentially fast, while if ${\cal E}_x=0$, ${\cal E}_y=0$ and $p\neq 3 $, $e^B \sim r^{-1-\frac{(p-3)^2}{16 (7-p)}}$ for large values of $r$. This behavior grants indeed a finite proper distance with the only, well--known, exception of BPS D3 branes.

The Ricci scalar diverges at all finite distance endpoints with the following exceptions:
\begin{itemize}
    \item $\beta_p=0$ and $z_1=0$.
    \item ${\cal E}_y>0$ and
    $ z_1  = - \, \frac{2}{p+1}\beta_p \sqrt{{\cal E}_y} $.
\end{itemize}
Moreover, computing the squared Riemann tensor, $R_{MNPQ}\,R^{MNPQ}$, one can see that the non--singular options reduce to
\begin{itemize}
    \item $\beta_p=0$ and ${\cal E}_x={\cal E}_y = z_1=0$.

    This corresponds to the D3 brane background, which is altogether free of singularities.
    \item $p=0$, ${\cal E}_{y}>0$ and
   $ z_1 \ = \ \frac{3}{2} \sqrt{{\cal E}_y} = \frac{3}{2} \,\sqrt{{\cal E}_x}$.

   This corresponds to the charged dilatonic black 0-branes of~\cite{hor_strom}. This case is recovered since for $p=0$ our ansatz becomes the one for black branes, and as $r\to\infty$ one approaches the outer horizon, which shields the true singularity.
\end{itemize}

\subsubsection[\sc The Uncharged Brane Limit]{\sc The Uncharged Brane Limit}
\label{sec:uncharged_brane_limit}

The uncharged branes of section~\ref{sec:curvature} can be recovered from the general charged solutions with ${\cal E}_x>0$ and ${\cal E}_y\geq0$.
From the expression of the charge in eq.~\eqref{eq:tension_and_charge_nonBPS}, one can see that the limit $Q_p\to 0$ is approached as $\left|{\cal F}\left({\cal E}_y, r_1\right)\right|\to\infty$, and the tension can remain finite provided $|r_1|\to\infty$ while ${\cal E}_y$ is kept finite. The charged brane background in eqs.~\eqref{eq:deformation_backgroun_general} then becomes
\bea
ds^2 &=&  \ e^{-\frac{2\,r}{R}}dx_{p+1}^2 \nonumber \ + \ e^{\frac{2(p+1)\,r}{\left(D-p-3\right)R}} \left[\frac{(D-p-3)\sigma}{\ell} \,\sinh{\left(\frac{r}{\sigma}\right)}\right]^{-\,\frac{2\left(D-p-2\right)}{D-p-3}} dr^2 \nonumber \\ &+& e^{\frac{2(p+1)\,r}{\left(D-p-3\right)R}}\left[ \frac{(D-p-3)\sigma}{\ell} \,\sinh{\left(\frac{r}{\sigma}\right)}  \right]^{-\,\frac{2}{D-p-3}} \ell^2 \, d\Omega_{D-p-2}^2 \ , \nonumber\\
\phi & = & \phi_0 \ + \ \phi_1 r \ ,
\eea
with
\bea
\sigma & = & \frac{1}{\sqrt{{\cal E}_x}} \ , \qquad \frac{1}{R} \ = \  \frac{4(D-p-3)}{\Delta}\,\sqrt{{\cal E}_y}\ - \ \frac{4(D-2)\beta_p}{\Delta} \, z_1 \ ,  \nonumber \\
\phi_1 &=& - \ \frac{(D-2)^2\beta_p}{\Delta}\, \sqrt{{\cal E}_y} \  - \ \frac{4(D-2)(p+1)}{\Delta}\, z_1 \ .
\eea
The Hamiltonian constraint of eq.~\eqref{eq:deformed_charged_z1} can then be cast in the form
\beq
\frac{1}{R^2} \ \frac{(p+1)(D-2)}{D-p-3} \ + \ \frac{4\,\phi_1^2}{D-2} \ = \ \frac{1}{\sigma^2} \ \frac{D-p-2}{D-p-3} \ ,
\eeq
which is the Hamiltonian constraint of the uncharged case in eq.~\eqref{ham_red}.
The limiting form of the background depends on three real parameters, which determine the asymptotic value of the dilaton, $\phi'(0)$ and the tension of the brane, and the solution is indeed the background profile of uncharged branes presented in eqs.~\eqref{eq:uncharged_harmonic}.

\subsubsection[\sc Some special cases]{\sc Some special cases}

For the reader's convenience, let us conclude this section by briefly summarizing some interesting options that arise in ten dimensions, the case relevant for String Theory.
\begin{itemize}
\item To begin with, in ten dimensions and for generic values of $p$ the backgrounds reduce to
\bea \label{eq:deformation_backgroun_general_sum}
ds^2 \!\!&=&\!\! \left|\frac{{\cal F}\left({\cal E}_y, r+r_1 \right)}{{\cal F}\left({\cal E}_y, r_1 \right)}\right|^{-\frac{7-p}{8}} e^{\beta_p z_1 r }  dx_{p+1}^2 \nonumber \\  \!\!&+&\!\! \left|\frac{{\cal F}\left({\cal E}_y, r+r_1 \right)}{{\cal F}\left({\cal E}_y, r_1 \right)}\right|^{\frac{p+1}{8}} \!\!e^{-\frac{(p+1)\beta_p}{7-p} z_1 r } \left(\frac{dr^2}{\left|\frac{(7-p)}{\ell } \, {\cal F}\left({\cal E}_x, r \right) \right|^\frac{2(8-p)}{7-p}}  +  \frac{\ell^2\ d\Omega_{8-p}^2}{\left|\frac{(7-p)}{\ell } \, {\cal F}\left({\cal E}_x, r \right) \right|^\frac{2}{7-p}} \right) \ , \nonumber \\
e^\phi \!\!&=&\!\!  e^{\phi_0}\ e^{-\frac{p+1}{2} z_1 r}  \left|\frac{{\cal F}\left({\cal E}_y, r+r_1 \right)}{{\cal F}\left({\cal E}_y,r_1 \right)}\right|^{-\beta_p} \ , \nonumber  \\
{\cal H}_{p+2} \!\!&=&\!\! - \ \epsilon \  e^{\beta_p \phi_0} \ \left|{\cal F}\left({\cal E}_y, r_1 \right)\right|^{-1}   \left|\frac{{\cal F}\left({\cal E}_y, r+r_1 \right)}{{\cal F}\left({\cal E}_y, r_1 \right)}\right|^{-2} \ dx^0 \wedge \ldots \wedge dx^p \wedge dr \ .
\eea
\item For the special case of deformed D3 branes in ten dimensions, taking the self--duality into account, the preceding expression change slightly, as we have explained, and take the form
\bea
ds^2 \!\!&=&\!\! \left|\frac{{\cal F}\left({\cal E}_y, r+r_1 \right)}{{\cal F}\left({\cal E}_y, r_1 \right)}\right|^{-\frac{1}{2}}  \!\!dx_{4}^2 \,+\, \left|\frac{{\cal F}\left({\cal E}_y, r+r_1 \right)}{{\cal F}\left({\cal E}_y, r_1 \right)}\right|^{\frac{1}{2}} \!\! \left(\frac{dr^2}{\left|\frac{4}{\ell } \, {\cal F}\left({\cal E}_x, r \right) \right|^\frac{5}{2}}  +  \frac{\ell^2\ d\Omega_{5}^2}{\left|\frac{4}{\ell } \, {\cal F}\left({\cal E}_x, r \right) \right|^\frac{1}{2}} \right) \ , \nonumber \\
e^\phi \!\!&=&\!\!  e^{\phi_0}\ e^{-2 z_1 r}   \ , \nonumber  \\
{\cal H}_{5} \!\!&=&\!\! - \ \frac{\epsilon}{\sqrt{2}} \   \left|{\cal F}\left({\cal E}_y, r_1 \right)\right|^{-1}   \left[\left|\frac{{\cal F}\left({\cal E}_y, r+r_1 \right)}{{\cal F}\left({\cal E}_y, r_1 \right)}\right|^{-2} \ dx^0 \wedge \ldots \wedge dx^3 \wedge dr \ + \  \text{vol}_{S^5}\right] \ ,
\eea
with the standard normalization for the kinetic term of the form field. In this case the expressions for tension and charge reduce to
\beq
{\cal T}_3 \ = \   \frac{\Omega_{5}}{2\kappa_{10}^2} \,\ell^{\,5}\,\frac{{\cal F}'\left({\cal E}_y, r_1\right)}{{\cal F}\left({\cal E}_y, r_1\right)} \ , \qquad
Q_3  \ = \ \frac{\epsilon}{\sqrt{2}} \ \frac{\Omega_{5}}{2\kappa_{10}^2} \,\ell^{\,5}\,\left|{\cal F}\left({\cal E}_y, r_1\right)\right|^{-1} \ ,
\eeq
and in general
\beq
{\cal T}_3{}^2 \ - \ {2} \ {\sign}\left({{\cal E}_y}\right) \,Q_3{}^2 \ = \ \left(\frac{\Omega_{5}\,\sqrt{\left|{{\cal E}_y}\right|}}{2\kappa_{10}^2}\right)^2 \, \ell^{\,10} \ .
\eeq
Note that only the dilaton profile depends on $z_1$, whose value does not affect tension and charge in this case.
\item For $p=0$ the background reduces to
\bea
ds^2 \!\!&=&\!\! -\left|\frac{{\cal F}\left({\cal E}_y, r+r_1 \right)}{{\cal F}\left({\cal E}_y, r_1 \right)}\right|^{-\frac{7}{8}} e^{-\frac{3}{4} z_1 r }  (dx^0)^2 \nonumber \\  \!\!&+&\!\! \left|\frac{{\cal F}\left({\cal E}_y, r+r_1 \right)}{{\cal F}\left({\cal E}_y, r_1 \right)}\right|^{\frac{1}{8}} \!\!e^{\frac{3}{28} z_1 r } \left(\frac{dr^2}{\left|\frac{7}{\ell } \, {\cal F}\left({\cal E}_x, r \right) \right|^\frac{16}{7}}  +  \frac{\ell^2\ d\Omega_{8}^2}{\left|\frac{7}{\ell } \, {\cal F}\left({\cal E}_x, r \right) \right|^\frac{2}{7}} \right) \ , \nonumber \\
e^\phi \!\!&=&\!\!  e^{\phi_0}\ e^{-\frac{1}{2} z_1 r}  \left|\frac{{\cal F}\left({\cal E}_y, r+r_1 \right)}{{\cal F}\left({\cal E}_y,r_1 \right)}\right|^{\frac{3}{4}} \ , \nonumber  \\
{\cal H}_{2} \!\!&=&\!\! - \ \epsilon \  e^{-\frac{3}{4} \phi_0} \ \left|{\cal F}\left({\cal E}_y, r_1 \right)\right|^{-1}   \left|\frac{{\cal F}\left({\cal E}_y, r+r_1 \right)}{{\cal F}\left({\cal E}_y, r_1 \right)}\right|^{-2} \ dx^0 \wedge  dr \ .
\eea
\end{itemize}
These are deformations of the BPS D0 brane, for which $z_1=w=0$ and
\beq
r_1 \ =\  e^{\frac{3}{4}\phi_0} \, \Omega_8 \, \frac{\ell}{(2\pi)^7} \ .
\eeq
The only option, among them, with a non--singular outer horizon, where the dilaton approaches a constant value, was identified by Horowitz and Strominger in~\cite{hor_strom}. It corresponds to $z_1>0$ and $w \, = \, \frac{23}{9} \, z_1^2$, so that $ {\cal E}_x \, = \,  {\cal E}_y \,= \, \frac{4}{9} \, z_1^2 $, and the resulting expressions for tension and charge read
\bea
{\cal T}_0 &=& e^{\frac{3}{4} \phi_0} \ \frac{\Omega_{8}}{2\kappa_{10}^2} \, \ell^{\,8}\,   \frac{2}{3} \, z_1  \left[ \coth\left(\frac{2}{3} \, z_1  \, r_1\right) \ + \ \frac{9}{7} \right] \ , \nonumber \\
Q_0 &=& \epsilon \  e^{\frac{3}{4} \phi_0} \ \frac{\Omega_{8}}{2\kappa_{10}^2}  \, \ell^{\,8}\, \frac{2}{3} \, z_1 \left[\sinh{\left(\frac{2}{3} \, z_1  \, r_1\right)}\right]^{-1} \ .
\eea
As $r \to \infty$, in this case one approaches indeed a horizon, where the harmonic coordinates end, which shields a true singularity. Note that $\left|{\cal T}_0\right|>\left|Q_0\right|$ for all these solutions.

\section[Vacua with Bulk Tadpoles and Internal Spheres]{\sc Vacua with Bulk Tadpoles and Internal Spheres} \label{sec:bulk_tadpole}

In this section we include the tadpole contribution in the equations, while turning off $p$-form fluxes. Our aim is to explore new types of vacua that are allowed, in the presence of a tadpole potential~\eqref{eq:tadpole_potential}, when the internal space has a positive curvature, thus complementing previous results obtained in~\cite{ms21_2}. In addition, the ensuing discussion can also have some bearing on the limiting behaviors of uncharged branes in the presence of tadpole potentials, in regions where the axisymmetric backgrounds discussed in~\cite{mrs24_2} leave way to spherically symmetric ones.
As explained in Section~\ref{sec:harmonic}, this analysis still relies on the ansatz of eqs.~\eqref{eq:ansatz} used in the previous two sections to discuss branes in Minkowski backgrounds, so that one can naturally address the issue here. The result is nonetheless a complicated coupled non-linear system, so that simple exact solutions along the lines of the preceding sections do not exist. However, one can identify analytically several interesting limiting behaviors, which suffice to illuminate the overall shapes of the solutions when supplemented by numerical analysis. As usual, the results obtained in this fashion are directly significant for String Theory only insofar as the string coupling is sufficiently small and the spacetime curvature is also bounded. Nonetheless, as we shall see, the asymptotics of the vacua possess some neat and instructive features.

\subsection[\sc The Equations]{\sc The Equations}

Although we shall eventually focus on the interesting case $D=10$, let us begin by formulating the system for generic values of $D$, resorting to the combinations
\bea
X&=&(p+1)A\ + \ (D-p-3)C \ , \nonumber  \\
W&=&(p+1)A\ +\ (D-p-2)C \ + \ \frac{\gamma}{2}\, \phi \ , \nonumber  \\
K&=&\phi\ + \ \gamma \ \frac{(D-2)^2}{8}\, A \ .
\eea
In terms of these, one can recover the original functions of eqs.~\eqref{Eqs_back} as
\bea \label{eq:XW_ABCphi}
A &=& \frac{16 (D-p-2)}{\Xi}\, X  \ - \ \frac{16 (D-p-3)}{\Xi}\, W \ + \ \frac{8(D-p-3)\gamma}{\Xi}\, K \ , \nonumber \\
B &=& \frac{(D-2)^2(D-p-2)\gamma^2}{\Xi} \,X \ + \ \frac{16 (p+1)}{\Xi}\, W \ - \ \frac{8(p+1)\gamma}{\Xi} \,K \ , \nonumber \\
C &=& \frac{\left((D-2)^2 \gamma^2 - 16 (p+1)\right)}{\Xi} \,X \ + \ \frac{16(p+1)}{\Xi} \,W \ - \ \frac{8(p+1)\gamma }{\Xi} \,K \ , \nonumber \\
\phi &=& -\frac{2(D-2)^2(D-p-2)\gamma}{\Xi} \,X \ + \ \frac{2(D-2)^2 (D-p-3)\gamma }{\Xi} \,W \ + \ \frac{16(p+1)}{\Xi} \,K \ ,
\eea
where we have defined
\beq
\Xi \ = \ 16(p+1)\ + \ (D-2)^2(D-p-3)\gamma^2 \ .
\eeq
The equations of motion thus become
\bea \label{eq:tadpole-curvature_system}
X''&=&\frac{(D-p-3)^2}{\ell^2}\ e^{2X}\ -\ T \ e^{2W} \ , \nonumber  \\
W''&=&\frac{(D-p-2)(D-p-3)}{\ell^2}\ e^{2X}\ + \ \frac{D-2}{16}\left(\gamma^2 -\gamma_c^2\right)T\ e^{2W} \ ,  \nonumber \\
K''&=&0 \ ,
\eea
where the critical value of $\gamma$ is
\beq
\gamma_c\ =\ \frac{4\sqrt{D-1}}{D-2} \ ,
\eeq
while the corresponding Hamiltonian constraint reads
\bea \label{eq:XW_ham_constr}
0 &=& \frac{(D-p-3)(D-p-2)}{\ell^2}\,e^{2X} \ -\ T e^{2W} \ + \ \frac{1}{\Xi}\Big[ 16(D-2)(D-p-3) (W')^2   \nonumber \\
&-&  32(D-2)(D-p-2)W'X'-(D-p-2) (D-2)^2 \left(\gamma^2-\gamma_c^2\right)(X')^2 \nonumber \\
&+& \frac{64 (p+1)}{D-2}(K')^2\Big] \ .
\eea

The equation for $K$ is thus solved by a linear function, which is determined by the Hamiltonian constraint up to an additive constant, and one is left with two coupled non--linear equations for $X$ and $W$.
Note that the last coefficient in the second equation vanishes for the critical value of $\gamma$ that pertains to the orientifold models. For this value of $\gamma$ the Hamiltonian constraint also simplifies, and becomes
\bea
0&=&\frac{(D-p-3)(D-p-2)}{\ell^2}\, e^{2X}\ - \ T \,e^{2W}\ -\ 2X'W'\ +\ \frac{D-p-3}{D-p-2}\,(W')^2 \nonumber \\
&+&\frac{4(p+1)}{(D-2)^2 (D-p-2)}\,(K')^2 \ .
\eea

\subsection[\sc Exact Results]{\sc Exact Results}

The above system does admit a simple class of exact solutions, which can be obtained demanding that $X$ and $W$ differ by a constant and read
\bea \label{eq:XsimW}
X &=& -\ \log\left[\sqrt{\frac{(D-p-3)\Xi}{\ell^2 \, (16-(D-2)^2 \gamma^2)}}\, \rho \,\cosh\left(\frac{r}{\rho}\right)\right] \ , \nonumber \\
W&=& X \ + \ \frac{1}{2}\,\log\left[\frac{16(D-p-3)(D-2)}{\ell^2\, T \left(16-(D-2)^2 \gamma^2\right)}\right] \ , \nonumber \\
K &=& k_0 \ + \ \sqrt{\frac{(D-2)\left[(D-2)^3 \gamma^2 + p (16-(D-2)^2 \gamma^2)\right]}{64(p+1)}}\ \frac{r}{\rho} \ ,
\eea
where $\rho$ and $k_0$ are integration constants. These solutions exist provided
\beq\label{eq:gamma_less}
\gamma \ < \ \frac{4}{D-2} \ ,
\eeq
or $\gamma< \frac{1}{2}$ in ten dimensions, which is instrumental to guarantee their reality.
Unfortunately, eq.~\eqref{eq:gamma_less} excludes the tadpole potentials of ten--dimensional non--supersymmetric strings, for which $\gamma \geq \frac{3}{2}$, and lies below the sphere level in all dimensions. Moreover, there are no scaling solutions of the type
\beq
X \ = \ - \ \log\left(\frac{r}{\rho}\right) \ + \ c_x \ , \qquad W \ = \ - \ \log\left(\frac{r}{\rho}\right) \ + \ c_w
\eeq
with internal spheres, but cosmological scaling solutions are possible with negative internal curvature.

In general, one can understand the limiting behavior of the solutions by taking a close look at the possible endpoints, where divergences of $X$ or $W$ occur.
From eqs.~\eqref{eq:tadpole-curvature_system}, for all values of $\gamma$ that are relevant to String Theory, and in particular for $\gamma\geq\gamma_c$,
\beq
(W-X)''\ = \ \frac{D-p-3}{\ell^2}\,e^{2X}\ + \ \frac{(D-2)}{16}\, \left(\gamma^2\,-\,\frac{16}{(D-2)^2}\right)\,T \,e^{2W}\,>\,0 \ , \label{convexity_WX}
\eeq
and moreover
\beq
W'' \ > \ 0 \ , \label{convexity_W}
\eeq
in view of eqs.~\eqref{eq:tadpole-curvature_system}.

Taking these results into account, one can now classify the possible asymptotic behaviors.
The convexity of $W-X$, together with the lack of solutions with $X$ and $W$ differing by a constant in the cases relevant to String Theory, suffice to identify the few available options.

For brevity, in the following we shall ignore all integration constants that do not affect the leading dependence on $r$. In this fashion, all constant limits will be effectively replaced by vanishing ones. Moreover, we shall use the freedom of shifting and reflecting the $r$ coordinate, thus focusing on the two limits $r\to 0^+$ and $r\to+\infty$. Any other case is equivalent to these up to reflections and translations.

\subsection[\sc Limiting Behaviors where the Tadpole is Sub--Dominant]{\sc Limiting Behaviors where the Tadpole is Sub--Dominant}
We can now begin our analysis, starting from cases where $e^X\gg e^W$. When this inequality holds, one can neglect the tension--dependent terms in eqs.~\eqref{eq:tadpole-curvature_system} and the system reduces to
\bea \label{eq:XW_XWK}
X''&=&\frac{(D-p-3)^2}{\ell^2}\ e^{2X} \ , \nonumber  \\
W''&=&\frac{(D-p-2)(D-p-3)}{\ell^2}\ e^{2X} \ , \nonumber \\
K''&=& 0 \ ,
\eea
while the Hamiltonian constraint reduces to
\bea
0 &=& \frac{(D-p-3)(D-p-2)}{\ell^2}\,e^{2X}  \ + \ \frac{1}{\Xi}\Big[ 16(D-2)(D-p-3) (W')^2   \nonumber \\
&-&  32(D-2)(D-p-2)W'X'-(D-p-2) (D-2)^2 \left(\gamma^2-\gamma_c^2\right)(X')^2 \nonumber \\
&+& \frac{64 (p+1)}{D-2}(K')^2\Big] \ .
\eea
The first of eqs.~\eqref{eq:XW_XWK} can be turned into
\beq
(X')^2 \ = \ \frac{(D-p-3)^2}{\ell^2} e^{2X} \ + \  {\cal E}_X \ ,
\eeq
while the equations for $W$ and $K$ are solved by
\beq
W=\frac{D-p-2}{D-p-3}X + w_1 r  \ , \qquad K=k_1 r  \ ,
\eeq
with $w_1$, $k_1$ real constants. The different constants are related by the
Hamiltonian constraint according to
\beq \label{eq:EX_ham_cnstr}
{\cal E}_X \ = \ \frac{64(D-p-3)(p+1)k_1^2 \ + \ 16 (D-2)^2 (D-p-3)^2 w_1^2}{(D-2)(D-p-2)\Xi} \ ,
\eeq
so that ${\cal E}_X =\frac{1}{\rho^2}\geq 0$, and consequently within regions of this type the solution for $X$ is approximately
\beq
X\ =\ - \ \log\left({\frac{D-p-3}{\ell}\, \rho \ \sinh\frac{r}{\rho}}\right)
\eeq
if ${\cal E}_X>0$, while the case ${\cal E}_X=0$ can be recovered from this expression as $\rho \to \infty$.
The condition that the tadpole potential be subdominant reads\beq \label{eq:XW_condition}
\frac{\rho}{\ell}\,\sinh\left(\frac{r}{\rho}\right)  e^{-(D-p-3) w_1 r }\ \gg \ 1 \ .
\eeq
It can only hold as $r\to\infty$, and one should distinguish two cases.
\begin{enumerate}
    \item[a. \ ] If ${\cal E}_X=0$, the Hamiltonian constraint sets $w_1=k_1=0$ and the condition of eq.~\eqref{eq:XW_condition} is automatically satisfied as  $r\to\infty$. This first type of behaviour, however, is somewhat trivial. In this case
        \bea
        X &\sim & - \ \log \frac{r}{\ell} \ , \nonumber \\
        W &\sim & - \ \frac{D-p-2}{D-p-3}\log \frac{r}{\ell} \ , \nonumber \\
        K &\sim& 0\ ,
        \eea
        but these limiting forms describe flat space in harmonic coordinates, as seen in eq.~\eqref{flat_limit}.
        Contrary to what the harmonic gauge coordinate $r$ might suggest, this asymptotic solution merely captures the values of the fields at a regular point, where the dilaton also approaches a constant value.
    \item[b. \ ] If ${\cal E}_x>0$ a second, more interesting, type of behaviour is possible. It concerns a region of space that is encountered as $r\to\infty$, provided eq.~\eqref{eq:XW_condition} holds, which demands that
        \beq
        \left(D-p-3\right)\rho \,w_1 \ < \ 1 \label{eq:ineq_w1} \ .
        \eeq
        In this case
        \bea
        X &\sim & - \ \frac{r}{\rho} \ , \nonumber \\
        W &\sim & \left(w_1 - \frac{D-p-2}{\left(D-p-3\right)\rho}\right) r \ . \nonumber \\
        K &\sim& k_1 r\ ,
        \eea
        and the background has the limiting behavior
        \beq \label{eq:linear_metric}
        ds^2 \ \sim \ e^{2 a r} \,dx_{p+1}^2 \ + e^{2 b r} dr^2 \ + \ e^{2 c r} \, \ell^2 \, d\Omega_{D-p-2}^2 \ , \qquad \phi \ \sim \ \phi_1 r \ ,
        \eeq
        with
        \bea
        a &=& \frac{8(D-p-3) (\gamma k_1 \ - \ 2 w_1)}{\Xi} \ , \nonumber \\
        b &=& - \ \frac{D-p-2}{(D-p-3)\rho} \ - \ \frac{8(p+1) (\gamma k_1 - 2 w_1)}{\Xi} \ , \nonumber \\
        c&=&  - \ \frac{1}{(D-p-3)\rho} \ - \ \frac{8(p+1)(\gamma k_1 \ - \ 2 w_1)}{\Xi}\ , \nonumber \\
        \phi_1 &=& \frac{16(p+1)k_1 \ + \ 2(D-2)^2 (D-p-3) \gamma\,w_1 }{\Xi} \ .
        \eea

        This background reduces to eq.~\eqref{eq:uncharged_near_singularity}, and recovers the tadpole--free near--singularity regions of the uncharged branes of Section~\ref{sec:curvature}, up to the identifications
        \bea \label{eq:uncharged_branes_constant_matching}
        \sigma &=& \rho \ , \nonumber \\
        \cos\alpha &=&  \frac{8\left(D-p-3\right)\rho\left(\gamma\,k_1\,-\,2\,w_1\right)}{\Xi} \ \sqrt{\frac{(p+1)(D-2)}{(D-p-2)}} \ .
        \eea
        A closer look at the behavior of $e^B$ and $e^\phi$ shows that, in the relevant case of ten dimensions, the singularity always lies at a finite proper distance. The string coupling can diverge, vanish or approach a finite value there, depending on whether the combination
        \beq
        8\left(p+1\right) k_1 \ + \ \left(D-2\right)^2\left(D-p-3\right) \gamma\,w_1 \ ,
        \eeq
        which is proportional to $\sin\alpha$, as defined in eq.~\eqref{eq:uncharged_branes_constant_matching}, is positive, negative or zero. The inequality~\eqref{eq:ineq_w1}, for which this behavior is valid, only constraints the angle $\alpha$ to lie in the interval
        \beq
        \alpha_0 \ - \ \pi \ - \ \arcsin\left(\frac{1}{\cal A}\right) \  < \  \alpha \  < \  \alpha_0 \ + \ \arcsin\left(\frac{1}{\cal A}\right) \ ,
        \eeq
        where
        \beq
        {\cal A} \ = \ \frac{\Xi^{\frac{3}{2}}}{64(p+1)} \ \sqrt{\frac{D-p-2}{D-2}}\ , \qquad \tan \alpha_0 \ = \ \frac{4}{(D-2)\gamma} \ \sqrt{\frac{p+1}{D-p-3}} \ .
        \eeq
    \end{enumerate}

\subsection[\sc Limiting Behaviors where the Tadpole is Dominant]{\sc Limiting Behaviors where the Tadpole is Dominant}
\label{sec:dominant_tadpole}

In the opposite regime, where $e^W\gg e^X$, the system of equations reduces to
\bea \label{eq:WXeqs}
X''&=& -\ T \ e^{2W} \ , \nonumber  \\
W''&=& \frac{D-2}{16}\left(\gamma^2 -\gamma_c^2\right)T\ e^{2W} \ ,  \nonumber \\
K''&=&0 \ ,
\eea
together with the Hamiltonian constraint
\bea
0 &=&-\ T e^{2W} \ + \ \frac{1}{\Xi}\Big[ 16(D-2)(D-p-3) (W')^2  \ - \ 32(D-2)(D-p-2)W'X' \nonumber \\
&-&(D-p-2) (D-2)^2 \left(\gamma^2-\gamma_c^2\right)(X')^2 \ + \ \frac{64 (p+1)}{D-2}(K')^2\Big] \ .
\eea
These equations coincide with those analyzed in~\cite{ms21_2}, where the internal space was a torus rather than a sphere. Therefore, the asymptotics that we are discussing capture limiting behaviors of some of the exact solutions found there, near one or the other end.
Since the behaviors at the two ends in~\cite{ms21_2} do not necessarily imply that $e^W\gg e^X$, as must be the case in the asymptotics we are after, only the limiting regions where this inequality holds are relevant to the present analysis. In the following we shall identify these regions explicitly.
The general lesson is that the singularity always lies at a finite proper distance, and the string coupling can diverge or vanish there. We can now examine in detail the available options.

In all these cases the second of eqs.~\eqref{eq:WXeqs} can be turned into
\beq
(W')^2 \ = \ \frac{D-2}{16}\left(\gamma^2-\gamma_c^2\right) T e^{2W} \ + \ {\cal E}_W \ ,
\eeq
while
\beq
K \ = \ k_1\, r  \ ,
\eeq
and when $\gamma \neq \gamma_c$
\beq
X \ = \ - \ \frac{16}{(D-2)\left(\gamma^2 - \gamma_c^2\right)} W \ + \ x_1 r \ , \qquad K \ =\  k_1 r \ ,
\eeq
while the Hamiltonian constraint reads
\beq
{\cal E}_W= \left(\gamma^2 - \gamma_c^2\right) \frac{(D-2)^3 (D-p-2)\left(\gamma^2 - \gamma_c^2\right) x_1^2 \ - \ 64(p+1)k_1^2}{16\,\Xi} \ ,
\eeq
where $x_1$, $k_1$ are real constants.
The condition $e^W\gg e^X$ becomes
\beq\label{eq:WX_condition}
\frac{\gamma^2 - \frac{16}{(D-2)^2}}{\gamma^2 - \gamma_c^2}\ W \ - \ x_1 \,r \  \gg \ 0\ .
\eeq

In general, the allowed limiting behaviors depend on the range of $\gamma$. We thus begin by considering the range $\gamma>\gamma_c$.

For $\gamma=\gamma_c$, as we have stressed, some of the preceding steps do not apply, and we shall return to it at the end.

\subsubsection[\sc Asymptotic Behaviors for $\gamma>\gamma_c$]{\sc Asymptotic Behaviors for $\gamma>\gamma_c$}

The solution for $W$ depends on the sign of ${\cal E}_W$. Letting
\beq
\frac{1}{r_0} \ = \  \sqrt{\frac{D-2}{16}\left|\gamma^2-\gamma_c^2\right| T} \ , \qquad \frac{1}{\rho^2} \ = \ \left|{\cal E}_W\right| \ ,
\eeq
one finds
\bea
W &= \ - \ \log\left[\frac{\rho}{r_0} \,\sinh\left(\frac{r}{\rho}\right) \right] \qquad & \mathrm{if \ \ } {\cal E}_W > 0 \ , \nonumber \\
W &= \ - \ \log\left(\frac{r}{r_0} \right) \qquad\qquad\qquad & \mathrm{if \  } {\cal E}_W = 0 \ , \nonumber \\
W &= \ - \  \log\left[\frac{\rho}{r_0} \,\sin\left(\frac{r}{\rho}\right) \right] \qquad & \mathrm{if \ }{\cal E}_W < 0 \ .
\eea
    \begin{enumerate}
    \item[c. \ ] In all three cases $W$ is dominated near $r=0$ by an expression of the form
    \beq
    W \ \sim \ - \ \log \frac{r}{\ell}   \ .
    \eeq
    Here $\gamma>\gamma_c$, and therefore it is a fortiori larger than $\frac{4}{D-2}$, so that eq.~\eqref{eq:WX_condition} always holds near $r=0$, and we are thus led to identify another type of asymptotic behavior:
        \bea
        X &\sim & r_0^2\,T\,\log\frac{r}{\ell} \ , \nonumber \\
        W &\sim &  - \ \log\frac{r}{\ell}\ , \nonumber \\
        K &\sim& 0 \ .
        \eea
        These expressions recover the singular asymptotics of the vacua described in Section 3.2.2 of~\cite{ms21_2}. In our case the metric and the dilaton thus approach the limiting forms
        \bea
        ds^2 &\sim& \left(\frac{r}{\ell}\right)^{2|\lambda|}\left(dx_{p+1}^2 \ + \ \ell^2 d\Omega_{D-p-2}^2\right) \ + \ \left(\frac{r}{\ell}\right)^{2(D-1)|\lambda|} dr^2    \ , \nonumber \\
        e^\phi &\sim&  \left(\frac{r}{\ell}\right)^{-\frac{2(D-1)\gamma |\lambda|}{\gamma_c^2}}  \ ,
        \eea
        with
        \beq
        \lambda \ = \ \frac{1}{D-1} \left(1-\frac{\gamma^2}{\gamma_c^2}\right)^{-1} \ ,
        \eeq
        and $r=0$ hosts curvature and string--coupling singularities.

        \item[d. \ ] The fourth type of behaviour is encountered as $r\to\infty$ if ${\cal E}_W\geq 0$. If ${\cal E}_W>0$, the asymptotic behavior of the solutions as $r\to\infty$ is captured by
        \bea
        X &\sim & \left(x_1 \ + \ r_0^2\,T \right) r \ , \nonumber \\
        W &\sim &  - \ \frac{r}{\rho}\ , \nonumber \\
        K &\sim& k_1 r \ , \label{eq:cased}
        \eea
        with $x_1$ given in terms of $\rho$ and $k_1$ as~\footnote{Only this sign for $x_1$ is compatible with the inequality~\eqref{eq:XW_condition}.}
        \beq
        x_1 \ = \  - \ \frac{4}{\left(D-2\right)^2}\left[\frac{\left(D-2\right) \Xi}{(D-p-2)\left(\gamma^2 -\gamma_c^2\right)^2 \rho^2}\ + \ \frac{4(p+1)\left(D-2\right) k_1^2}{(D-p-2)(\gamma^2 - \gamma_c^2)}\right]^{\frac{1}{2}} \ ,
        \eeq
        and provided the inequality
        \beq \label{eq:k1rhoineq}
        64 (p+1)\left(\gamma^2 - \gamma_c^2\right)k_1^2 \ > \ \left[(D-2)^3 (D-p-2)\left(\gamma^2 - \frac{16}{(D-2)^2}\right)^2-16 \, \Xi\right] \frac{1}{\rho^2}
        \eeq
        holds, so that the dominance condition~\eqref{eq:WX_condition} is satisfied in the limit.
        The metric is of the form~\eqref{eq:linear_metric}, but the coefficients $a,b,c,\phi$ are now more involved.
        This case captures deformations of the vacua with positive ``energy'' studied in~\cite{ms21_2}. In fact, since $W\to-\infty$, the solutions approach a tensionless Kasner behavior as $r \to \infty$. This also captures the asymptotics of the uncharged branes of Section~\ref{sec:curvature} close to their cores, with parameters $(\sigma,\alpha)$ that are determined by $(\rho,k_1)$.

        The ${\cal E}_W\to 0$ limit can be recovered as the $\rho\to\infty$ limit of the preceding results.
        $W$ has a logarithmic dependence on $r$, but this is subleading with respect to the linear profiles of $X $ and $K$, which must be non-vanishing since $k_1^2>0$ from eq.~\eqref{eq:k1rhoineq}.
        In this fashion, one connects to the zero ``energy'' solutions considered in Section 3.2.2 of~\cite{ms21_2}.
    \end{enumerate}

\subsubsection{\sc Asymptotic Behaviors for $\gamma<\gamma_c$}

If $e^W\gg e^X$ and $\gamma<\gamma_c$, a range that is not directly relevant for ten--dimensional strings but is interesting nonetheless, the solution for $W$ is
\beq
W \ = \ - \log\left[\frac{\rho}{r_0}\cosh\left(\frac{r}{\rho}\right)\right] \ ,
\eeq
and now ${\cal E}_W$ is bound to be positive.
    \begin{enumerate}
        \item[e. \ ] In this case, one finds a single family of asymptotics as $r\to\infty$, for which
        \bea
        X &\sim & \left(x_1 \ - \ \frac{16}{(D-2)\left(\gamma_c^2-\gamma^2\right)\rho}\right) r \ , \nonumber \\
        W &\sim &  - \ \frac{r}{\rho}\ , \nonumber \\
        K &\sim& k_1 r \ .
        \eea
        The Hamiltonian constraint links the various constants according to
        \beq
        \frac{1}{\rho^2} \ = \ \frac{\gamma_c^2 - \gamma^2}{16\, \Xi}\left[64 (p+1)k_1^2 \ +\  (D-2)^3 (D-p-2)\left(\gamma_c^2 - \gamma^2\right)x_1^2\right] \ ,
        \eeq
        and the inequality
        \beq
        \frac{\gamma^2 - \frac{16}{(D-2)^2}}{\left(\gamma_c^2 - \gamma^2\right)\rho} \ > \ x_1
        \eeq
        must hold to guarantee that $e^W\gg e^X$.
        The limiting form of the background is the same as in case d, and captures again an asymptotic limit of the vacua discussed in Section 3.2.1. of~\cite{ms21_2}. These solutions approach a tensionless Kasner behavior for large values of $r$, as is the case the uncharged branes of Section~\ref{sec:curvature} close to their cores.
        \end{enumerate}

\subsubsection{\sc Asymptotic Behaviors for $\gamma=\gamma_c$}

The case $e^W\gg e^X$ with $\gamma=\gamma_c$ is more subtle, since the large exponential disappears altogether from the equation for $W$. The systems reduces to
\bea
X''&=& -\ T \ e^{2W} \ , \nonumber  \\
W''&=& 0 \ ,  \nonumber \\
K''&=&0 \ ,
\eea
and the last two equations are solved by
\bea
W &=& w_1\,r \ , \nonumber \\
K &=& k_1\,r \ ,
\eea
but the solution for $X$ depends crucially on whether or not $w_1=0$. At the same time, the Hamiltonian constraint reduces to
\beq
0\ = \ - \ T \,e^{2W}\ -\ 2 w_1 \,X'\ +\ \frac{D-p-3}{D-p-2}\,w_1^2 \ + \ \frac{4(p+1)}{(D-2)^2 (D-p-2)}\,k_1^2 \ .
\eeq

One can see that in all cases $X \to -\infty$ in the asymptotic region, which is approached as $r\to\infty$, but it is still convenient to distinguish three types of behavior.
    \begin{enumerate}
        \item[f. \ ] The first two types of asymptotic behavior emerge when $w_1 \neq 0$, and are captured by
        \bea
        X &\sim & -\ \frac{T}{4 w_1^2}\ e^{2 w_1 r} \ +\ \frac{1}{2\,w_1}\left[\frac{\left(D-p-3\right) w_1^2}{\left(D-p-2\right)}\ + \  \frac{4(p+1)\,k_1^2}{(D-2)^2 (D-p-2)}\right] r \ , \nonumber \\
        W &\sim &  w_1 r \ , \nonumber \\
        K &\sim& k_1 r \ ,
        \eea
        but the sign of $w_1$ has a major effect on the asymptotics.

        If $w_1>0$ only the exponential matters in $X$ and $e^W$ is automatically much larger than $e^X$. In this case, letting
        \beq
        U(r) \ = \ -\ \frac{T }{8(D-2)w_1^2}\ e^{2 w_1 r} \ ,
        \eeq
        the background approaches
        \bea \label{eq:critical_dmlike}
        ds^2  &\sim& e^{2 U(r)} \left( dx_{p+1}^2 \ + \  \ell^2\, d\Omega_{D-p-2}^2\right) \ + \  e^{2\left(D-1\right) U(r)}\ dr^2 \nonumber \\
        e^{\phi} &\sim& e^{-\, \left(D-2\right)\, \sqrt{D-1}\, U(r)}\ .
        \eea
        This limiting behavior is akin to that of the original nine--dimensional solution of~\cite{dm_vacuum}, but the present background is actually conformal to the direct product of a $(p+2)$--dimensional Minkowski space and an internal sphere $S^{D-p-2}$. Referring to eqs.~\eqref{ExA1phi1}, this solution also captures the near--core behavior of uncharged branes, with
        \beq
        \cos\alpha \ = \ - \ \sqrt{\frac{p+1}{(D-2)(D-p-2)}} \ ,
        \eeq
        while $\sigma$ can be set to any finite value in this asymptotic region by a rescaling and a shift the variable $r$, which have no other effect.

        On the other hand, if $w_1<0$ the exponential term can be neglected altogether and the original hierarchy $e^W\gg e^X$ is guaranteed provided
        \beq
        w_1^2 \ < \ \frac{4(p+1)\, k_1^2}{(D-2)^2 (D-p-1)} \ .
        \eeq
        The three metric functions $A$, $B$ and $C$ are then linear in $r$, and one recovers a tensionless Kasner behavior, or the near--core region of uncharged branes.
        In all cases, the end result is a deformation of the vacua of~\cite{ms21_2}, and $w_1$ is related to the parameter $\beta$ of that paper.
        \item[g. \ ] The last type of behavior is encountered as $r\to\infty$ with $w_1=0$, so that
        \bea
        X &\sim & - \ \frac{2(p+1)\, k_1^2}{(D-2)^2(D-p-2)}\, r^2 \ , \nonumber \\
        W &\sim &  0 \ , \nonumber \\
        K &\sim& k_1 r \ ,
        \eea
        and the linear terms introduced by $k_1$ are subdominant. This limiting behavior captures the critical $\beta\to 0$ case of~\cite{ms21_2}, and letting
        \beq
        U(r) \ = \ -\ \frac{(p+1)k_1^2}{(D-2)^3 (D-p-2)} \ r^2 \ ,
        \eeq
        it approaches again the form in eqs.~\eqref{eq:critical_dmlike}.
    \end{enumerate}

In conclusion, the different types of behavior that we have identified recover the asymptotics of the toroidal compactifications of~\cite{ms21_2}, or the limiting behaviors of the uncharged brane solutions of Section~\ref{sec:curvature} their near--core regions.
The issue is linking to one another two asymptotic behaviors in the absence of exact solutions. Some help, in this respect, comes from the fact the profile of $K$ is exactly linear. Therefore the values of $k_1$ must coincide at both ends, which leads to a number of restrictions summarized in Appendix~\ref{app:asymptotics} and is also a convenient tool in the numerics.
\begin{figure}[ht]
\centering
\begin{tabular}{cc}
\includegraphics[width=63mm]{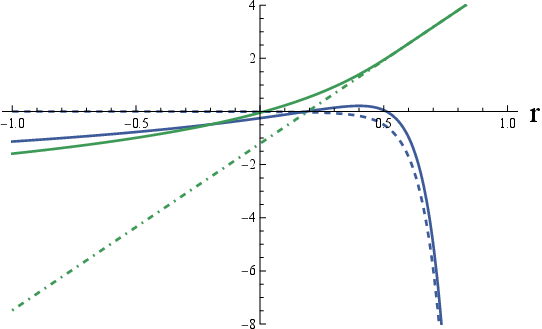} \qquad \qquad &
\includegraphics[width=63mm]{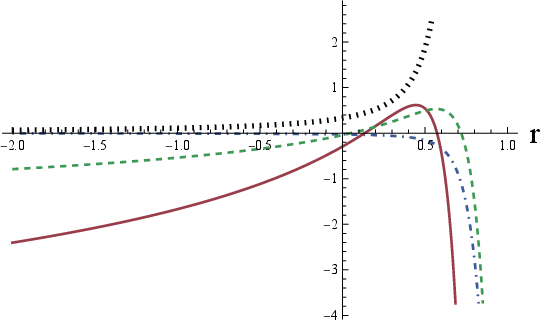} \\
\end{tabular}
\caption{\small The left panel illustrates the two functions $X$ (blue, solid curve) and $W$ (green, solid curve) for a solution corresponding to $D=10$ and $\gamma=\frac{3}{2}$ that starts from the limiting behavior of type $a$ for $r$ large and negative and approaches the asymptotic limit of type $f$ (dashed and dot-dashed curves). The right panel illustrates the corresponding behaviors of $A(r)$ (blue, dot-dashed), $B(r)$ (red, solid), $C(r)$ (green, dashed) and $\phi(r)$ (black, dotted).}
\label{fig:orientifold_vacuum_1}
\end{figure}
\begin{figure}[ht]
\centering
\begin{tabular}{cc}
\includegraphics[width=52mm]{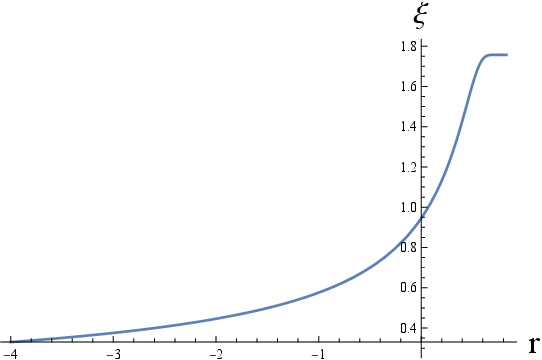} \qquad \qquad &
\includegraphics[width=30mm]{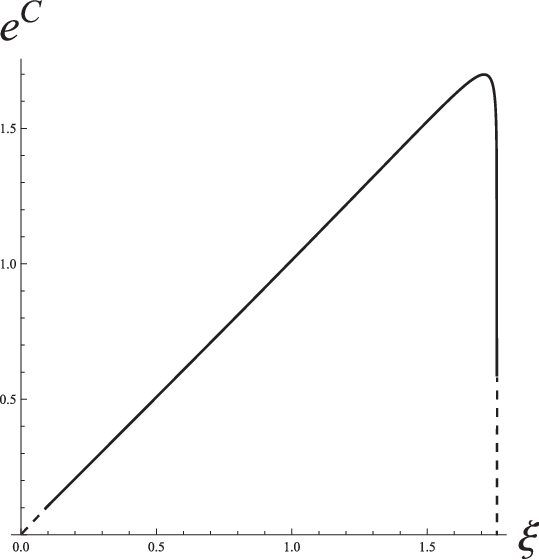} \\
\end{tabular}
\caption{\small The left panel illustrates how the proper radial distance $\xi$ depends on $r$ for the solution illustrated in figure~\ref{fig:orientifold_vacuum_1}, which starts from the limiting behavior of type $a$ for $r$ large and negative. The right panel illustrates the behavior of $e^C$, which determines the radii of the orthogonal spheres, as a function of the proper distance $\xi$. The dashed regions have been completed by hand, since the limiting behavior is clear but obtaining it directly would be too demanding in terms of computer time. The orthogonal spheres grow as in flat space before the tadpole contribution suddenly takes over.}
\label{fig:orientifold_vacuum_2}
\end{figure}

\subsection{\sc Numerical Results} \label{sec:numerics}

We have explored different options for $\gamma=\frac{3}{2}$, the value corresponding to the ten--dimensional orientifolds of ~\cite{susy95,sugimoto}, performing numerical tests for $D=10$ and for $p=0,\ldots,6$. Figs.~\ref{fig:orientifold_vacuum_1} and \ref{fig:orientifold_vacuum_2} describe our findings for $p=5$, but the results are qualitatively similar in all cases. Note that the tension $T$ can be identified with $\frac{1}{\ell^2}$ up to a shift of $W$ in eqs. \eqref{eq:tadpole-curvature_system}, and then $\frac{1}{\ell^2}$ can be absorbed defining a dimensionless radial variable, so that the numerical results apply
to both ten--dimensional orientifolds.

A general lesson is that in all cases the spacetime closes abruptly, at a finite distance from the origin, when one departs from it along the radial $r$ direction, due to the tadpole potential. Independently of whether the spacetime is flat near the origin or a brane is contained there, the behavior near the radial end is always along the lines of~\cite{dm_vacuum}, or of its extensions in~\cite{ms21_2}, since the curvature has a subleading role in the region where spacetime ends abruptly.

The left panel in fig.~\ref{fig:orientifold_vacuum_1} illustrates the $r$-dependence of the metric functions $X$ and $W$ for a \emph{vacuum} solution. This is flat around the origin, which is approached, in harmonic coordinates, as $r \to - \infty$ but closes, due to the tadpole potential, at a finite distance from it. The results are also compared with the asymptotic limit of type f of the previous section, corresponding to the dashed and dot-dashed curves, which are closely approached near the radial end. The right panel in fig.~\ref{fig:orientifold_vacuum_1} describes the corresponding behavior of $A(r)$, $B(r)$, $C(r)$ and $\phi(r)$. Fig.~\ref{fig:orientifold_vacuum_2} illustrates, for this solution, the $r$-dependence of the proper distance $\xi$ from the origin and the behavior of $e^C$, which characterizes the size of the transverse spheres, as a function of $\xi$.

\begin{figure}[ht]
\centering
\begin{tabular}{cc}
\includegraphics[width=63mm]{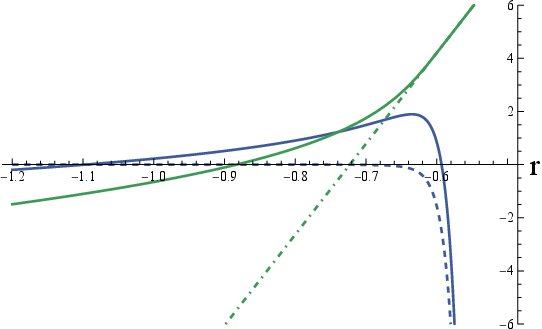} \qquad \qquad &
\includegraphics[width=63mm]{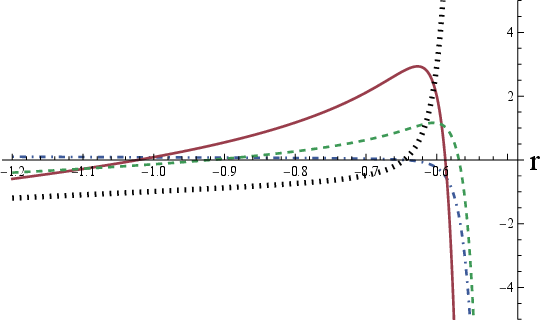} \\
\end{tabular}
\caption{\small The left panel illustrates the two functions $X$ (blue, solid curve) and $W$ (green, solid curve) for a solution corresponding to $D=10$, $\gamma=\frac{3}{2}$ and $p=5$ that starts from the limiting behavior of type $b$ with $k_1=0$ for $r$ large and negative and approaches the asymptotic limit of type $f$ (dashed and dot-dashed curves). The right panel illustrates the corresponding behavior of $A(r)$ (blue, dot-dashed), $B(r)$ (red, solid), $C(r)$ (green, dashed) and $\phi(r)$ (black, dotted).}
\label{fig:orientifold_vacuum_3}
\end{figure}
\begin{figure}[ht]
\centering
\begin{tabular}{cc}
\includegraphics[width=50mm]{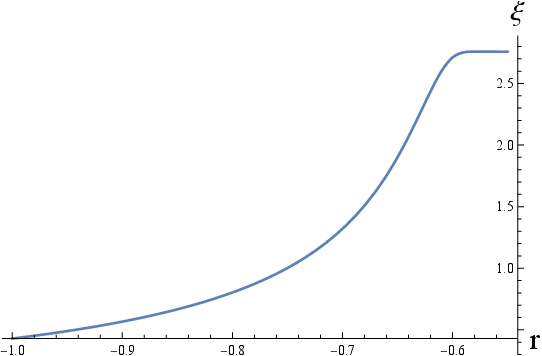} \qquad \qquad &
\includegraphics[width=25mm]{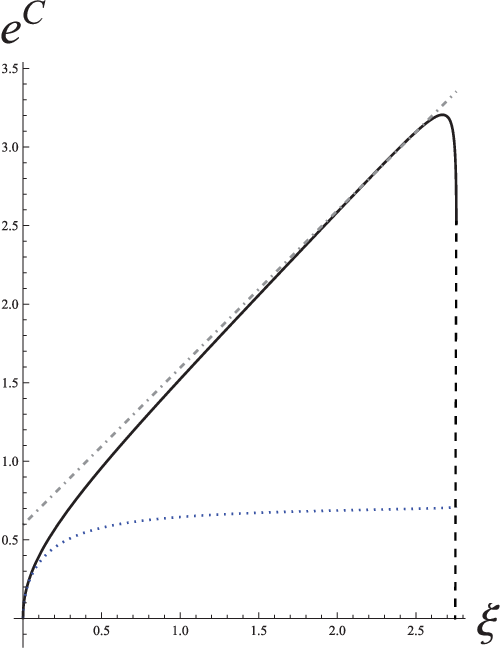} \\
\end{tabular}
\caption{\small The left panel illustrates how the proper radial distance $\xi$ depends on $r$ for the solution of fig.~\ref{fig:orientifold_vacuum_3}, which starts from the limiting behavior of type $b$ for $r$ large and negative and readily approaches the behavior of type $f$. The right panel illustrates the behavior of $e^C$ (black, solid line), which determines the radii of the orthogonal spheres, as a function of the proper distance $\xi$. The dashed region on the right of this curve was again completed by hand, since the limiting behavior is clear but obtaining it directly would be too demanding in terms of computer time. Note that, differently from the case in fig.~\ref{fig:orientifold_vacuum_2}, the curve in the right panel only approaches a linear flat--space behavior (gray, dot-dashed line) after a while. As $\xi \to 0$, $e^C \sim \xi^\frac{7}{15}$, which is the behavior near the core for an uncharged brane of Section~\ref{sec:curvature} (blue dotted line) consistent with the chosen asymptotics.}
\label{fig:orientifold_vacuum_4}
\end{figure}
Figs.~\ref{fig:orientifold_vacuum_3} and \ref{fig:orientifold_vacuum_4} describe similar findings, for $p=5$, when one of the uncharged branes of Section~\ref{sec:curvature} is contained at the origin. As $r \to - \,\infty$ the metric approaches in this case eq.~\eqref{dsalpha} with $\Gamma=\frac{5}{4}$ and $\sigma=\frac{3}{2}$, so that $e^C \sim \xi^{7/15}$, but the spacetimes still closes, away from the brane, as in the preceding example. There also solutions with $\gamma=\frac{3}{2}$ and $k_1=0$, which interpolate between two asymptotic behaviors of type f, so that the tadpole dominates at both ends.

\begin{figure}[ht]
\centering
\begin{tabular}{cc}
\includegraphics[width=63mm]{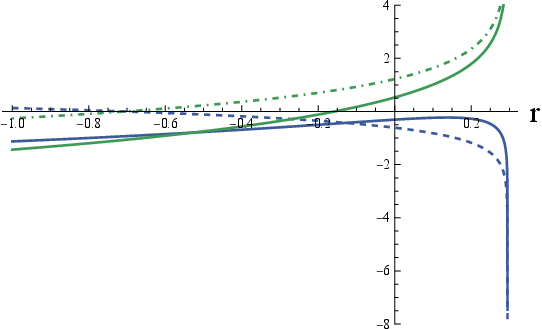} \qquad \qquad &
\includegraphics[width=63mm]{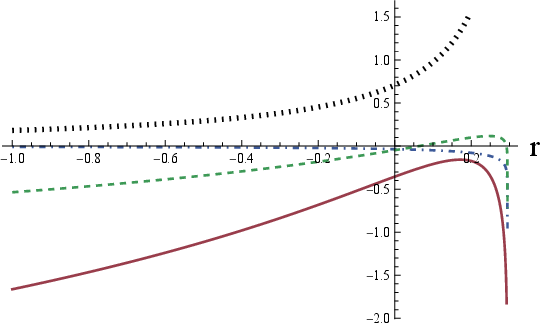} \\
\end{tabular}
\caption{\small The left panel illustrates the two functions $X$ (blue, solid curve) and $W$ (green, solid curve) for a solution corresponding to the SO(16) $\times$ SO(16) model and $p=5$ that start from the limiting behavior of type $a$ for $r$ large and negative and approach the asymptotic limit of type $c$ for a finite value of $r$ (dashed and dot-dashed curves). The right panel illustrates the corresponding behavior of $A(r)$ (blue, dot-dashed), $B(r)$ (red, solid), $C(r)$ (green, dashed) and $\phi(r)$ (black, dotted).}
\label{fig:heterotic_vacuum_1}
\end{figure}
\begin{figure}[!ht]
\centering
\begin{tabular}{cc}
\includegraphics[width=50mm]{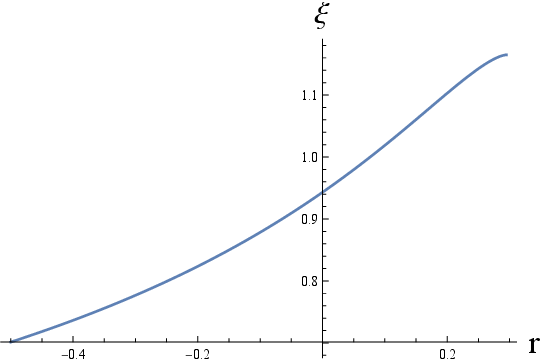} \qquad \qquad &
\includegraphics[width=23mm]{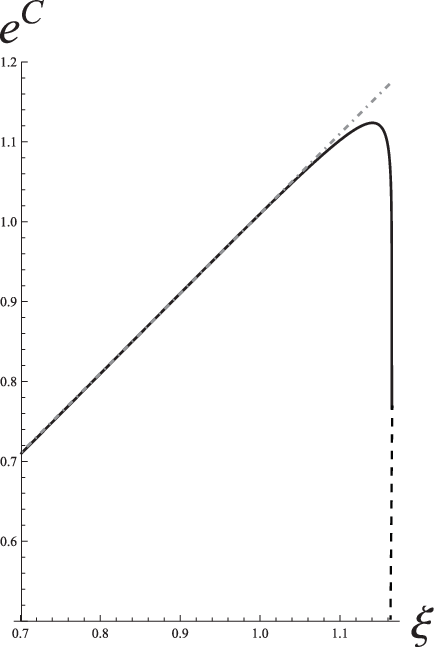} \\
\end{tabular}
\caption{\small The left panel illustrates how the proper radial distance $\xi$ depends on $r$ for the solution corresponding to the SO(16) $\times$ SO(16) model and $p=5$ that starts from the limiting behavior of type $a$ for $r$ large and negative. The right panel illustrates the behavior of $e^C$, which determines the radii of the orthogonal spheres, as a function of the proper distance $\xi$. The dashed region has been completed by hand, since the limiting behavior is clear but obtaining it directly would be too demanding in terms of computer time. The gray dot-dashed line corresponds to a flat--space growth of the orthogonal spheres that holds before the vacuum energy contribution takes over.}
\label{fig:heterotic_vacuum_2}
\end{figure}
We have also explored these types of behavior for $\gamma=\frac{5}{2}$, the value corresponding to the  SO(16) $\times$ SO(16) heterotic string of~\cite{so1616}. Qualitatively, one finds the similar results, but the faster growth of the tadpole term restricts the region captured by the solutions of~\cite{ms21_2} and has the overall effect of making the numerical results less accurate. Figs.~\ref{fig:heterotic_vacuum_1} and \ref{fig:heterotic_vacuum_2} illustrate a solution that approaches the vacuum as the proper distance $\xi \to 0$, and is thus the counterpart of those in figs.~\ref{fig:orientifold_vacuum_1} and \ref{fig:orientifold_vacuum_2}. This case has the interesting feature of involving a collapse of type c, the only case where, as explained in~\cite{ms21_2}, the collapse is directly sensitive to the tadpole potential. On the other hand, for $\gamma \leq \frac{3}{2}$ the collapse is always dominated by a tadpole--free behavior, even in strong--coupling regions, a property that for $\gamma> \frac{3}{2}$ only holds in asymptotic weak--coupling regions (type d).

\section[Vacua with Bulk Tadpoles, Fluxes and Internal Spheres]{\sc Vacua with Bulk Tadpoles, Fluxes and Internal Spheres} \label{sec:bulk_tadpole_flux}

In this section we finally address the most general setups depending on a single variable $r$, with $\mathfrak{h}_{p+1}=0$. To this end, we begin by defining the convenient combinations
\bea
X&=&(p+1)A\ + \ (D-p-3)C \ , \nonumber  \\
Y&=& (p+1)A\ + \ \beta_p \, \phi \ ,  \nonumber \\
W&=&(p+1)A\ +\ (D-p-2)C \ + \ \frac{\gamma}{2}\, \phi \ ,
\eea
which are, in eqs.~\eqref{Eqs_back}, the exponents associated to curvature, flux and tadpole in the harmonic gauge.
These quantities determine the original functions as
\bea \label{eq:XYW_ABCphi}
A &=& \frac{2(D-p-2)\beta_p}{\Theta}\, X  \ + \ \frac{(D-p-3)\gamma}{\Theta}\, Y \ - \ \frac{2(D-p-3)\beta_p}{\Theta}\, W \ , \nonumber \\
B &=& \frac{(p+1)(D-p-2)\gamma}{\Theta}\, X \ - \ \frac{(p+1)\gamma}{\Theta} \, Y \ + \ \frac{2(p+1)\beta_p}{\Theta}\, W \ , \nonumber \\
C &=& \frac{(p+1)(\gamma-2\beta_p)}{\Theta}\, X  \ - \ \frac{(p+1)\gamma}{\Theta}\, Y \ + \ \frac{2(p+1)\beta_p}{\Theta}\, W \ , \nonumber \\
\phi &=& - \, \frac{2(p+1)(D-p-2)}{\Theta}\, X  \ + \ \frac{2(p+1)}{\Theta}\, Y \ + \ \frac{2(D-p-3)(p+1)}{\Theta}\, W \ ,
\eea
where
\beq
\Theta \ = \ (p+1)\,\Big[2\beta_p + (D-p-3)\gamma\Big] \ ,
\eeq
Defining the three combinations
\bea
\Delta &=& 4(D-p-3)(p\,+\,1) \ +\  (D-2)^2 \beta_p{}^2  \ ,  \nonumber \\
\Xi &=& 16(p+1)\ + \ (D-2)^2(D-p-3)\gamma^2 \ , \nonumber \\
\gamma_c &=& \frac{4\sqrt{D-1}}{D-2} \ ,
\eea
when expressed in terms of $X$, $Y$ and $W$ the background equations become
\bea \label{eq:tadpole-curvature-flux_system}
X''&=&\frac{(D-p-3)^2}{\ell^2}\ e^{2X}\ -\ T \ e^{2W} \ , \nonumber  \\
Y''&=&\frac{\Delta}{4(D-2)} \ \frac{H_{p+2}^2}{2} \  e^{2Y} \ + \ \left(\frac{D-2}{8} \,\beta_p \,\gamma \,-\,\frac{p+1}{D-2}\right) \,T\ e^{2W} \ ,  \nonumber \\
W''&=&\frac{(D-p-2)(D-p-3)}{\ell^2}\ e^{2X}\ + \ \left(\frac{D-2}{8} \,\beta_p \,\gamma \, - \,\frac{p+1}{D-2}\right) \ \frac{H_{p+2}^2}{2} \ e^{2Y} \nonumber  \\
    &+&  \frac{D-2}{16}\left(\gamma^2 -\gamma_c^2\right)\,T\ e^{2W}  \ ,
\eea
while the Hamiltonian constraint reads
\bea
0 &=& \frac{(D-p-3)(D-p-2)}{\ell^2}\,e^{2X} \ -\ T \, e^{2W} \ -  \ \frac{H_{p+2}^2}{2} \ e^{2Y} \ + \ \frac{p+1}{(D-2) \, \Theta^2} \, \Bigg\{ \nonumber \\
    &-& (D-p-2)\left[32(p+1)\left(\frac{D-2}{8} \,\beta_p \,\gamma \, - \,\frac{p+1}{D-2}\right) \,+\, \frac{p+1}{D-2}\, \Xi \, - \, 4\frac{D-1}{D-2}\, \Delta \right] (X')^2 \nonumber \\
    &+&  \Xi \  (Y')^2 \, + \, 4 (D-p-3)\Delta \ (W')^2 \,-\, 8(D-p-2)\Delta \ X'W' \nonumber \\
    &+& 32(D-2)\left(\frac{D-2}{8} \,\beta_p \,\gamma \, - \,\frac{p+1}{D-2}\right) Y' \left[(D-p-2)X'- (D-p-3)W'\right]   \Bigg\} \ .
\eea

Note that performing the redefinitions
\bea\label{eq:shifts}
\widetilde{X} &=& X \ +  \ \frac{1}{2}\log\left[(D-p-3)(D-p-2)\right] \ , \nonumber \\
\widetilde{Y} &=& Y \ + \  \frac{1}{2}\log\left(\frac{H_{p+2}^2\, \ell^2}{2}  \right) \ , \nonumber \\
\widetilde{W} &=& W \ + \  \frac{1}{2} \log\left(T \, \ell^2 \right) \ ,
\eea
and working in dimensionless units by rescaling the radial variable according to $r\to\frac{r}{\ell}$, one can recast the whole system in the simpler form
\bea \label{eq:tadpole-curvature-flux_system_reduced}
X''&=&\frac{D-p-3}{D-p-2}\ e^{2X}\ - \ e^{2W} \ , \nonumber  \\
Y''&=&\frac{\Delta}{4(D-2)}  \  e^{2Y} \ + \ \left(\frac{D-2}{8} \,\beta_p \,\gamma \,-\,\frac{p+1}{D-2}\right)\ e^{2W} \ ,  \nonumber \\
W''&=& e^{2X}\ + \ \left(\frac{D-2}{8} \,\beta_p \,\gamma \, - \,\frac{p+1}{D-2}\right) \ e^{2Y} \ + \  \frac{D-2}{16}\left(\gamma^2 -\gamma_c^2\right) \ e^{2W}  \ ,
\eea
where, or brevity, here and in the following, we are leaving the ``$\sim$'' implicit on all redefined quantities.  This result, together with the corresponding Hamiltonian constraint, which becomes
\bea
0 &=&  e^{2X} \, -  \, e^{2W} \, -  \,  e^{2Y} \, + \, \frac{p+1}{(D-2) \, \Theta^2} \, \Bigg\{-(D-p-2)\Bigg[32(p+1)\left(\frac{D-2}{8} \,\beta_p \,\gamma \, - \,\frac{p+1}{D-2}\right) \nonumber \\
&+& \frac{p+1}{D-2}\, \Xi \, - \, 4\frac{D-1}{D-2}\, \Delta \Bigg] (X')^2 \,+ \, \Xi \,  (Y')^2 \, + \, 4 (D-p-3)\Delta \, (W')^2 \,-\, 8(D-p-2)\Delta \, X'W' \nonumber \\
    &+& 32(D-2)\left(\frac{D-2}{8} \,\beta_p \,\gamma \, - \,\frac{p+1}{D-2}\right) Y' \left[(D-p-2)X'- (D-p-3)W'\right]   \Bigg\} \ ,
\eea
are a convenient starting point to discuss the qualitative features of the solutions.

Inverting the system~\eqref{eq:tadpole-curvature-flux_system}, or alternatively the system~\eqref{eq:tadpole-curvature-flux_system_reduced} and taking into account that the exponential functions have a definite sign one can derive three linear inequalities involving $X''$, $Y''$ and $W''$. Their indications, however, are less clear than in the preceding section, since they mix all three variables.

\subsection{\sc A Special Case}

A special case presents itself when
\beq
\frac{D-2}{8} \,\beta_p \,\gamma \, - \,\frac{p+1}{D-2}  \ = \ 0 \ . \label{eq:conditionY}
\eeq
This is only possible, in ten--dimensional strings, for the orientifold D5 brane.
When eq.~\eqref{eq:conditionY} holds, $Y$ decouples and the corresponding equation can be integrated once in terms an energy-like quantity, obtaining
\beq
(Y')^2 \ = \ \frac{\Delta}{4(D-2)} \ e^{2Y} \ + \ {\cal E}_y \ ,
\eeq
while the remaining system becomes
\bea
X''&=&\frac{D-p-3}{D-p-2}\ e^{2X}\ - \ e^{2W} \ , \nonumber  \\
W''&=& e^{2X}\ + \  \frac{D-2}{16}\left(\gamma^2 -\gamma_c^2\right) \ e^{2W}  \ ,
\eea
and the Hamiltonian constraint reduces to
\bea
0 &=& e^{2X} \ -\  e^{2W} \ + \ \frac{4(D-2)}{\Delta} {\cal E}_y  \  + \ \frac{1}{\Xi}\Big[ 16(D-2)(D-p-3) (W')^2   \nonumber \\
&-&  32(D-2)(D-p-2)W'X'-(D-p-2) (D-2)^2 \left(\gamma^2-\gamma_c^2\right)(X')^2 \Big] \ .
\eea
When the contributions involving $e^{2X}$, which arise from the curvature, are neglected, this case affords a class of exact solutions that were discussed in detail in~\cite{ms21_2}.

The present results are similar to those captured by eqs.~\eqref{eq:tadpole-curvature_system} and~\eqref{eq:XW_ham_constr} for $X$ and $W$, where ${\cal E}_y$ is replaced by $(K')^2$, up to a positive proportionality constant.
There is a key difference, however, since ${\cal E}_y$ can have any sign while $(K')^2$ is inevitably non-negative. Undoing the redefinitions performed in eqs.~\eqref{eq:shifts}, the explicit solution for $Y$ is finally
\beq
Y \ = \ - \log\left|\sqrt{\frac{\Delta}{8(D-2)}} \, H_{p+2} \ {\cal F}\left({\cal E}_y, r+r_1 \right)\right| \ ,
\eeq
where ${\cal F}$ was defined in eq.~\eqref{cases_F}.

When eq.~\eqref{eq:conditionY} holds, starting from any solution $(A_0, B_0, C_0, \phi_0)$ without fluxes
one can build a solution $(A, B, C, \phi)$ with fluxes. Indeed, eq.~\eqref{eq:conditionY} implies that replacing $Y$ in eqs.~\eqref{eq:XYW_ABCphi} with $\beta_p \,K$ reproduces eqs.~\eqref{eq:XW_ABCphi}. The detailed correspondence is
\bea
A &=& A_0 \ + \ \frac{(D-p-3)\gamma}{\Theta}\, \left(Y \ - \ \beta_p \,k_1\, r \right) \ , \nonumber \\
B &=& B_0  \ - \ \frac{(p+1)\gamma}{\Theta} \, \left(Y \ - \ \beta_p \,k_1\, r \right) \ , \nonumber \\
C &=& C_0  \ - \ \frac{(p+1)\gamma}{\Theta}\, \left(Y \ - \ \beta_p \,k_1\, r \right) \ , \nonumber \\
\phi &=& \phi_0  \ + \ \frac{2(p+1)}{\Theta}\, \left(Y \ - \ \beta_p \,k_1\, r \right) \ ,
\eea
with
\beq
{\cal E}_y \ = \ \beta_p^2 \ k_1^2  \ .
\eeq
In this case, the form field strength is determined by eq.~\eqref{eq:ansatz}, and reads
\beq
{\cal H}_{p+2} \ = \  \frac{8(D-2)}{\Delta \, H_{p+2}}\left[ {\cal F}\left({\cal E}_y, r+r_1 \right)\right]^{-2} \ dx^0 \wedge \ldots \wedge dx^p \wedge dr \ ,
\eeq
but $Y$ generically introduces another singular point at $r\,=\,-\,r_1$, where the solution must end.

When ${\cal E}_y\geq0$, one can thus reconstruct asymptotics with fluxes starting from the corresponding ones without fluxes of Section~\ref{sec:bulk_tadpole}.
When ${\cal E}_y<0$, the asymptotic behavior of the system can be deduced as in the proceeding section, up to the replacement of $k_1^2$ with $- \,k_1^2$, but the range of $r$ will be confined to the interval between two consecutive zeros of ${\cal F}\left({\cal E}_y, r+r_1 \right)$, so that one will never reach the $r\to\infty$ regions.

\begin{figure}[ht]
\centering
\begin{tabular}{cc}
\includegraphics[width=63mm]{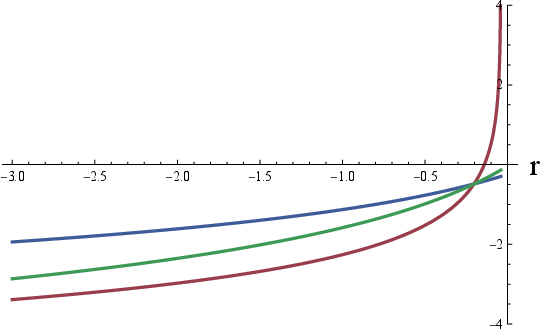} \qquad \qquad &
\includegraphics[width=63mm]{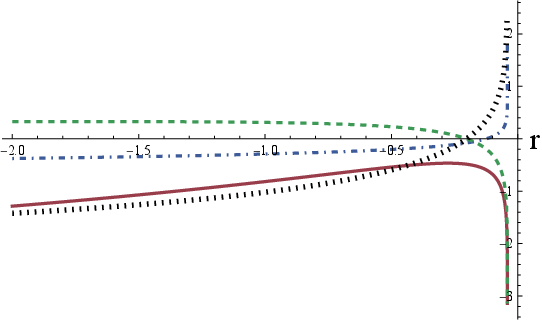} \\
\end{tabular}
\caption{\small The left panel illustrates the three functions $X$ (blue), $Y$ (red) and $W$ (green) for a solution corresponding to $D=10$, $\gamma=\frac{3}{2}$ and $p=5$ that starts from the core of a BPS D5 brane for $r$ large and negative (asymptotic limit of type $5$  in the orientifold classification of Section~\ref{sec:orientifold_classification}) and approaches an asymptotic limit of type $7$ in the same classification. The right panel illustrates the corresponding behavior of $A(r)$ (blue, dot-dashed), $B(r)$ (red, solid), $C(r)$ (green, dashed) and $\phi(r)$ (black, dotted). The spacetime closes as in the preceding cases, but the tensor lines of force converge on the outer surface, where the field strength becomes singular. The behavior of the proper distance $\xi(r)$ and of $e^{C(\xi)}$ is along the lines of the preceding cases.}
\label{fig:D5_tadpole}
\end{figure}
A physically interesting sub-case obtains when ${\cal E}_y=0$, so that $k_1=0$. Here, $Y$ has the logarithmic profile
\beq
Y \ = \ - \log\left|\sqrt{\frac{\Delta}{8(D-2)}} \, H_{p+2} \ (r+r_1)\right| \ ,
\eeq
and only a few cases of Section~\ref{sec:bulk_tadpole} with $k_1=0$ apply: case a, case b with the condition in eq.~\eqref{eq:k1w1_values}, and case f with $w_1>0$ if $\gamma=\gamma_c$, case c if $\gamma>\gamma_c$ and case e with the condition in eq.~\eqref{eq:k1x1_values} if $\gamma<\gamma_c$.
Focusing on the cases that are directly relevant for ten-dimensional strings, the condition in eq.~\eqref{eq:conditionY} only holds, as we have seen, if $\gamma=\gamma_c$ and $p=5$. Then, for the asymptotics corresponding to cases b and f, the logarithmic contribution makes $Y$ subleading with respect to $X$, and one is effectively led back to the flux-less asymptotics.
The only genuinely new limiting behavior corresponds to case a, for which $X$, $Y$ and $W$ have, as $r\to\infty$, the logarithmic profiles
\beq
X \ \sim \ - \ \log\frac{r}{\ell} \ , \qquad Y \ \sim \ - \ \log\frac{r}{\ell} \ , \qquad W \ \sim \ - \ \frac{3}{2}\log\frac{r}{\ell} \ ,
\eeq
up to constants.
Equivalently, the  metric coefficients, the dilaton and the seven--form field strength approach
\bea
A & \sim & - \ \frac{1}{8}\log\frac{r}{\ell} \ , \qquad  B \ \sim \  -  \ \frac{9}{8}\log\frac{r}{\ell} \ ,\qquad C \ \sim \ - \ \frac{1}{8}\log\frac{r}{\ell} \ ,\nonumber \\
\phi & \sim & - \ \frac{1}{2}\log\frac{r}{\ell} \ , \qquad {\cal H}_7 \ \sim \ \frac{\ell}{r^2} \, dx^0\wedge \ldots \wedge dx^5\wedge dr \ .
\eea
This result is apparently  along the lines of case a of Section~\ref{sec:bulk_tadpole}, but its meaning is quite different. Rather that describing a generic non--singular point, these expressions capture the leading behavior near the singularity of a BPS D5 brane, as can be seen comparing them with eqs.~\eqref{eq:BPS_near_horizons}.
Figure~\ref{fig:D5_tadpole} illustrates some numerical results on this case.

\subsection{\sc A Class of Exact Solutions}\label{sec:exact_solutions}

The system of eqs.~\eqref{eq:tadpole-curvature-flux_system_reduced} has a class of exact solutions where $X$, $Y$ and $W$ differ by a constant, provided
\beq\label{eq:XYW_condition}
-\frac{D-p-3}{2}\gamma<\beta_p<0 \ .
\eeq
In terms of the shifted fields and of the dimensionless radial variable $r$, the solution reads
\bea
X &=& - \log\left[\sqrt{\frac{(p+1)(2\beta_p+(D-p-3)\gamma)}{(D-p-2)((p+1)\gamma-2\beta_p)}} \  r \right]\ , \nonumber  \\
Y &=& - \log\left[\sqrt{\frac{(p+1)(2\beta_p+(D-p-3)\gamma)}{(D-2)\gamma}} \  r \right]  \ , \nonumber \\
W &=& - \log\left[\sqrt{\frac{(p+1)(2\beta_p+(D-p-3)\gamma)}{-2(D-2)\beta_p}} \  r \right]   \ .
\eea
These expressions capture, as special case, the $AdS_3\times S^7$ orientifold vacuum and the $AdS_7 \times S^3$ heterotic vacuum of~\cite{gm_02,ms_16}. If these solutions are perturbed, one finds that the resulting linearized system has power--like solutions that, however, can blow up at both ends of the $r>0$ range. Therefore, these solutions do not behave as attractors in the whole range, and are actually unstable at one end or the other.
The analogy with the standard $AdS_5 \times S^5$ case~\cite{maldacena} suggested a possible correspondence~\cite{Basile_D1} between the $AdS_3\times S^7$ orientifold vacuum and the near-horizon region of the D1 brane. However, the strong coupling present in the region makes the argument at most suggestive.

We can now turn to the analysis of the asymptotic regimes, which can be done on rather general grounds, but for convenience we confine our attention to the ten--dimensional non--supersymmetric string and treat separately the two orientifolds and the heterotic model.

\subsection{\sc The Ten--Dimensional Orientifolds}
\label{sec:orientifold_classification}

For the two ten--dimensional orientifolds   $\gamma=\frac{3}{2}$ and $\beta_p=\frac{p-3}{4}$, with $p=1,3,5$ so that, using the redefinitions in eqs.~\eqref{eq:shifts}, the equations become
\bea
X'' &=& \frac{7-p}{8-p}\,e^{2X}\ - \ e^{2W}\ , \nonumber\\
Y'' &=& 2 \,e^{2Y}\ - \ \frac{5-p}{4} \,e^{2W}\ , \nonumber\\
W'' &=& e^{2X} \ - \ \frac{5-p}{4} \,e^{2Y} \ , \nonumber\\
0 &=&  e^{2X}\ -\ e^{2W} \ -\ e^{2Y}\ -\ \frac{1}{(9-p)^2(p+1)}\Big[ -32(7-p) (W')^2\nonumber \\
&-&(8-p)\left((5-p)X'-4Y'\right)^2\ +\ 8 W' \left(8(8-p)X'-(7-p)(5-p)Y'\right)\Big] \ . \label{gen_system}
\eea
When $p\neq5$ this system is still rather complicated. We thus confine our attention to the \emph{possible} asymptotic behaviors, although the lack of conserved quantities and convenient convexity conditions will not allow to match pairs of them as in previous sections.

There are in principle eight different options for the limiting behavior of the above system.
\begin{enumerate}
    \item In regions where the three functions $X$, $Y$ and $W$ tend to $-\infty$, only the second derivatives are left, and they all approach a linear dependence on $r$, as pertains to Kasner-like solutions.
    \item If the three functions $X$, $Y$ and $W$ differ by a constant, one is led to the solutions described in the preceding section, and the inequality~\eqref{eq:XYW_condition} selects $p=1$ for the orientifolds. However, the instability implies that this behavior is fine tuned and not generic.
    \item If $Y$ and $W$ differ by a constant, and both dominate over $X$, the $W$ equation becomes
    \beq\label{eq:orient_YW}
    Y''\ = \ - \ \frac{5-p}{4}  \ e^{2Y}\ ,
    \eeq
    while consistency with the $Y$ equation requires that $p < 5$, and eq.~\eqref{eq:orient_YW} implies that the asymptotic profiles of $Y$ are linear. Consequently, the same is true for $W$, and in fact even for $X$, on account of the limiting form of the equations. This is again a Kasner--like asymptotics.
    \item The case when $X$ and $W$ dominate and differ by a constant leads to no solutions, as we have already saw in the flux-less case of Section~\ref{sec:bulk_tadpole}.
    \item If $X$ and $Y$ are the leading contributions, there is always a Kasner-like behavior, together with one asymptotics as $r \to 0$ when $p=1$ and one as $r \to \infty$ for $p=5$.
    In the last two cases, the limiting form of the solutions approach the BPS $r_1\to 0$ cases of eqs.~\eqref{eq:BPS_near_horizons}.
    The latter limit captures the near--horizon region of the D5 brane, where the string coupling is weak, as can be seen from the behavior of the dilaton in eqs.~\eqref{backgrounds_reduced_betanot0}.
    \item When $W$ is the leading contribution, its asymptotic behavior is linear, $W ~\sim \ w_1 r$, but one must distinguish three cases. If $w_1<0$, the solution approaches a Kasner-like behavior, if $w_1=0$ it approaches case g of section~\ref{sec:dominant_tadpole}, and finally if $w_1>0$ it approaches case f of section~\ref{sec:dominant_tadpole}.
    \item When $Y$ is the leading contribution, there are Kasner-like asymptotics, as above, and also dipole-like asymptotics, for all values of $p$, with
    \beq
    X \ \sim \ 0 \ , \qquad Y \ \sim \  -\log r\ , \qquad W \ \sim \ \frac{5-p}{8}\,\log r \ ,
    \eeq
    as $r\to 0$. This corresponds to the $r= -r_1$ singularities of section~\ref{sec:charged_branes}, as discussed after eqs.~\eqref{backgrounds_reduced_betanot0} and in section~\ref{sec:string_theory_BPS} in connection with orientifolds.
    \item When $X$ is the leading contribution, aside from the usual Kasner-like asymptotics there is only another option, a regular point, as in the flux-free case.
\end{enumerate}
Note that all these orientifold asymptotics, with the exception of the $AdS_3 \times S^7$ case, are those of the tadpole--free theory.
This result is along the lines of what happens for the original nine--dimensional vacuum of~\cite{dm_vacuum}, or for its lower dimensional counterparts of~\cite{ms21_2}.

\subsection{\sc The Ten--Dimensional SO(16) $\times$ SO(16) Model}

For the heterotic SO(16) $\times$ SO(16) model $\gamma=\frac{5}{2}$ and $\beta_p=\frac{3-p}{4}$, with $p=1,5$ so that, using the redefinitions in eqs.~\eqref{eq:shifts}, the equations become
\bea
X''&=&\frac{7-p}{8-p}\ e^{2X}\ - \ e^{2W} \ , \nonumber  \\
Y''&=& 2 \  e^{2Y} \ + \ \frac{7-3p}{4}\ e^{2W} \ ,  \nonumber \\
W''&=& e^{2X}\ + \ \frac{7-3p}{4} \ e^{2Y} \ + \  2 \ e^{2W}  \ ,\\
0 &=&  e^{2X} \, -  \, e^{2W} \, -  \,  e^{2Y} \, + \, \frac{1}{(19-3p)^2 (p+1)} \, \Bigg[32(7-p) (W')^2  \nonumber \\
&-& 3(8-p)(5-p)(1+3p) (X')^2 + 8(8-p)(7-3p) X'Y' + 16(22-3p)(Y')^2 \nonumber \\
&+& 8 W' \left(-8(8-p)X'-(7-p)(7-3p)Y'\right)\Bigg] \ .
\eea

Again, there are in principle eight different options for the limiting behavior of the above system
\begin{enumerate}
    \item In regions where the three functions $X$, $Y$ and $W$ tend to $-\infty$, only the second derivatives are left, and they all approach a linear dependence on $r$, as pertains to Kasner-like solutions.
    \item If the three functions $X$, $Y$ and $W$ differ by a constant, one is led once more to the exact solutions described in Section~\ref{sec:exact_solutions}, and the inequality~\eqref{eq:XYW_condition} selects $p=5$ for the heterotic string. However, the instability implies that this behavior is also fine tuned and not generic.
    \item If $Y$ and $W$ differ by a constant, and both dominate over $X$, for $p=1$ one finds the asymptotic solution
    \beq \label{eq:YW_exact_solution_1}
    X  \ \sim \ \frac{1}{3}\ \log r \ , \qquad Y \ \sim \ - \ \log r \ , \qquad W \ \sim \ - \ \log r \ ,
    \eeq
    around $r \to 0$. Note that this is actually part of an exact solution of the system of eqs.~\eqref{eq:tadpole-curvature-flux_system} in the absence of curvature as in~\cite{ms21_2}, which reads
    \bea
    X &=& \frac{1}{3}\ \log\left(\sqrt{3} \,\rho\, \sinh\frac{r}{\rho} \right) \ \pm \ \frac{8\, r}{\sqrt{63}}\ , \nonumber \\
    Y &=& - \ \log\left(\sqrt{3} \,\rho \,\sinh\frac{r}{\rho}  \right) \ , \nonumber \\
    W &=& - \ \log\left(\sqrt{3} \,\rho \,\sinh\frac{r}{\rho}  \right) \ .
    \eea
For $p=5$, there is another asymptotic solution for which $X$ is the dominant negative contribution, now as $r\to +\infty$, with
   \beq \label{eq:YW_exact_solution_2}
   X\ \sim\  -\ \frac{e^{2 y_1 r}}{4\, y_1^2} \ ,
   \eeq
   with $y_1$ a positive constant.
   This is again part of a special exact solution of the full system in the absence of curvature as in~\cite{ms21_2}, which reads
   \beq
   X \ = \ -\ \frac{ e^{2 y_1 r}}{4 y_1^2}\ + \ \frac{19 y_1 r}{24}\ , \qquad Y \ = \  y_1 r \ , \qquad W  \ = \  y_1 r \ .
   \eeq
   $y_1$ can have any sign in this exact solution, but in the asymptotic regime of interest, when the curvature is present, the sign is fixed to be positive by the dominance condition.
    \item The case when $X$ and $W$ dominate and differ by a constant leads to no solutions, as in the orientifold case.
    \item If $X$ and $Y$ are the leading contributions and differ by a constant, there is always a Kasner-like behavior, together with one asymptotics as $r \to 0$ when $p=5$ and one as $r \to \infty$ for $p=1$.
    In the last two cases, as in the orientifold settings, the limiting forms of the solutions approach the BPS $r_1\to 0$ cases of eqs.~\eqref{eq:BPS_near_horizons}.
    The $p=1$ case captures the near--core region of fundamental strings, where the string coupling is weak, as can be seen from the behavior of the dilaton in eqs.~\eqref{backgrounds_reduced_betanot0}.
     \item When $W$ is the leading contribution, there are Kasner-like asymptotics, together with a limiting behavior of type c, from the classification of section~\ref{sec:dominant_tadpole}, for all the relevant values of $p$, for which
     \beq
     X \ \sim \ \frac{1}{2}\log r \ , \qquad Y \ \sim \ \frac{3p-7}{8}\log r \ , \qquad W \ \sim \ - \ \log r \ ,
     \eeq
     as $r\to 0$.
    \item When $Y$ is the leading contribution, there are Kasner-like asymptotics, as above, and also dipole-like asymptotics, for all values of $p$, with
    \beq
    X \ \sim \ 0 \ , \qquad Y \ \sim \  - \ \log r\ , \qquad W \ \sim \ \frac{3p-7}{8}\,\log r \ ,
    \eeq
    as $r\to 0$. This corresponds to the $r= -r_1$ singularities of section~\ref{sec:charged_branes}, as discussed after eqs.~\eqref{backgrounds_reduced_betanot0}.
    \item When $X$ is the leading contribution, aside from the usual Kasner-like asymptotics there is only another option, a regular point, as in the flux-free case.
\end{enumerate}
Contrary to what we have seen for the orientifolds, here some of the asymptotics are sensitive to the tadpole. This occurs for the $AdS_7 \times S^3$ solution, for case c of Section~\ref{sec:dominant_tadpole}, which is also a lower--dimensional counterpart of the original nine--dimensional solution of~\cite{dm_vacuum} but with an internal sphere, and lastly for the genuinely new case of eq.~\eqref{eq:YW_exact_solution_1}, only when $p=1$.

\section*{\sc Acknowledgments}
\vskip 12pt
We are grateful to Craig Clark, Emilian Dudas and Tom Westerdijk for stimulating discussions. AS was supported in part by Scuola Normale and by INFN (IS GSS-Pi). JM is grateful to Scuola Normale Superiore for the kind hospitality while this work was in progress. AS is grateful to Universit\'e  Paris Cit\'e for the kind hospitality, while this work was in progress.

\newpage
\begin{appendices}

\section[Matched Flux--Free Asymptotics]{\sc Matched Flux--Free Asymptotics}
\label{app:asymptotics}

In this Appendix we summarize the restrictions that the exact linear behavior
\beq
K(r) \ = \ k_1\,r \ + \ k_0
\eeq
places on the possible types of asymptotics for the vacua of Section~\ref{sec:bulk_tadpole}.
    \begin{itemize}
    \item[1. \ ] For $\gamma > \gamma_c$, the convexity of $W-X$, together with the unique value of $k_1$ throughout the range of $r$ imply that if the solution behaves at one end as in case a, it must behave at the other as in case c. Alternatively, if the solution behaves at one end as in case b, with
    \beq
    k_1\ =\ 0 \,, \qquad w_1\,\rho \ =\  -\  \frac{1}{(D-p-3)}\ \sqrt{\frac{(D-p-2)\,\Xi}{16(D-2)}} \ ,
    \label{eq:k1w1_values}
    \eeq
    it is again bound to behave as in case c at the other. There is a third option, however, which we cannot exclude in this fashion: the solution might behave as c at both ends. Note that the $r$ coordinate must terminate at a finite value in case c, and only in this case. Moreover, the asymptotic behavior of type b with $k_1\neq 0$ at one end can only combine with the behavior of type d at the other. Finally, asymptotic behaviors of type d are both ends are not excluded by the present analysis.
     \item[2. \ ] For $\gamma = \gamma_c$, if the solution behaves as in case a at one end, it must behave as in case f at the other, with $k_1=0$ and $w_1>0$. Alternatively, if the solution behaves as in case b at one end, with $k_1$ and $w_1$ as in eq.~\eqref{eq:k1w1_values}, it must behave as in case f at the other, with $k_1=0$ and $w_1>0$. Moreover, if at one end the asymptotic behavior is of type b, with $k_1 \neq 0$, there are several options at the other end, since it can be of types d, f or g, with the same value of $k_1$. Finally, the asymptotic behavior could be of any of the three type d, f, g at both ends.
      \item[3. \ ] For $\gamma < \gamma_c$, the conservation of $k_1$ implies that if the solution behaves as in case a at one end, it must behave as in case e at the other, with
      \beq
       k_1\ =\ 0 \,, \qquad x_1\,\rho \ =\  -\  \frac{1}{\gamma_c^2-\gamma^2}\ \sqrt{\frac{16\,\Xi}{(D-2)^3 (D-p-2)}} \ .
       \label{eq:k1x1_values}
      \eeq
      Alternatively, if the solution behaves as in case b at one end, with $k_1$ and $w_1$ as in eq.~\eqref{eq:k1w1_values}, it must again behave as in case c at the other, with $k_1$ and $x_1$ as in eq.~\eqref{eq:k1x1_values}. Moreover, the asymptotic behavior of type b with $k_1\neq 0$ can only combine with case e, with the same value of $k_1$, at the other end. Finally, asymptotic behaviors of type e at both ends are not excluded by the present analysis.
\end{itemize}

We have thus identified a limited number of options for the combined asymptotics at the two ends:
\begin{itemize}
    \item $\gamma>\gamma_c$
    \beq
    k_1 = 0 \ \text{ and } \ \left\{  \quad \text{\underline{a}} \quad ; \quad \text{\underline{b} \ with \ eq.~\eqref{eq:k1w1_values}} \quad ; \quad \text{c} \quad \right\} \nonumber
    \eeq
    \beq
    k_1\neq0 \ \text{ and } \ \left\{ \quad \text{\underline{b}} \quad ; \quad \text{d}  \quad \right\} \nonumber
    \eeq
    \item $\gamma=\gamma_c$
    \beq
     k_1 = 0 \ \text{ and } \ \left\{  \quad \text{\underline{a}} \quad ; \quad \text{\underline{b} \ with \  eq.~\eqref{eq:k1w1_values}} \quad ; \quad \text{f \ with \ }  w_1 >0 \quad  \right\} \nonumber
    \eeq
    \beq
    k_1\neq0 \ \text{ and } \ \left\{ \quad \text{\underline{b}} \quad ; \quad \text{d}  \quad ; \quad \text{f} \quad ; \quad \text{g}  \quad\right\} \nonumber
    \eeq
    \item $\gamma<\gamma_c$
    \beq
    k_1 = 0 \ \text{ and } \ \left\{  \quad \text{\underline{a}} \quad ; \quad \text{\underline{b} \ with \ eq.~\eqref{eq:k1w1_values}} \quad ; \quad\text{\underline{e} \ with \ eq.~\eqref{eq:k1x1_values}} \quad \right\} \nonumber
    \eeq
    \beq
    k_1\neq0 \ \text{ and } \ \left\{ \quad \text{\underline{b}} \quad ; \quad \text{e}  \quad \right\} \nonumber
    \eeq
\end{itemize}
Within each of the six groups above, one can combine any pairs of asymptotics, with the condition that the underlined options be taken at most once. For example, in the first group, which refers to $\gamma>\gamma_c$ and $k_1=0$, there are only three possible pairs:
$(\text{{a}},\text{{c}})$, $(\text{{b}},\text{{c}})$, $(\text{{c}},\text{{c}})$.

\newpage

\end{appendices}

\end{document}